\documentclass[a4paper,fleqn,usenatbib]{mnras}

\usepackage{newtxtext,newtxmath}

\usepackage[T1]{fontenc}
\usepackage{ae,aecompl}

\usepackage{bm}

\usepackage{mleftright}
\let\left\mleft
\let\right\mright

\usepackage{amsmath}

\usepackage{graphicx}	
\usepackage{subcaption}
\usepackage{siunitx}
\usepackage{mwe}

\renewcommand{\arraystretch}{1.1}

\newcommand{\Div}{\nabla \cdot }
\newcommand{\Lap}{\nabla^2 }

\graphicspath{{./Figures/}}

\title[Magneto-Thermal Instability In Galaxy Clusters]{Magneto-Thermal Instability In Galaxy Clusters II: Three-Dimensional Simulations}

\author[L. M. Perrone et al.]{
Lorenzo M. Perrone,$^{1}$\thanks{E-mail: lmp61@cam.ac.uk}
Henrik Latter$^{1}$
\\
$^{1}$Department of Applied Maths and Theoretical Physics, University of Cambridge, UK\\
}

\date{Accepted XXX. Received YYY; in original form ZZZ}

\pubyear{2020}

\begin{document}
\label{firstpage}
\pagerange{\pageref{firstpage}--\pageref{lastpage}}
\maketitle

\begin{abstract}
	In the intracluster medium (ICM) of galaxies, exchanges of heat across magnetic field lines are strongly suppressed. This anisotropic heat conduction, in the presence of a large-scale temperature gradient, destabilizes the outskirts of galaxy clusters via the magneto-thermal instability (MTI), and thus supplies a source of observed ICM turbulence. In this paper we continue our investigation of the MTI with 3D simulations using the Boussinesq code SNOOPY. We redress two issues  intrinsic to our previous 2D study: an inverse energy cascade and the impossibility of dynamo action. Contrary to 2D simulations, we find inconsequential transport of energy across scales (most energy is dissipated at the same scale as its injection), and that turbulent eddies are vertically elongated at or below the thermal conduction length, but relatively isotropic on larger scales. 
    Similar to 2D, however, the saturated turbulent energy levels and the integral scale follow clear power-laws that depend on the thermal diffusivity, temperature gradient, and buoyancy frequency.
	We also show that the MTI amplifies magnetic fields, through a fluctuation dynamo, to equipartition strengths provided that the integral scale of MTI turbulence is larger than the viscous dissipation scale. Finally, we show that our scaling laws are consistent with extant observations of ICM turbulence if the thermal conductivity is reduced by a factor of $\sim 10$ from its Spitzer value, and that on global cluster scales the stable stratification significantly reduces the vertical elongation of MTI motions.
\end{abstract}


\begin{keywords}
galaxies: clusters: intracluster medium -- plasmas -- instabilities -- turbulence
\end{keywords}



\section{Introduction}

Galaxy clusters are filled with hot and rarefied plasma, the intra-cluster medium (ICM), which sustains significant levels of subsonic turbulence \citep{Schuecker2004,Churazov2012,Zhuravleva2014b}. Turbulent motions contribute to the transport of heat and momentum, and hence can control the large-scale structure of clusters. \citep{Narayan2001,Zakamska2003,Voigt2004,Zhuravleva2019}. It has long been known that the ICM is threaded by magnetic fields, as measured by various techniques, such as synchrotron emission  \citep[e.g.,][]{Giovannini1993}, inverse Compton X-ray emission \citep{Fusco-Femiano2004}, and Faraday rotation \citep{Vogt2005,Bonafede2010}. The field strength is typically of the order of $\sim 0.1-1 \si{\mu G}$ up to distances of $1\si{Mpc}$ from the cluster core, with coherence scales of $\sim 10 \si{kpc}$ \citep{Carilli2002}. While the magnetic pressure is subdominant to thermal pressure, with typical values of the plasma beta $\sim 100$ \citep{Vogt2005},
particle collisions are rare and the plasma is strongly magnetized \citep{Sarazin1988}. As a consequence, the ICM is highly anisotropic with respect to the local direction of the magnetic field: in particular, transport of heat and momentum perpendicular to field lines is greatly reduced \citep{Braginskii1965a}.

This anisotropy transforms not only transport but also the large-scale structure of the ICM by triggering new buoyancy instabilities. On account of the ICM's thermally stratification -- as shown by \textit{XMM-Newton} and \textit{Chandra} \citep[see, e.g.,][]{Vikhlinin2005,Leccardi2008,Ghirardini2019} --  the periphery of galaxy clusters should be unstable to the \textit{magneto-thermal} instability (MTI) \citep{Balbus2000,Balbus2001}, which could provide one source of turbulence on intermediate scales in the ICM \citep{Parrish2008,McCourt2011}.
Other competing sources of turbulence may include mergers \citep{ZuHone2011}, cool-core sloshing \citep{ZuHone2010,Roediger2012} and AGN feedback \citep{Churazov2002,McNamara2007,Fabian2012}.

In Perrone \& Latter (2021; hereafter "Paper 1"), we re-examined the MTI theoretically and numerically, focusing on the asymptotic regime of weak magnetic fields and low viscosity and resistivity. Using the Boussinesq code SNOOPY, modified to include anisotropic heat conduction, we performed an extensive set of simulations of the MTI in 2D, finding that its saturation exhibits a conspicuous separation between small scales, into which the MTI injects energy, and large-scales, where this energy travels via a 2D inverse cascade and subsequently excites large-scale g-modes. Mean turbulent properties follow clear power law scalings, the most important of which is the turbulent kinetic energy, which is $\sim \chi \omega_T^3 / N^2$, where $\omega_T$ is the maximum MTI growth rate, $\chi$ is thermal diffusivity, and $N$ is the 
Brunt-V\"ais\"al\"a frequency. In addition, the larger the $\chi$, the larger the characteristic turbulent length scales. The numerically derived scalings are in agreement with theoretical predictions based on a simple "Epstein drag" argument, by which g-modes on large scales act as a sink of energy.

However, the numerical simulations in Paper 1 are somewhat compromised by the peculiar properties of 2D turbulence, namely the inverse cascade of energy \citep{Kraichnan1967,Batchelor1969}, and the absence of a magnetic dynamo \citep{Zeldovich1957}. Both may distort the true dynamics of the MTI at saturation: the inverse cascade accumulates energy at large scales, which may not be the case in 3D, whereas a dynamo can produce stronger magnetic fields at small scales that could hinder the MTI. In this work, we perform a systematic numerical study of the 3D MTI in order to address these questions and to better gauge the applicability of the MTI to the ICM.

\subsection{The MTI and the fluctuation dynamo}\label{sec:mti_smallscale_dynamo}

In galaxy clusters the observed magnetic fields are dynamically important, possessing energies close to equipartition with those of the turbulent motions \citep{Govoni2004}. These observations, as well as the absence of a large-scale mean field, suggest that ICM magnetic fields are internally generated, probably by a fluctuation (or small-scale) dynamo \citep[see, e.g.,][]{Schekochihin2004}. 

Broadly speaking, there are two classes of fluctuation dynamo, and these appear in the asymptotic limits of either high or low magnetic Prandtl number $\text{Pm} = \nu / \eta$ (where $\nu$ and $\eta$ are the viscosity and magnetic diffusivity, respectively). The two dynamos exhibit distinct features owing to their contrasting hierarchies of viscous ($l_{\nu}$) and resistive scales ($l_{\eta}$). If $\text{Pm} \gg 1$, then $l_{\nu} \gg l_{\eta}$ and the magnetic field is randomly stretched by the fast turbulent eddies at the viscous scale. In the limit of $\text{Pm} \ll 1$, on the other hand, the resistive scale lies between the scale of energy injection ($L$) and viscous dissipation, and thus the magnetic field grows on the inertial-range. Because of the computational challenge in resolving $L \gg l_{\eta} \gg l_{\nu}$ simultaneously, the low-$\text{Pm}$ dynamo has been demonstrated only recently, and determining the exact physical mechanism for growth is a matter of ongoing research \citep{Iskakov2007,Schekochihin2007,Brandenburg2011,Malyshkin2010,Schober2012,Brandenburg2018,Bott2020}.

Previous numerical simulations of the MTI in 3D showed that it is capable of generating equipartition magnetic fields via dynamo action \citep{Parrish2007,Parrish2008,McCourt2011}. Its properties, however, have not been studied in any depth. Since thermal diffusivity and stratification control the MTI turbulent saturation (as shown in Paper 1), we expect that they will also determine its associated fluctuation dynamo, with the (ordinary) Prandtl number $\text{Pr}=\nu / \chi$ and $N$ featuring prominently alongside $\text{Pm}$.  We also anticipate the strong fields generated (and the resulting magnetic tension) to reshape the saturated state of the MTI itself. This is in contrast to our 2D weak-field simulations in Paper I, where the magnetic field was limited to conducting heat, and little else.

\subsection{The effect of micro-instabilities}

In galaxy clusters, $\text{Pr}$ and $\text{Pm}$ may be estimated using the Spitzer values for collisional transport \citep{Spitzer1962}: the Prandtl number is fixed and roughly given by the square root of the electron to ion mass ratio, $\text{Pr} \approx 0.02$ \citep{Spitzer1962,Braginskii1965a}, which means that thermal conduction is significantly faster than the viscous transport of momentum. Meanwhile, the magnetic Prandtl number can reach enormous values, $\text{Pm} \sim 10^{29}$ \citep{Schekochihin2005}.

Observations of the ICM, however, suggest that small-scale diffusive processes depart from the Spitzer picture, owing to 
the presence of fast kinetic instabilities working on scales smaller than the particle mean free path. These are excited by the pressure anisotropy between directions parallel and perpendicular to the local magnetic field \citep{Schekochihin2006,Schekochihin2008}. Numerical \citep{Kunz2014,Riquelme2016,Melville2016} and observational evidence from the solar wind (similar in many respects to the ICM) \citep{Bale2009,Chen2016}  indicate that these micro-instabilities act in such a way as to counter the pressure anisotropy and to enhance the scattering of charged particles, leading to a suppression of the effective viscosity and thermal conductivity
\citep{Bale2013,Roberg2016,Komarov2018,Roberg2018,Drake2020,Berlok2021}.
A quantitative prediction for the effect of the microinstabilities is currently unavailable and is the subject of intense theoretical and numerical investigation. 
But the combined effect of suppressed viscosity and thermal conductivity will surely transform the values of the Prandtl and magnetic Prandtl numbers quoted above. Nevertheless, the disparity of scales between the viscous and the resistive scale inferred from observations supports the idea that $\text{Pm} \gtrsim 1$ remains the appropriate regime for the ICM \citep{Schekochihin2005}, and it is expected that the effective $\text{Pr}$ will still be less than unity \citep{Zhuravleva2019}.

In the light of these theoretical and observational arguments, we construct simulations adopting parameters $\text{Pm} \gtrsim 1 $, $\text{Pr} \lesssim 1$. This means that to investigate MTI-driven dynamos it is necessary to resolve three different scale separations $L \gtrsim l_{\chi} \gtrsim l_{\nu} \gtrsim l_{\eta}$, where $L$ is now the box size and $l_{\chi} = \sqrt{\chi/\omega_T}$ is a typical conduction length. This is one more separation than in standard simulations of high-$\text{Pm}$ dynamos with forced turbulence, and thus poses a serious computational challenge; it certainly limits the parameter range that we can explore.

\subsection{Summary and outline of the paper}

The aim of this work is to perform an in-depth study of the MTI in three dimensions in the Boussinesq approximation. We focus on the similarities and differences between 3D and 2D, the latter covered in Paper I. Furthermore, we look at the emergence of the MTI's fluctuation dynamo and its effect on the properties of the saturated state. We finally attempt to confront our numerical results with ICM observations of turbulence and magnetic fields, while fitting our approach into previous and current theoretical work. 

The outline of the paper is as follows. In Section~\ref{sec:equations_numerical} we describe our model equations, as well as the numerical methods and diagnostics adopted. Our results are presented in Section~\ref{sec:results}: in Sections \ref{sec:fiducial_dynamo}-\ref{sec:non-dynamo-runs} we analyse the properties of fiducial dynamo and non-dynamo runs; Section~\ref{sec:scaling_laws} shows the results of a large parameter sweep, which determine scaling laws for the mean turbulent properties and the criterion for dynamo action. Finally, we discuss the relevance of this work for the real ICM and summarize our results in Sections~\ref{sec:applications}-\ref{sec:conclusions}.

\section{Equations and numerical setup}\label{sec:equations_numerical}

\subsection{Boussinesq model}

The subsonic and small-scale dynamics of the ICM at a given spherical radius $R_0$ can be described by the Boussinesq approximation, in which we assume the presence of a large-scale radial gradient of both entropy and temperature, and include anisotropic thermal conduction. The model represents a small block of cluster gas in Cartesian geometry, with $z$ pointing in the positive radial direction (see Appendix A in Paper 1). The governing equations are
\begin{gather}
	\Div \bm u = \Div \bm B = 0 ,\label{eq:div_eq} \\
	\left( \partial_t + \bm u \cdot \nabla \right) \bm u = - \frac{\nabla p_{tot}}{\rho_0} - \theta \bm e_z + \left( \bm B \cdot \nabla \right) \bm B + \nu \Lap \bm u , \label{eq:mom_eq} \\
	\left( \partial_t + \bm u \cdot \nabla \right) \bm B = \left( \bm B \cdot \nabla \right) \bm u + \eta \Lap \bm B, \label{eq:bfield_eq} \\
	\left( \partial_t + \bm u \cdot \nabla \right) \theta = N^2 u_z + \chi \Div \left[ \bm b \left( \bm b \cdot \nabla \right)\theta \right]	+ \chi \omega_T^2 \Div \left( \bm b b_z\right). \label{eq:buoyancy_eq}
\end{gather}
 Here $\rho_0$, $p_0$, and $T_0$ denote the background density, pressure, and temperature of the cluster at the reference point $R_0$. In Eq.~\eqref{eq:div_eq}--\eqref{eq:buoyancy_eq}, $\theta \equiv g_0 \rho'/\rho_0$ is the buoyancy variable -- proportional to the density fluctuation $\rho'$ -- with $g_0$ the (constant) gravitational acceleration, and $p_{tot}$ is the sum of thermal and magnetic pressures. Finally, $N^2$ and $\omega_T^2$ are the squared Brunt-V\"ais\"al\"a and MTI frequency, respectively, defined through
\begin{align}
	N^2 = \frac{g_0}{\gamma} \left. \frac{\partial \ln p \rho^{-\gamma}  }{\partial R} \right|_0, 
	\qquad \omega_T^2 = - g_0  \left. \frac{\partial \ln T}{\partial R} \right|_0 .
\end{align}
We have rescaled the magnetic field so that it has the dimensions of a velocity, and the thermal diffusivity $\chi$ (in units of [$\si{cm^2.s^{-1}}$]) absorbs a factor of $(\gamma -1)/\gamma$. 
Analogously to Paper I, we decide to neglect anisotropic (Braginskii) viscosity in Eqs.~\eqref{eq:div_eq}--\eqref{eq:buoyancy_eq} and instead use isotropic viscosity. Our choice is driven by both practical considerations, as we wish to focus (initially) on the essential features of the MTI, free of additional complications, and by previous work \citep[e.g.][]{Kunz2011,Parrish2012a,Kunz2012} that suggested Braginskii viscosity only weakly affects the properties of MTI turbulence at saturation. For the derivation of Eqs.~\eqref{eq:div_eq}--\eqref{eq:buoyancy_eq} and further discussion, please refer to Paper 1.  

\subsection{Numerical methods}

In this section we describe the numerical methods that we have used and outline our diagnostics. In addition to the pseudospectral code SNOOPY employed in Paper I -- to which we refer the reader for additional information, giving  below only a brief description --  we also ran a number of MTI simulations using PLUTO \citep{Mignone2007}, a finite-volume, Godunov code in order to compare our results. The details of the MTI simulations with PLUTO and their numerical setup can be found in Appendix \ref{app:PLUTO_runs}.

\subsubsection{\textsc{SNOOPY} code}

Our simulations were performed with the latest version of SNOOPY \citep{Lesur2015}, a pseudo-spectral 3D code suitable for incompressible MHD simulations, which we have modified so as to include anisotropic heat conduction both with explicit integration at each timestep, and with a super time-stepping (STS) algorithm \citep{Alexiades1996}. In the SNOOPY code, a Fourier transform is applied to the fluid equations so as to go from position space to Fourier space, and the Fourier coefficients are then advanced in time with a three-step Runge-Kutta algorithm. 

\subsubsection{Initial conditions and parameters}

We adopt units so that $\omega_T=1$ and $L=1$, where $L$ is the box size. The main physical parameters that characterize our simulations are the Peclet number, defined as $\text{Pe}=L^2 \omega_T / \chi$, the reduced squared Brunt-V\"{a}is\"{a}l\"{a} frequency  $\tilde{ N}^2 = N^2 / \omega_T^2$, the Prandtl number $\text{Pr} = \nu / \chi$, and the Roberts number $q=\eta / \chi$ (although we will often refer to the magnetic Prandtl number $\text{Pm} = \nu/\eta = \text{Pr}/q$ in our discussion). Finally, we vary the strength of the imposed horizontal magnetic field, $B_0$.

The majority of the runs are initialized with random white noise at all wavenumbers in the velocity field, with amplitude $\lvert \delta u_i \rvert  = 10^{-4}$ for $i=x,y,z$, and with a uniform horizontal magnetic field of strength $B_0 = 10^{-5}$ aligned in the $x$ direction. This ensures that, during the exponential phase of the instability, magnetic fields are dynamically unimportant, but appear indirectly through the anisotropic heat conduction term. 

We performed most simulations with either $(256)^3$ or $(512)^3$ Fourier modes in 3D, in triply-periodic domains with an aspect ratio of $1:1:1$. As we explore relatively large values of thermal conductivity (low $\text{Pe}$), $(512)^3$ simulations are rather expensive to run, and this inevitably limits the parameter range that we can explore. A list of all the runs undertaken, together with their physical and numerical parameters, can be found in Table~\ref{tab:table_runs}.

\subsection{Diagnostics}

\subsubsection{Energetics}

We follow the time evolution of the system's kinetic, magnetic, and potential energy, defined by $K = \langle u^2 \rangle /2$, $M = \langle B^2 \rangle /2$, and $U = \langle \theta^2  \rangle /2 N^2$, respectively, where the angle brackets denote a volume average (see Paper I for the full set of equations). 

The total energy $E_{tot} = K + M + U$ obeys:
\begin{align}
	\frac{d }{dt}E_{tot} = -\epsilon_{\nu} - \epsilon_{\eta} - \epsilon_{\chi}  + \epsilon_I , \label{eq:energy_balance}
\end{align}
where the first three terms on the right-hand side are viscous,  Ohmic,  and thermal dissipation, respectively, while the last term accounts for the MTI forcing. Their mathematical definitions are
\begin{align}
	\epsilon_{\nu} &= \nu \langle \lvert \nabla \bm{u} \rvert^2 \rangle, \quad \epsilon_{\eta} = \eta \langle \lvert \nabla \bm{B} \rvert^2 \rangle, \label{eq:dissipation_injection_defs_1}\\
	\epsilon_{\chi} &= \frac{\chi}{N^2} \langle \left| \bm b \cdot \nabla \theta \right|^2 \rangle, \quad \epsilon_I = -  \frac{\chi \omega_T^2}{N^2}  \langle  b_z \bm b   \cdot \nabla \theta \rangle \label{eq:dissipation_injection_defs_2}.
\end{align}

The box-averaged advected and conductive heat fluxes are defined through
\begin{align}
	Q_\text{adv}  &= -\langle \theta u_z \rangle, \\
	Q_\text{cond} &= \chi \langle  b_z \bm b \cdot \nabla \theta \rangle + \chi \omega_T^2  \langle b_z^2 \rangle,
\end{align}
and we define the total turbulent Nusselt number as the sum of the advective and conductive Nusselt numbers,
\begin{equation}\label{eq:nusselt_num}
	\text{Nu} = \text{Nu}_\text{adv} + \text{Nu}_\text{cond} = \frac{Q_\text{adv} + Q_\text{cond}}{Q_{0}}.
\end{equation}
The fluxes have been scaled by $Q_0= \chi \omega_T^2$, which is the amount of heat that would be carried through the box if there was a purely vertical magnetic field or, equivalently, if conduction was isotropic.

\subsubsection{Fourier spectra}

Alongside the evolution equations for the volume-averaged energies, we derive equations for the energy contained in each Fourier mode $\bm k = k_x \bm{e}_x + k_y \bm{e}_y + k_z \bm{e}_z$:
\begin{align}
	\frac{1}{2}\frac{d \lvert \hat{u} (\bm k, t) \rvert^2}{d t} &= \hat{T}_{\bm u}^{\bm u}  - \hat{\Theta}  - \hat{L}  - \nu \hat{D}_{\nu}  ,  \label{eq:spectral_vel}\\
	\frac{1}{2}\frac{d \lvert \hat{B} (\bm k, t) \rvert^2}{d t} &= \hat{T}_{\bm B}^{\bm u}  + \hat{L}  - \eta \hat{D}_{\eta}  ,  \label{eq:spectral_mag}\\ 
	\frac{1}{2}\frac{d \lvert \hat{\theta} (\bm k, t) \rvert^2}{d t} &= \hat{T}_{\theta}^{\bm u} + N^2 \hat{\Theta}  - \chi \hat{D}_{\chi}  + \chi \omega_T^2 \hat{\mathcal{F}} , \label{eq:spectral_temp}
\end{align}	
where $ \hat{T}_{\bm q}^{\bm u}$indicates the nonlinear transfer term of quantity $\bm q $ by the velocity field, $\hat{\Theta} $ is the buoyancy force, $\hat{L}$ is the Lorentz force, $\hat{\mathcal{F}}$ is the MTI forcing term, and finally $\hat{D}_{\nu}, \hat{D}_{\chi}$ and $\hat{D}_{\eta} $ are the diffusive sinks due to viscosity, conductivity and resistivity, respectively. Their definitions can be found in Appendix ~A in Paper 1. 

In all the simulations performed we track the evolution of the "shell-integrated" spectra (by summing over the modes with the same $k= \lvert \bm k \rvert$). For instance, for the kinetic spectral energy density $E_K(k)$ we have
\begin{equation}
	E_K(k,t) dk = \sum_{k \leq \lvert \bm k' \rvert < k + dk} \frac{1}{2} \lvert \hat{u} (\bm k', t) \rvert^2 ,
\end{equation}
and so forth for the other potential and magnetic energy densities $E_\theta$, and $E_M$.

One problem with the one-dimensional spectral densities is that the shell-averaging destroys information on the energy's distribution over wavevector orientation. While not an issue in isotropic turbulence, in stably-stratified runs the energy is usually distributed unequally between the vertical and horizontal modes. As we shall see, this is also the case in 3D MTI turbulence. We thus supplement our analysis with so-called \textit{directional} energy spectra, where the spherical shell of radius $k$ is divided into latitude bands starting from the equator and moving to the poles, with the averaging performed over each band individually. This approach has been recently employed in simulations of forced stably-stratified \citep{Lang2019,Skoutnev2021} and rotating turbulence \citep{Delache2014}. 

To compute directional spectra, we follow the approach of \cite{Lang2019} and divide the spherical shell into $6$ latitude bands of equal angular width, with the $i$-th band delimited by $|\theta| \in \left[  \theta_{i-1}, \theta_i \right) $ for $i=1,...,6$, where $\theta = \arcsin(k_z/k) $, and $\theta_i = i \times \pi / 12$. With this notation, the latitudinal bands go progressively from near-equatorial (for $m=1$), to near-polar ($m=6$), the latter including the mode $\bm{k} = k_z \bm{e}_z$. If we define the weights $m_i = 3 / \left[ \sin (\theta_i) - \sin (\theta_{i-1}) \right] $, then the average kinetic energy contained in the $i$-th latitudinal band is
\begin{align}
	E_{K,i} (k,t) dk = \sum_{|\theta| \in \left[  \theta_{i-1}, \theta_i \right)} \frac{1}{2m_i} \lvert \hat{u} (\bm k', t) \rvert^2 ,
\end{align}
with the total energy spectrum recovered from $E_K = \sum_i m_i 	E_{K,i} $. 

\subsubsection{Scales and dimensionless numbers}\label{sec:scales_numbers}

To characterize the turbulent state of the MTI, it is helpful to define the following scales and dimensionless numbers:
\begin{enumerate}
	\item the MTI integral scale, $k_i$, which is the scale where the kinetic power spectrum peaks, and -- contrary to simulations of forced turbulence -- is set self-consistently by the MTI:
	\begin{equation}\label{eq:buoy_scale}
		k_{i}^{-1} = \frac{\int k^{-1} E_k d k}{\int  E_k d k}.
	\end{equation}
	Correspondingly, we also introduce the Reynolds number at the integral scale $\text{Re}_i$, defined as $\text{Re}_i = \sqrt{2 K} / \nu k_i $;
	\item the forcing scale, $k_f$, where most of the energy is injected by the MTI; it corresponds to the wavenumber at which the net input energy flux $\chi \omega_T^2 \mathcal{\hat{F}} - \chi \hat{D}_{\chi}$ is largest;
	\item the Ozmidov scale, $k_{Oz} = (N^3 / \epsilon_{\nu})^{1/2}$, where the eddy turnover frequency is comparable to the Brunt-V\"ais\"al\"a frequency; classically it signals the transition between stratified and isotropic turbulence;
	\item the average correlation wavenumber of the buoyancy fluctuations along the magnetic field line, $k_{\parallel}^{\theta}$, defined through
	\begin{align}\label{eq:kpartheta}
		k_{\parallel}^{\theta} = \left( \frac{\langle |\bm b \cdot \nabla \theta|^2  \rangle }{\langle \theta^2 \rangle} \right)^{1/2};
	\end{align}
	it represents the largest distance at which two fluid elements are thermally connected;
	\item the viscous $k_{\nu}$ and resistive $k_{\eta}$ wavenumbers which we estimate assuming Kolmogorov scalings and taking $k_i$ to be the outer scale for the turbulent cascade; thus $ k_{\nu} = l_{\nu}^{-1} \sim \text{Re}_i^{3/4} k_i$ and $k_{\eta} = l_{\eta}^{-1} \sim \text{Pm}^{1/2} k_{\nu}$, and for large $\text{Pm}$ we have $k_{\nu} \ll k_{\eta}$.
\end{enumerate}

For large $\text{Re}_i$, the integral scale is longer than the viscous scale, while we find $k_f$ is only somewhat larger than $k_i$ (contrary to the 2D cases examined in Paper I). It is more difficult to estimate $k_{\parallel}^{\theta}$ \textit{a priori}, but it is reasonable to expect that, for turbulence driven on a timescale of $1/\omega_T$, the parallel coherence length of buoyancy is $\sim l_{\chi}$. Finally, based on the analogy with stably-stratified turbulence, we expect the scales above the Ozmidov scale to be dominated by stratification, and thus form a very long-scale regime, $L^{-1} \ll k \lesssim k_{Oz} \lesssim k_i$, if the box size $L$ is sufficiently large.

Finally, following \cite{Schekochihin2004}, we compute various magnetic scales in order to elucidate the fluctuation dynamo. We measure the average scale of variation of the magnetic field along ($k_{\parallel}$) and across ($k_{\bm B \times \bm J} $) itself, the variation of the field in the direction orthogonal to both $\bm B$ and $\bm B \times \bm J$ (denoted $k_{\bm B \cdot  \bm J}$), and the overall scale of variation of $\bm B$ (denoted $k_{rms}$), defined through
\begin{align}
	&k_{\parallel} = \left(  \frac{ \langle|\bm B \cdot \nabla \bm B |^2 \rangle}{\langle B^4 \rangle}\right)^{1/2}, \quad k_{\bm B \times \bm J} = \left(  \frac{ \langle|\bm B \times  \bm J |^2 \rangle}{\langle B^4 \rangle}\right)^{1/2}, \\
	&k_{\bm B \cdot  \bm J} = \left(  \frac{ \langle |\bm B \cdot  \bm J |^2 \rangle}{\langle B^4 \rangle}\right)^{1/2}, \quad k_{rms} = \left(  \frac{ \langle |\nabla \bm B  |^2 \rangle}{\langle B^2 \rangle}\right)^{1/2}.
\end{align}
In large $\text{Pm}$ fluctuation dynamos, the parallel scale of variation is generally the largest and is set by the viscous-scale motions \citep{Schekochihin2004}, while the overall variation of the field is comparable to the small resistive scale. A typical ordering of these wavenumbers is $k_{\parallel} \lesssim k_{\bm B \cdot  \bm J} \ll k_{\bm B \times \bm J} \sim k_{rms} $ for magnetic fields arranged in folded sheets, or $k_{\parallel} \ll k_{\bm B \cdot  \bm J} \lesssim k_{\bm B \times \bm J} \sim k_{rms} $ for folded ribbons \citep{StOnge2020}.

\section{Results}\label{sec:results}

\begin{table*}  
	\setlength{\tabcolsep}{0.35em} 
	\centering 
	\caption{Parameters and mean outputs from 3D simulations. Mean quantities have been averaged over $100 \omega_T$, for a total of $10^3$ samples. All runs have been initialized with a horizontal uniform magnetic field of strength $B_0 = 10^{-5}$, except for runs D0Bx1e-2, D0Bz1e-2, which have an initial field of strength $B_0 = 10^{-2}$ aligned in the $x$ and $z$ direction, respectively. The resolution is $(512)^3$ and the numerical domain is cubic. With the exception of run D0NoDyn, which has $\text{Pm} = 0.4$, all simulations have $\text{Pm} = 10$. Finally, in the column "dyn?" we indicate the presence or not of a dynamo. The dagger symbol, $^{\dagger}$, indicates the run has not reached saturation. For compactness we use the bracket notation to indicate uncertainties, with the error appearing at the last significant digit: e.g. $ 5.1(1) = 5.1 \pm 0.1$.}  
	\label{tab:table_runs} 
	\begin{tabular}{cccccccccccccc}  
		\hline 
		Run 	 &  $\text{Pe}$ 	 &  $\tilde{N}^2$ 	 &  $\text{Pr}$ 	 &  $K(\times 10^{4})$ 	 &  $M$ 	 &  $U(\times 10^{3})$ 	 & dyn? &  $ \langle | b_z | \rangle$ 	 &  $\phi$ 	 &  $\delta_K$ 	 &  $\delta_M$ 	 &  $\text{Nu}_\text{cond}$ 	 &  $\text{Nu}_\text{adv}$ 	  \\ 
		\hline 
		RsPe00 	    &  $ 6 \times 10^{2} $ 	   &  $ 0.10 $ 	    &  $ 0.012 $ 	    &  $ 5.1(1)  $ 	    &  $ 3.69(4) \times 10^{-4} $ 	    &  $ 1.63(1)  $ 	    & y &  $ 0.764(3) $ 	    &  $ 0.791(3) $ 	    &  $ 8.8(3) $ 	    &  $ 7.9(2) $ 	    &  $ 0.409(4) $ 	    &  $ 0.186(3) $ 	    \\ 
		RsPe01 	    &  $ 1 \times 10^{3} $ 	   &  $ 0.10 $ 	    &  $ 0.02 $ 	    &  $ 3.14(6) $ 	    &  $ 2.24(3) \times 10^{-4} $ 	    &  $ 1.15(1)  $ 	    & y &  $ 0.771(2) $ 	    &  $ 0.792(3) $ 	    &  $ 9.1(2) $ 	    &  $ 8.4(2) $ 	    &  $ 0.420(3) $ 	    &  $ 0.203(3) $ 	    \\ 
		RsPe02/D0 	    &  $ 3 \times 10^{3} $ 	   &  $ 0.10 $ 	    &  $ 0.06 $ 	    &  $ 1.10(1)  $ 	    &  $ 5.82(6) \times 10^{-5} $ 	    &  $ 0.507(2)  $ 	    & y  &  $ 0.759(1) $ 	    &  $ 0.784(1) $ 	    &  $ 8.4(1) $ 	    &  $ 7.43(9) $ 	    &  $ 0.412(2) $ 	    &  $ 0.232(2) $ 	    \\ 
		RsPe03 	    &  $ 6 \times 10^{3} $ 	   &  $ 0.10 $ 	    &  $ 0.12 $ 	    &  $ 0.544(3)  $ 	    &  $ 1.93(1) \times 10^{-5} $ 	    &  $ 0.2960(7)  $ 	    & y &  $ 0.749(1) $ 	    &  $ 0.787(4) $ 	    &  $ 7.9(1) $ 	    &  $ 6.67(5) $ 	    &  $ 0.408(1) $ 	    &  $ 0.244(1) $ 	    \\ 
		\hline 
		RsN200 	    &  $ 3 \times 10^{3} $ 	   &  $ 0.02 $ 	    &  $ 0.06 $ 	    &  $ 3.90(8)  $ 	    &  $ 2.43(2) \times 10^{-4} $ 	    &  $ 4.06(4)  $ 	    & y &  $ 0.761(2) $ 	    &  $ 0.789(5) $ 	    &  $ 9.5(3) $ 	    &  $ 7.4(2) $ 	    &  $ 0.456(3) $ 	    &  $ 0.512(6) $ 	    \\ 
		RsN201 	    &  $ 3 \times 10^{3} $ 	   &  $ 0.05 $ 	    &  $ 0.06 $ 	    &  $ 1.98(1)  $ 	    &  $ 1.20(1) \times 10^{-4} $ 	    &  $ 1.284(3)  $ 	    & y &  $ 0.7664(9) $ 	    &  $ 0.790(3) $ 	    &  $ 9.1(2) $ 	    &  $ 7.86(7) $ 	    &  $ 0.440(1) $ 	    &  $ 0.342(1) $ 	    \\ 
		RsN202 	    &  $ 3 \times 10^{3} $ 	   &  $ 0.20 $ 	    &  $ 0.06 $ 	    &  $ 0.562(3)  $ 	    &  $ 2.27(2) \times 10^{-5} $ 	    &  $ 0.1883(6)  $ 	    & y  &  $ 0.752(1) $ 	    &  $ 0.790(3) $ 	    &  $ 7.76(9) $ 	    &  $ 6.93(6) $ 	    &  $ 0.388(2) $ 	    &  $ 0.1460(6) $ 	    \\ 
		RsN203 	    &  $ 3 \times 10^{3} $ 	   &  $ 0.40 $ 	    &  $ 0.06 $ 	    &  $ 0.274(1)  $ 	    &  $ 4.05(3) \times 10^{-6} $ 	    &  $ 0.0656(1)  $ 	    & y &  $ 0.7339(6) $ 	    &  $ 0.789(3) $ 	    &  $ 6.59(5) $ 	    &  $ 5.78(3) $ 	    &  $ 0.3547(8) $ 	    &  $ 0.0856(2) $ 	    \\ 
		RsN204 	    &  $ 3 \times 10^{3} $ 	   &  $ 0.60 $ 	    &  $ 0.06 $ 	    &  $ 0.1718(5) $ 	    &  $ 1.13(4) \times 10^{-7} $ 	    &  $ 0.0354(1)  $ 	    & n &  $ 0.7328(5) $ 	    &  $ 0.787(3) $ 	    &  $ 6.19(4) $ 	    &  $ 5.40(3) $ 	    &  $ 0.3536(6) $ 	    &  $ 0.0619(2) $ 	    \\ 
		RsN205 	    &  $ 3 \times 10^{3} $ 	   &  $ 1.00 $ 	    &  $ 0.06 $ 	    &  $ 0.0963(3) $ 	    &  $ 5.94(5) \times 10^{-9} $ 	    &  $ 0.0172(1)  $ 	    & n &  $ 0.7599(6) $ 	    &  $ 0.7699(8) $ 	    &  $ 6.39(5) $ 	    &  $ 5.92(3) $ 	    &  $ 0.4011(9) $ 	    &  $ 0.0432(1) $ 	    \\ 
		\hline 
		D0NoDyn 	    &  $ 3 \times 10^{3} $ 	   &  $ 0.10 $ 	    &  $ 0.06 $ 	    &  $ 2.28(2)  $ 	    &  $ 7.7(1) \times 10^{-10} $ 	    &  $ 0.350(2) $ 	    & n &  $ 0.670(1) $ 	    &  $ 0.699(3) $ 	    &  $ 3.78(3) $ 	    &  $ 3.08(3) $ 	    &  $ 0.377(1) $ 	    &  $ 0.232(2) $ 	    \\ 
		D0Bx1e-2$^{\dagger}$ 	    &  $ 3 \times 10^{3} $ 	   &  $ 0.10 $ 	    &  $ 0.06 $ 	    &  $ 1.0(6) $ 	    &  $ 4.4(1) \times 10^{-3} $ 	    &  $ 6.2(2)  $ 	    & \dotfill &  $ 0.9441(6) $ 	    &  $ 0.468(3) $ 	    &  $ 19.3(8) $ 	    &  $ 97(3) $ 	    &  $ 0.699(3) $ 	    &  $ 0.57(3) $ 	    \\ 
		D0Bz1e-2 	    &  $ 3 \times 10^{3} $ 	   &  $ 0.10 $ 	    &  $ 0.06 $ 	    &  $ 0.70(1)  $ 	    &  $ 1.018(7) \times 10^{-4} $ 	    &  $ 0.341(2) $ 	    & \dotfill&  $ 0.892(1) $ 	    &  $ 0.786(2) $ 	    &  $ 10.6(2) $ 	    &  $ 21.6(5) $ 	    &  $ 0.715(2) $ 	    &  $ 0.139(2) $ 	    \\ 
		\hline 
& 
	\end{tabular} 
\end{table*}

\subsection{Basic properties of the fiducial 3D MTI}\label{sec:fiducial_dynamo}

We begin with a representative run, which we call D0, and follow its evolution into the saturated regime. Run D0 has $\text{Pe} = 3\times 10^3$, with a low viscosity and resistivity ($\text{Pr}=0.06$ and $\text{Pm} = 10$), and it is initialized with a uniform horizontal magnetic field in the $x$ direction of initial strength $B_0 = 10^{-5}$. While run D0 is characterized by a magnetic dynamo, not all our MTI runs produce one, though "realistic" choices of parameters for the ICM (i.e., large $\text{Pm}$, moderate conduction and entropy stratification) do. We postpone to Section~\ref{sec:non-dynamo-runs} the discussion of representative non-dynamo runs.

\subsubsection{Energetics}

\begin{figure}
	\centering
	\includegraphics[width=1.0\columnwidth]{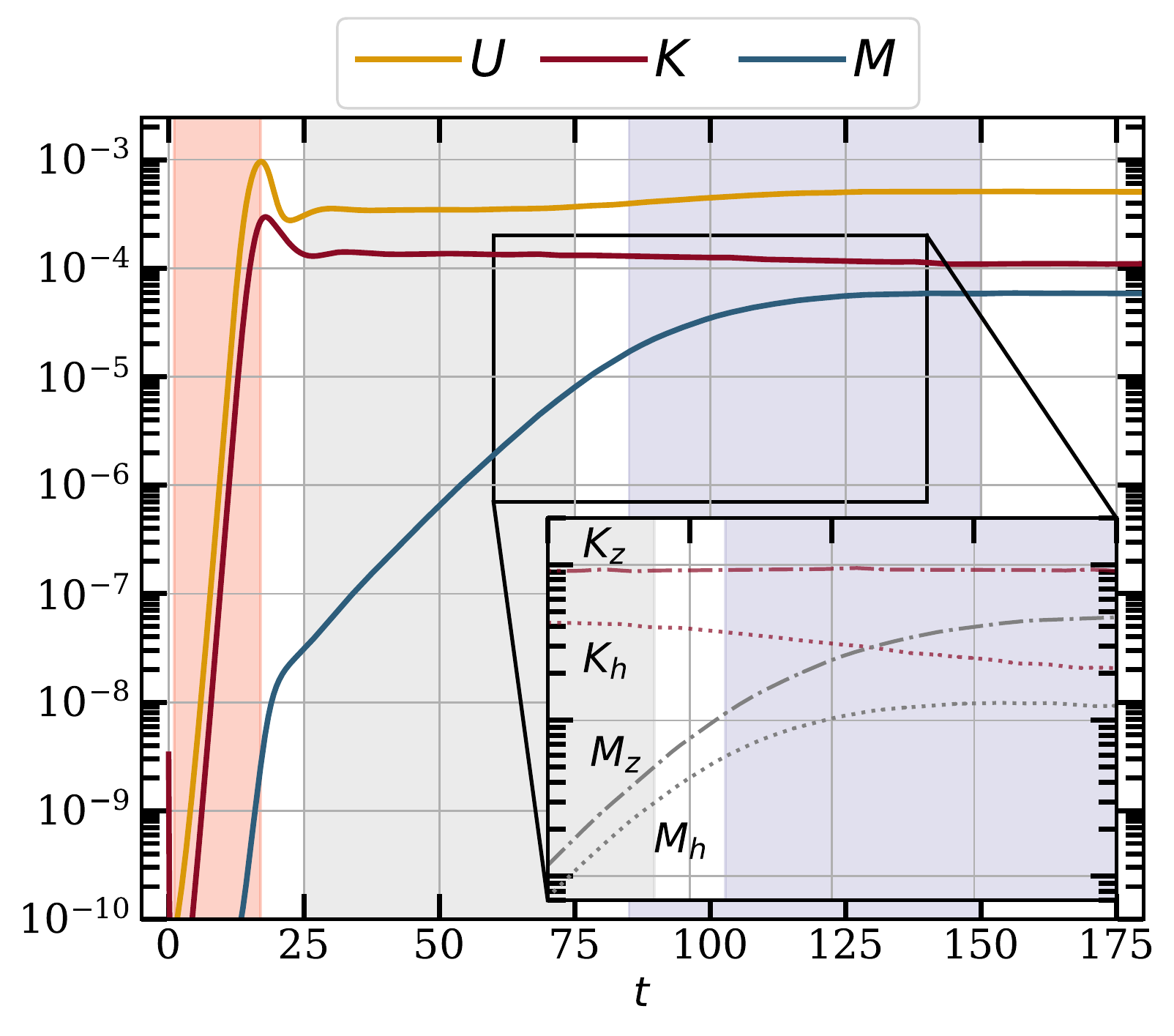}
	\caption{Kinetic ($K$), magnetic ($M$), and potential ($U$) energies in the fiducial 3D run D0. The the linear, kinematic dynamo, and nonlinear/dynamical dynamo phases are represented by red, grey, and purple shading respectively. In the inset, we show the vertical and horizontal components of the kinetic and magnetic energyies ($K_z$, $K_h$ and $M_z$, $M_h$, respectively) during the late kinematic and nonlinear dynamo phases.}
	\label{fig:3D_kin_pot_mag_energy}
\end{figure}

In Fig.~\ref{fig:3D_kin_pot_mag_energy} we show the time evolution of the kinetic, potential, and magnetic energies for run D0, highlighting three distinct phases: the MTI's initial exponential growth (light red), the kinematic phase of the dynamo (light grey), and the dynamical dynamo phase (light purple), after which the system has finally saturated. The initial growth phase bears a close resemblance to the 2D cases examined in Paper I, as the fastest growing modes are 2D and attacked by the same parasitic modes (for more details, see Section 4.1 in Paper I). 

Contrary to the 2D simulations, however, once the kinetic and potential energies saturate, around $t\sim 20$, the system enters a phase in which the magnetic field grows exponentially (i.e. a kinematic dynamo). When the Lorentz force starts to impede these fluid motions, visible in the slight decrease of $K$ around $t\sim 75$, the kinematic phase ends and is replaced by a nonlinear (or dynamical) dynamo phase. Subsequently, the magnetic growth continues, but at a slower rate, until it reaches about $50\%$ of the kinetic energy.
The growth of magnetic energy in this last phase is well approximated by a linear function of time, which is consistent with the picture of magnetic field stretching by velocity eddies on the scales still impervious to the rising Lorentz force \citep{Schekochihin2002,Schekochihin2004,Maron2004}. Interestingly, we also notice an increase in the potential energy by about $40\%$ of its value in the kinematic phase (see Section \ref{sec:vertical_aniso} for further discussion).
Finally, the magnetic field stops growing around $t \sim 150$ and the system reaches a quasi-steady saturated state. 

\begin{figure}
	\centering
	\includegraphics[width=0.95\columnwidth]{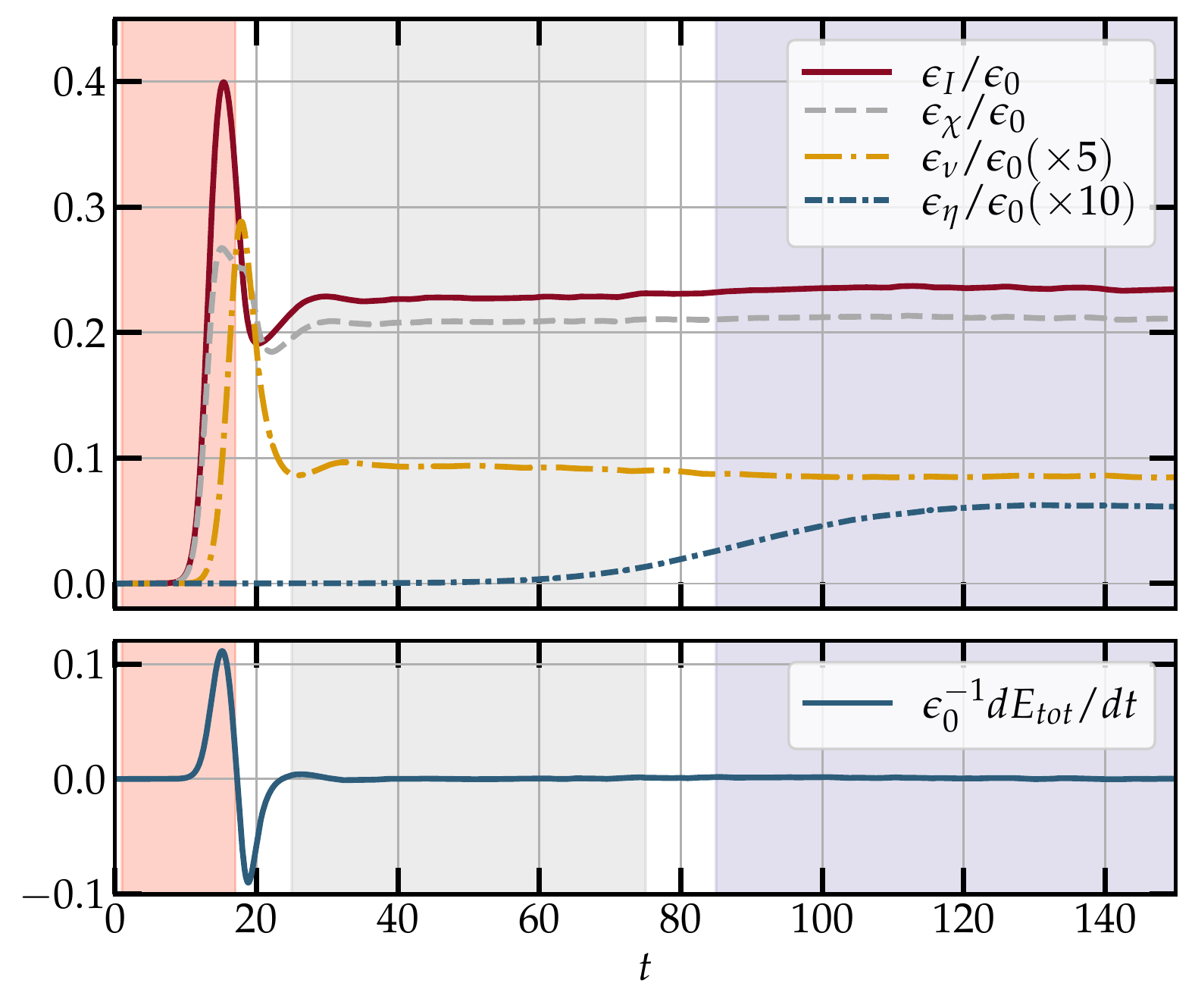}
	\caption{Volume-averaged energy rates normalized by $\epsilon_0 \equiv \chi \omega_T^4/N^2$ (top panel), and the net energy injection rate (bottom panel). The shaded regions are colour-coded as in Fig.~\ref{fig:3D_kin_pot_mag_energy}.}
	\label{fig:en_fluxes}
\end{figure}

In Fig.~\ref{fig:en_fluxes} we plot the volume-averaged energy rates for run D0, indicating the three evolutionary phases by the same colours in Fig.~\ref{fig:3D_kin_pot_mag_energy}. During the MTI linear growth phase, the instability injects energy into the system $(\epsilon_I)$, and it is incompletely balanced by thermal dissipation ($\epsilon_\chi$). The initial break-up of the MTI mode leads to a sudden but transient increase in viscous dissipation $(\epsilon_\nu)$; shortly afterwards, the net energy input rate $d E_{tot}/ dt$ quickly stabilizes around zero, with a small mean input of energy during the following phases, which is fueling dynamo growth. The energy rates -- with the obvious exception of Ohmic dissipation -- are only marginally affected by the emergence of a kinematic dynamo.  Finally, note that the energy injection rate at saturation is consistent with the theoretical value $ \sim \chi \omega_T^4 / N^2$, which we obtained in Paper I.

\begin{figure*}
		\centering
		\includegraphics[width=1.0\linewidth]{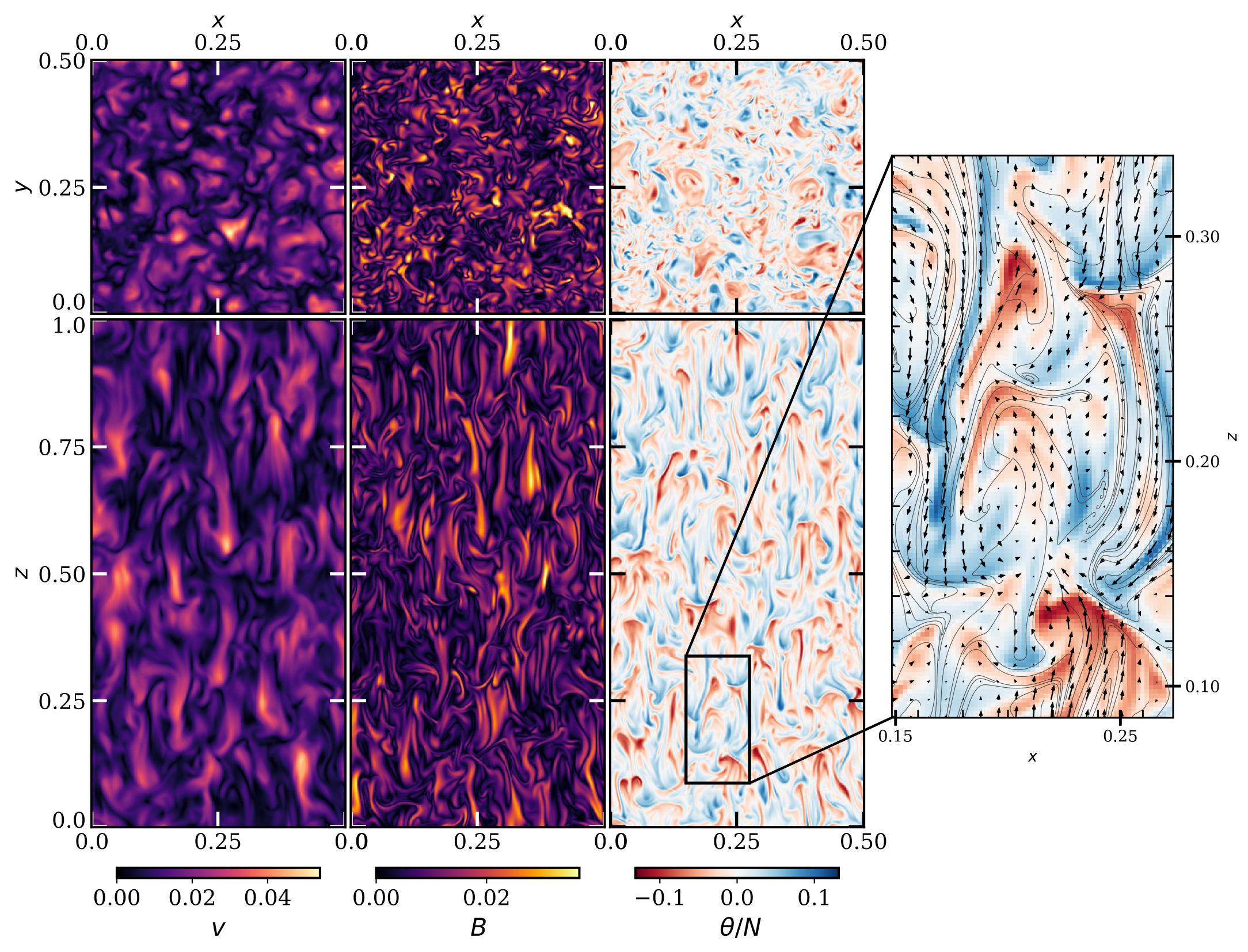}
	\caption{From left to right: amplitude of the velocity field $v = (v_x^2 + v_y^2 + v_z^2)^{1/2}$, amplitude of the magnetic field $B = (B_x^2 + B_y^2 + B_z^2)^{1/2}$, and density fluctuations of run D0 at saturation ($t=300$). The upper row shows a horizontal cut of the domain at $z=0$, with $x,y \in \left[ 0.0,0.5 \right]$, while in the bottom row we plot a vertical cut at $y=0$, with $x \in \left[ 0.0,0.5 \right]$ and $z \in \left[ 0.0, 1.0 \right]$. In the zoomed panel, we superimpose magnetic field lines and velocity vectors on the coloured density field. 
	}
	\label{fig:3D_b_v_temp_field}
\end{figure*}

\subsubsection{Flow field}

We show a representative snapshot of the velocity, magnetic, and density fluctuations at saturation in Fig.~\ref{fig:3D_b_v_temp_field}. For clarity, we only plot a portion of a vertical and horizontal cut of the domain. The saturated state of the MTI is characterized by remarkably self-similar turbulence, with strong correlations between the various fields, and a clear vertical anisotropy.
The velocity field displays a \textit{coarser} structure than $\bm B$ -- a consequence of the larger viscosity relative to resistivity. In the horizontal cut of the velocity field (upper left plot), we recognize a granulation pattern which is typical of simulations of Rayleigh-Benard convection, with the individual granules representing convective cells \cite[see, e.g.,][]{Lappa2009}.

In the right panel of Fig.~\ref{fig:3D_b_v_temp_field} we show in detail a composite view of the boxed region that cleanly illustrates the main features of the saturated 3D MTI. In addition to the density fluctuations (in colour), we plot the local direction and magnitude of the velocity field (arrows), as well as the magnetic field lines. We can clearly distinguish the plume-like structures, which represent rising and falling fluid elements: there is a strong correlation between fluid shaded in red (hot and less dense) and upward motion, while the opposite is true for fluid in blue, which is cold and heavier. Moreover, as the plumes rise and sink their fronts are effectively insulated by magnetic field lines, which prevent exchanges of heat between the plumes and their immediate neighborhood. This insulation reinforces the updrafts and downdrafts, which might otherwise re-equilibriate and dissolve. On the other hand, the density/temperature does not vary significantly along magnetic field lines.

\subsubsection{Vertical anisotropy}\label{sec:vertical_aniso}

At saturation the velocity and magnetic fields exhibit a clear vertical anisotropy in both their preferred spatial scales and in the relative magnitudes of their vector components. Of course, the two are linked in incompressible flows because of the two solenoidal conditions, \textit{viz.} $u_{\perp} / u_z \sim B_{\perp} / B_z \sim k_{z} / k_{\perp}$, with $ u_{\perp}$ and $B_{\perp}$ the components in the $xy$ plane. Thus the anisotropy in spatial variation is necessarily associated with an unequal partition of energy in the vertical and horizontal.

\begin{figure}
	\centering
	\includegraphics[width=1.0\columnwidth]{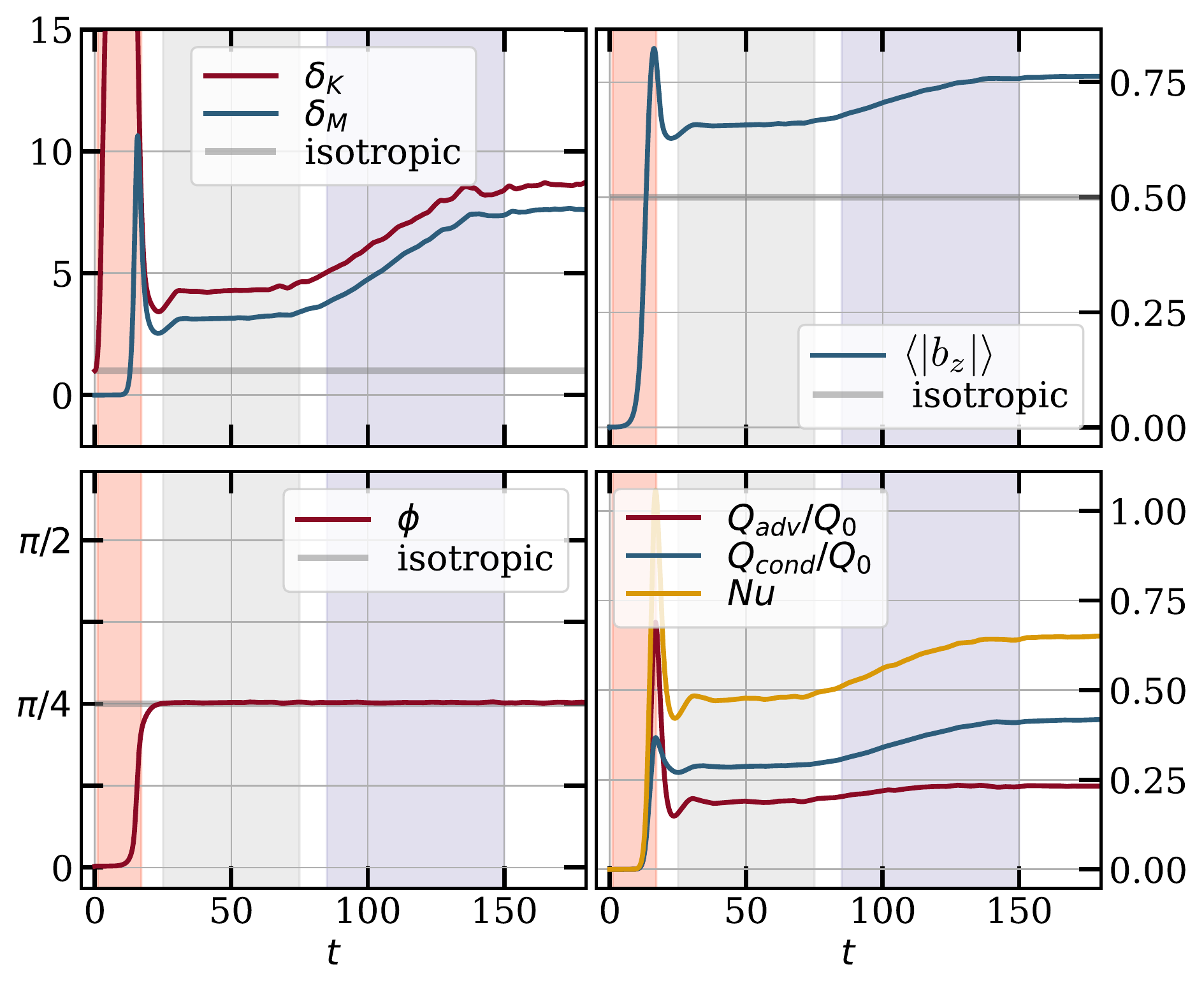}
	\caption{ Top left: velocity (dark red) and magnetic (blue) vertical biases as functions of time; the grey line represents statistical isotropy. Top right: the mean vertical component of the unit magnetic field. Bottom left: the mean horizontal angle $\phi$ of the magnetic field. 
   Bottom-right: advective and conductive heat fluxes normalized by $Q_0= \chi \omega_T^2$, and Nu.}
	\label{fig:3D_anisotropies_bias_angles_nusselt}
\end{figure}

To quantify the vertical  \textit{bias} in the components of the vector fields, we introduce the two ratios  $\delta_K = 2 K_z / (K_x + K_y) $, and similarly $\delta_M = 2 M_z / (M_x + M_y)$.
Isotropic turbulence produces ratios $\approx 1$ when averaged over time. In the top left panel of Fig.~\ref{fig:3D_anisotropies_bias_angles_nusselt} we plot the evolution of these two ratios. We observe an initial spike in $\delta_K$ and $\delta_M$ because of the exponential growth of the MTI modes (when the magnetic is progressively aligned in the vertical direction), but both the kinetic and magnetic biases remain well above $1$ during the kinematic dynamo phase, and further increase during the nonlinear dynamo phase, reaching a value of $\approx 8$ at saturation. There appears a slightly stronger bias in the velocities, which we discuss further in Section~\ref{sec:impact_dynamo_MTI}. 

Another commonly used diagnostic is the volume average of $\lvert b_z \rvert$, shown in the top-right panel of Fig.~\ref{fig:3D_anisotropies_bias_angles_nusselt}, which is well above the value indicative of statistical isotropy in 3D, i.e.\ $0.5$. This preferential alignment of the velocity and magnetic field in the vertical direction is a typical feature of MTI-driven turbulence in 3D, and sets it apart from 2D MTI (we postpone to Section~\ref{sec:comparison_2D} a more detailed comparison between 2D and 3D MTI).

While the system is vertically biased, it exhibits little anisotropy in the $x$ and $y$ directions, despite the (weak) net magnetic flux pointing along $x$. To quantify the horizontal anisotropy, we plot in Fig.~\ref{fig:3D_anisotropies_bias_angles_nusselt} the mean angle of the magnetic field in the $xy$ plane $\phi$ defined through $\cos\phi = \lvert b_x \rvert / (b_x^2 + b_y^2)^{1/2}$. For an isotropic field, we expect $\phi = \pi /4$, which is close to the value we observe in D0.

\subsubsection{Heat transport}

The vertical bias in the velocity and magnetic fields has a direct impact on the amount of heat that is carried through the system by both advection (via the velocity field) and conduction (along the magnetic field lines). In the bottom right panel of Fig.~\ref{fig:3D_anisotropies_bias_angles_nusselt} we show the advected and conductive heat fluxes for run D0, together with the Nusselt number, defined in Eq.\eqref{eq:nusselt_num}. The advected flux $Q_\text{adv}$ is positive throughout the simulation owing to the fact that density fluctuations are strongly anti-correlated with vertical velocity -- as illustrated in the detail plot of Fig.~\ref{fig:3D_b_v_temp_field}. At saturation, however, $Q_\text{cond}$ dominates and accounts for about $2/3$ of the total heat flux. While both fluxes remain stationary over the kinematic dynamo phase, we notice a further increase in the magnitude of $Q_\text{adv}$ and particularly of $Q_\text{cond}$ during the nonlinear dynamo phase. 

The increase in the conductive flux arises from the formation of coherent vertical magnetic fields at larger scales, which can then facilitate heat conduction across the box, while the higher levels of advective transport are instead due to the increase in the potential energy, which is in turn tightly related to the value of the correlation length of the temperature fluctuations (see Section~\ref{sec:impact_dynamo_MTI} for further discussion).
As a result, the total heat flux across the box reaches a value of $\approx 65\%$ of the maximum flux, as seen from the value of Nu, meaning that the MTI is extremely efficient at transporting heat even in the presence of purely anisotropic conduction and a highly disordered magnetic field geometry.

\subsubsection{Spectral densities}\label{sec:spectra}


\begin{figure}
	\centering
	\includegraphics[width=1.0\columnwidth]{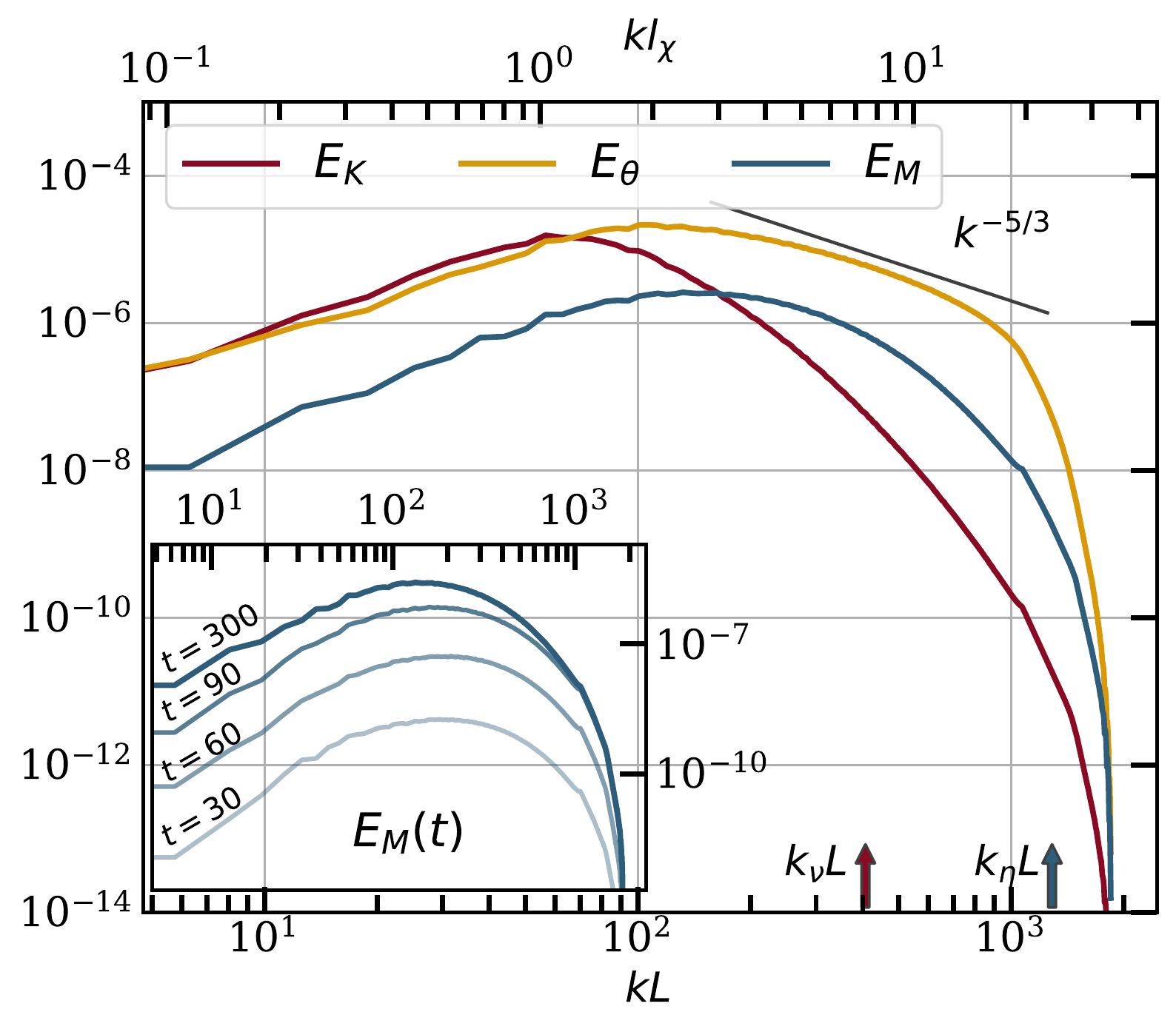}
	\caption{Power spectrum of the kinetic (solid red line), density (solid gold line), and magnetic (solid blue line) fluctuations for run D0 at saturation. For reference, we also plot the Kolmogorov $k^{-5/3}$ power-law and indicate the viscous and resistive wavenumbers $k_\nu$ and $k_\eta$ with arrows. The inset exhibits the time evolution of the magnetic power spectrum. }
	\label{fig:3D_energy_spectra}
\end{figure}
\begin{figure}
	\centering
	\includegraphics[width=1.0\columnwidth]{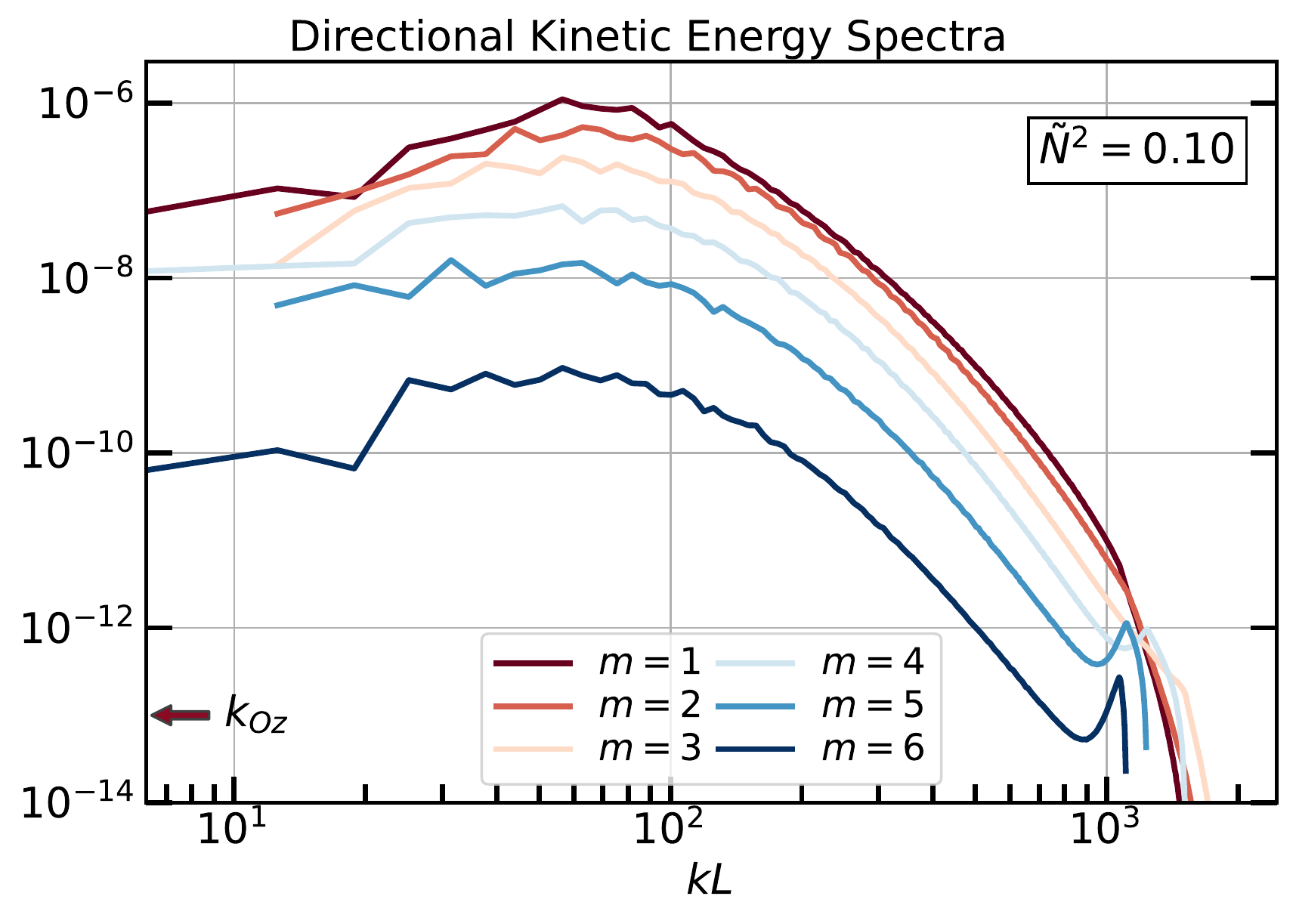}
	\caption{Directional power spectrum of the kinetic fluctuations for run D0 at saturation. The latitudinal band $m=1$ consists of horizontal wavevectors ($k_z \approx 0$), while band $m=6$ consists of nearly vertical wavevectors ($k \approx k_z$).
	}
	\label{fig:3D_directional_energy_spectra_D0}
\end{figure}

To understand the distribution of energy in spectral space we plot in Fig.~\ref{fig:3D_energy_spectra} the shell-averaged kinetic, potential, and magnetic energy spectra of run D0 at saturation. 
We note that the smaller scales are dominated by density fluctuations, 
whereas at longer scales the kinetic and potential energies are largely equivalent.
In contrast to the 2D cases examined in Paper I, it is hard to distinguish the presence of a well-defined power-law decay in $E_K$, likely due to the weak separation of scales between energy injection and viscous dissipation. We remark that the vertical components of the kinetic and magnetic energy spectra (not shown) are substantially larger than their horizontal counterparts at all scales, except near the cutoff. This signals a departure from  "standard" -- non-MTI -- stably-stratified turbulence, where the vertical motions are instead suppressed (see, e.g., \cite{Davidson2013}, or for a discussion focussed on the ICM, \cite{Mohapatra2020,Mohapatra2020b}).

Due to the dynamo, the relatively weak field generated at the end of the MTI growth phase is amplified and becomes larger than kinetic fluctuations at small scales. 
Indeed, the kinetic and magnetic power spectra in Fig.~\ref{fig:3D_energy_spectra} bear a strong resemblance to the classic picture of a high-$\text{Pm}$ fluctuation dynamo \citep{Kulsrud1992,Schekochihin2002,Schekochihin2004}, albeit in a situation where the separation between the viscous $k_{\nu}$ (red arrow) and the resistive scales $k_{\eta}$ (blue arrow) is not pronounced.
The growth of the magnetic power spectrum is plotted in the inset in Fig.~\ref{fig:3D_energy_spectra}: the magnetic field amplifications slows down as the Lorentz force begins to suppress the velocity motions (around $t\sim 90$), before eventually reaching saturation.

To quantify the spatial anisotropy of the turbulence, mentioned earlier, we plot in Fig.~\ref{fig:3D_directional_energy_spectra_D0} the directional spectra of the kinetic energy where, for each radius $k$, we divide the spherical shell into $6$ evenly-spaced latitudinal bands that go from near-equatorial ($m=1$) to near-polar ($m=6$). Thus, the band $m=1$ contains wavevectors with $k_z \approx 0$, while vertical wavevectors ($k \approx k_z $) are included in the $m=6$ band. Figure~\ref{fig:3D_directional_energy_spectra_D0} reveals a clear latitudinal anisotropy, as most of the energy is contained
in the horizontal wavenumbers (low $m$) at both large and small scales. This confirms the visual impression of Fig.~\ref{fig:3D_b_v_temp_field}, which depicts the MTI saturated fields as composed of thin, vertically-elongated plumes.

\begin{figure*}
	\centering
	\includegraphics[width=1.0\linewidth]{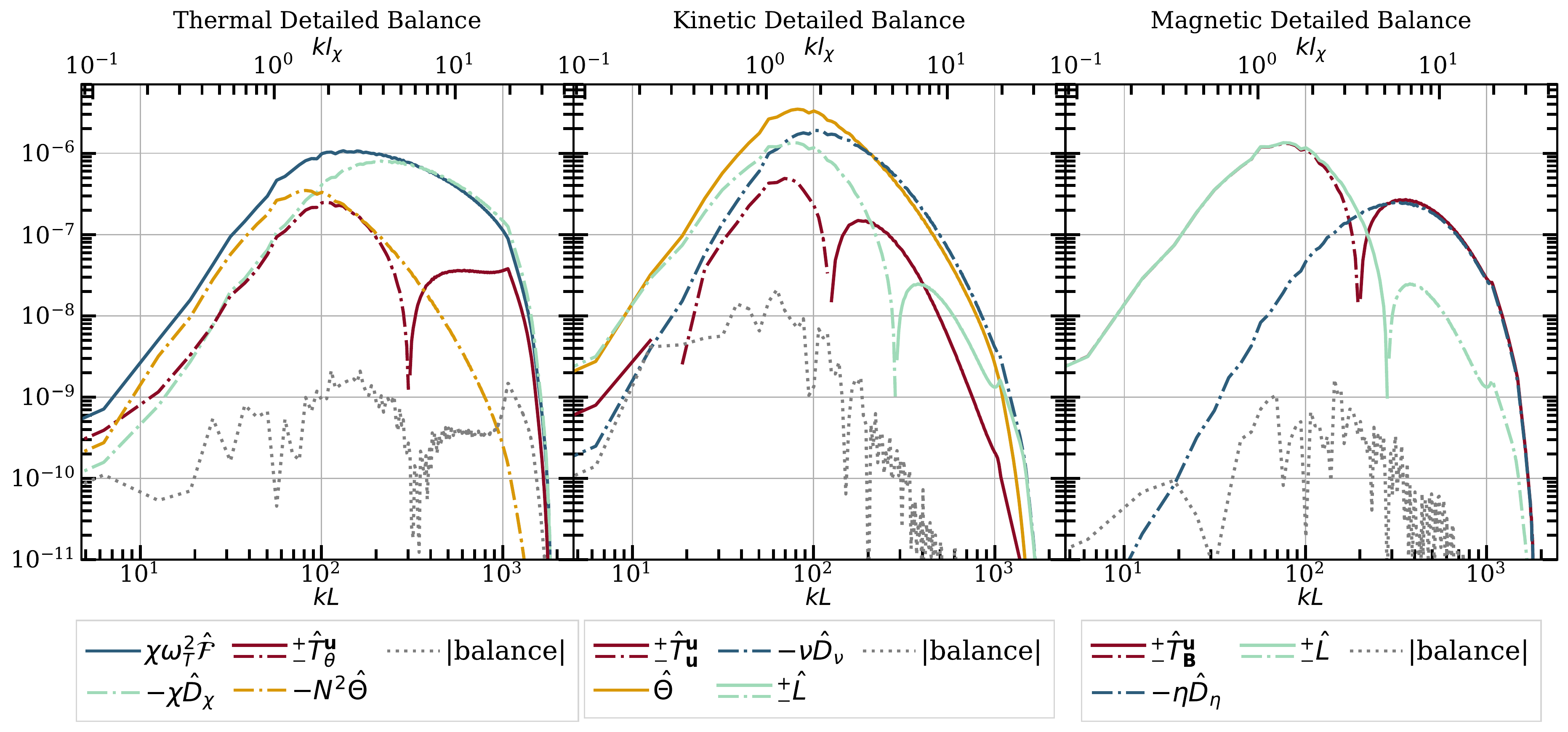}
	\caption{Detailed thermal (left), kinetic (middle), and magnetic (right) energy balance of run D0 at saturation. In all plots, the solid lines indicate where energy is being injected in the spherical shell of radius  $k$, while the dash-dotted lines represent sinks of energy and are plotted with the sign reversed. 
	}
	\label{fig:3D_scale_by_scale_energy}
\end{figure*}

Finally, if we compute the Ozmidov scale -- where the stable stratification starts to suppress vertical turbulent motions -- we find that it is larger than the box size, for these parameters. This, together with the fact that even at large scales the most energetic wavenumbers are those near-horizontal, suggests that the entropy stratification of the fiducial run is too weak to reach the very long-wavelength regime upon which the vertical anisotropy is diminished.

\subsection{Analysis of 3D MTI saturation}\label{sec:fiducial_analysis}

In the following section we take a closer look at the saturation of the 3D MTI. 
We concentrate on (a) the energy flows across the various scales and thereby construct a theory of the saturation mechanism , (b) the properties of the magnetic field and its topology in both the kinematic and nonlinear phases of the dynamo, and (c) the feedback of the dynamo on the MTI.

\subsubsection{Energy flows in spectral space}\label{sec:scale-by-scale}

The MTI turbulence is buoyancy-driven, in the sense that energy is injected into the flow on a wide range of scales in the form of density fluctuations. In this respect, it shares similarities with other buoyancy driven flows, such as Rayleigh-Benard convection or bubbly flows \citep[see, e.g.,][]{Kumar2014,Prakash2016,Verma2017}, where energy is transferred from buoyancy to kinetic fluctuations, rather than in the opposite direction, as would be the case in forced stably-stratified turbulence. The general picture that our simulations present can be summarised as follows. On small scales the energy injected by the MTI is locally dissipated by thermal diffusion, but on longer scales the injected energy is transferred, through the buoyancy coupling term, into kinetic fluctuations. These, in turn, are either locally dissipated by viscosity, or converted into magnetic fluctuations that cascade down to the very short resistive scale. We now show evidence supporting this picture.

The transfers of energy that undergird the saturation of the 3D MTI can be viewed
in Fig.~\ref{fig:3D_scale_by_scale_energy}, where we plot the thermal, kinetic and magnetic scale-by-scale (or detailed) energy balance in $k$ space, with the spectral fluxes defined in Eq.~\eqref{eq:spectral_vel}-\eqref{eq:spectral_temp}, for run D0. In the thermal balance (left panel), the anisotropic dissipation (light dotted-dashed blue)  balances forcing (solid blue) at small scales. But at intermediate to large scales it is the buoyancy transfer into kinetic energy (gold) that mostly balances MTI forcing. The net effect of MTI forcing and anisotropic dissipation is to inject energy into the system across a wide range of scales, roughly $20 \lesssim kL \lesssim 300$, with a peak around $kL \approx 100$ (which is approximately equal to the wavenumber of fastest linear growth), while the buoyancy term always removes energy from density fluctuations at all scales. 

Next we turn to the kinetic energy balance (middle panel): here energy is injected by the buoyancy term (solid gold) over roughly the same scales as those excited by the MTI, and the majority of the energy so injected is either viscously dissipated (blue dotted-dashed) on small scales or converted into magnetic fluctuations through the Lorentz tension term (light blue) at larger scales. 

There are signatures of a direct cascade in the detailed energy balance for the magnetic energy (right panel), where on large-scales the advective transfer term (red )matches the injection term due to the Lorentz force (light blue), while on small scales Ohmic dissipation removes the energy transported by the advection term across the inertial range.

It is worth stressing that in 3D the injection and removal of kinetic and thermal energy are local processes in $k$ space, in the sense that only a small portion of the energy partakes in the direct cascade from larger to smaller scales, and therefore our fiducial run lacks a proper inertial range. In fact, the nonlinear advection terms (red) in the thermal and kinetic energy scale-by-scale balances are about one order of magnitude smaller than the injection terms at practically all scales\footnote{In the middle panel of Fig.~\ref{fig:3D_scale_by_scale_energy} the nonlinear advection term becomes positive at very large scales, but a careful analysis of the energy spectrum up to $t \approx 300$ shows slow variability but no build-up of energy at large-scales \citep[unlike, e.g.,][]{Cattaneo2001,Rincon2005}}.

\begin{figure}
	\centering
	\includegraphics[width=1.0\columnwidth]{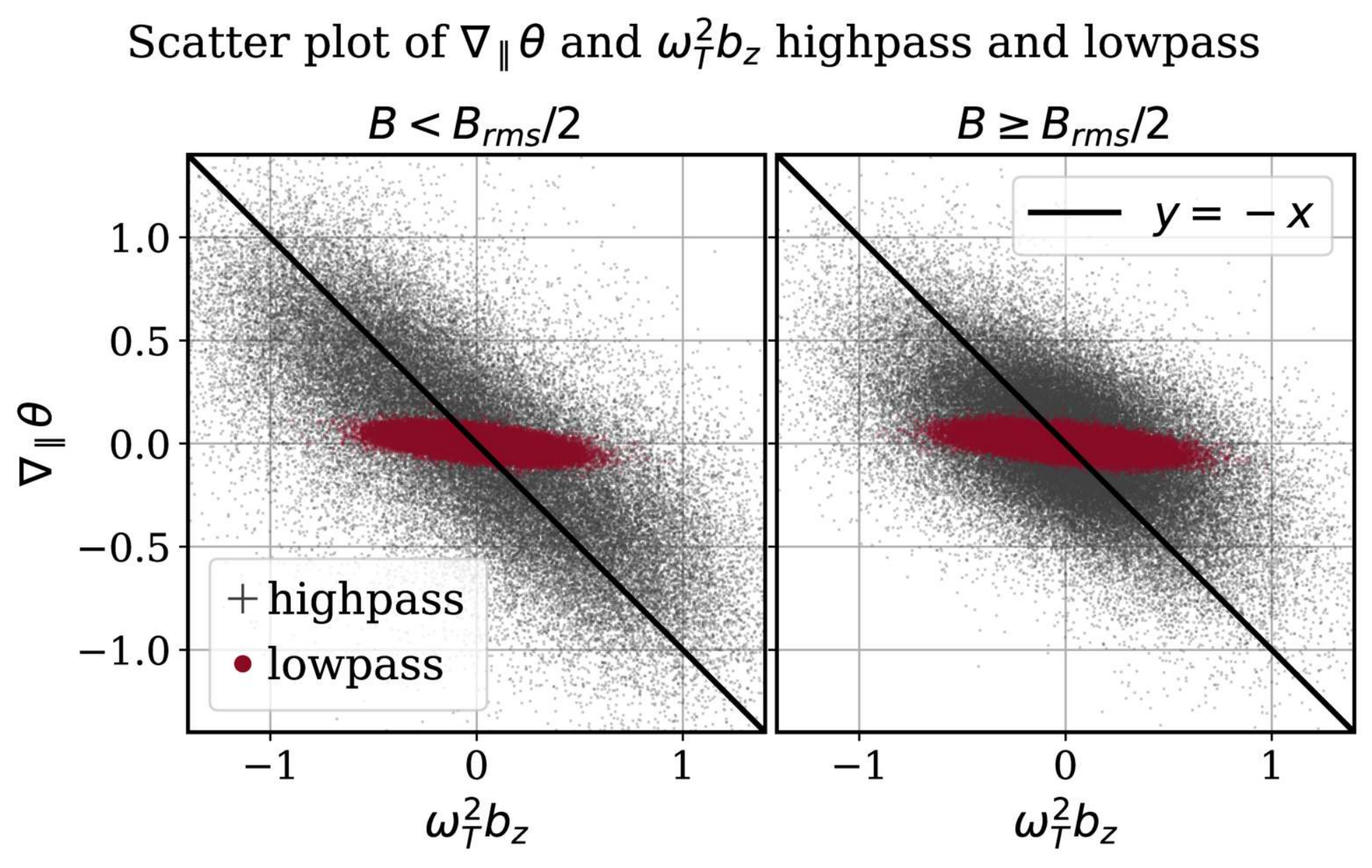}
	\caption{Scatter plot of $\nabla_{\parallel} \theta$ vs $\omega_T^2 b_z$ at saturation. Left panel shows points for which $B<B_{rms}/2$, and the right panel stronger fields, $B>B_{rms}/2$.
	Black crosses represent (high-pass filtered) small scales, and red dots (low-pass filtered) large scales. The line represents perfect anticorrelation, as predicted by Eq.~\eqref{eq:small_scale_balance}. The data has been thinned by a factor $10^3$.}
	\label{fig:3D_scatter_bz_bgradth_low_high_B}
\end{figure}

\subsubsection{Balancing MTI injection and dissipation at small scales}\label{sec:small_scales}

Since the two dominant terms at the small scales in the thermal energy balance are the MTI forcing and the anisotropic dissipation, it is natural to seek a balance between these two terms. Both terms take the form of a divergence of a flux, thus one possible balance is
\begin{align}\label{eq:small_scale_balance}
	\nabla_{\parallel} \theta \approx - \omega_T^2 b_z,
\end{align}
which shows that these two quantities are \textit{anticorrelated} at small scales. Using the definitions of $\omega_T^2$, $\theta$, and the linearised equation of state,  Eq.~\eqref{eq:small_scale_balance} may be rewritten as 
$\nabla_\parallel T' \approx - b_z (dT/dR)_0 $, which tells us that
the gradient of the temperature fluctuation arranges itself so as to counterbalance the background temperature gradient, along a given magnetic field line. In other words, the turbulence wants to bring the atmosphere to isothermality, and thus to MTI \emph{marginal stability} (though it is constrained to do so only along field lines). This drive to marginality was already discussed in Sec. 2.2.2 in Paper I, but in a far less precise box-averaged sense. It also generalizes the prediction that the linear Lagrangian temperature perturbation vanishes in the limit of rapid conduction \citep[see, e.g.,][]{Quataert2008}.

We numerically confirm this picture in Fig.~\ref{fig:3D_scatter_bz_bgradth_low_high_B}, which shows a scatter plot of $\nabla_{\parallel} \theta$ and $\omega_T^2 b_z$ from our fiducial run at saturation. To separate the large scales from the small scales, we filter our data by applying a high-pass mask to the wavenumbers above $k = 300$ in Fourier space and transform back into real space. We also split the datapoints based on the value of the \textit{total} (unfiltered) magnetic field at their location, showing the points where $B < B_{rms}/2$ ($B \geq B_{rms}/2$) in the left (right) panel. As we can see, the highpass-filtered terms (i.e. small scales) are distributed around the trend suggested by Eq.~\eqref{eq:small_scale_balance} (solid black line), particularly so in the group with weaker magnetic fields. On the contrary, the lowpass-filtered terms (retaining wavenumbers below $k = 80$) do not show any significant trend and are rather clustered around $\nabla_{\parallel} \theta \approx 0$ irrespective of the value of $b_z$, both for weak and strong magnetic fields.

\subsubsection{Balancing forcing and buoyancy at the integral scale}\label{sec:large_scales}

At small scales, the turbulent motions are mostly unaffected by the stable entropy stratification, and, in the equation for $\theta$, the buoyancy term is negligible compared to the forcing and dissipation. Moving to larger scales, however, the stratification becomes gradually more important until a point where the buoyancy is comparable to the forcing. We call the scale at which this transition happens $l^*$, characterized by the balance $ N^2 u_z \sim \chi \omega_T^2 \nabla \cdot (\bm b b_z)$. We interpret $l^*$ as both the largest scale that is excited by the MTI, and where the buoyancy coupling between thermal and velocity fluctuations is effective. This is consistent with the fact that the peaks of the kinetic and of the potential energy spectra are close to each other, see Fig.~\ref{fig:3D_energy_spectra}.

Because the forcing term has the dimensions of an inverse length, we approximate $\nabla \cdot (\bm b b_z) \sim (l^*)^{-1}$ at the transition scale $l^*$. 
Balancing buoyancy and forcing at $l^*$ then yields $u_{z}^* l^* \approx \chi \omega_T^2/N^{2}$, where $u_{z}^*$ is the typical vertical velocity at scale $l^*$. This relation allows us to define a Reynolds number at the transition scale $Re^*$
\begin{align}\label{eq:rey_buoy}
	Re^* \equiv \frac{u_{z}^* l^*}{\nu}  \sim \frac{\chi \omega_T^2}{\nu N^2 }.
\end{align}

From our numerical simulations, the transition length roughly coincides with the integral scale, and thus $l^*\sim l_i$, where $l_i=1/k_i$. We then take $u_z^*$ to be the velocity at the peak of the kinetic energy spectrum (which is approximately equal to $ u_{rms}$, since $u_z \gg u_x, u_y$). Thus $\text{Re}^* \approx \text{Re}_i$, the Reynolds number at the integral scale (note $\text{Re}_i \approx 11$ in run R0, see Table~\ref{tab:table_runs_scales}), for which Eq.~\eqref{eq:rey_buoy} gives an expression in terms of the MTI parameters.

The expression for the product $u_{rms}  l_i$ can be combined with the theoretical scaling for the total energy injection rate, $\epsilon_{I} \sim
\chi \omega_T^4 / N^2$, to obtain an expression for $u_{rms}$ and $l_i$ separately. In fact, assuming that a fixed fraction of $\epsilon_{I}$ goes into kinetic energy, we have on dimensional grounds $\epsilon_{I} \sim
u_{rms}^3/l_i$ at the integral scale, yielding
\begin{align}\label{eq:scalings_K_lB}
	K \sim  \frac{\chi \omega_T^3}{N^2}, \, \, \, \, \, \,  l_i \sim \frac{(\chi \omega_T)^{1/2}}{N}.
\end{align}
The scaling relationships derived here will be verified in Sec.~\ref{sec:energetics_scalings}-\ref{sec:buoyancy_scale}  
Remarkably, these scalings are exactly the same as in 2D MTI turbulence, despite the radically different saturation routes involved (this is discussed further in Sec.~\ref{sec:comparison_2D}).

\begin{figure}
	\centering
	\includegraphics[width=1.0\columnwidth]{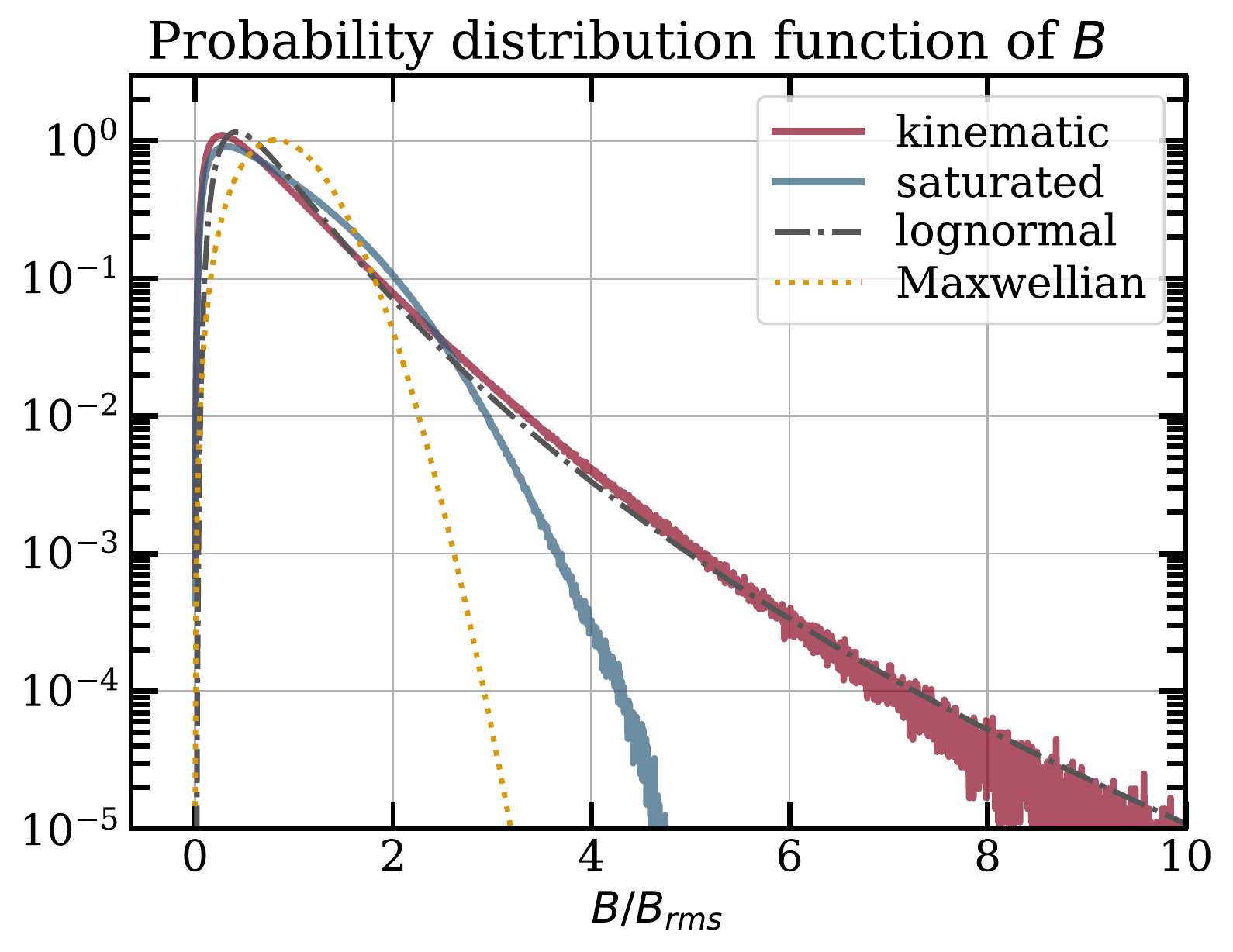}
	\caption{Probability distribution function of the normalized magnetic field strength $B/B_{rms}$ for run D0 during the kinematic phase (red solid line) and at saturation (blue solid line).  The black dash-dotted line represents the lognormal fit from \citet{Schekochihin2004}. The gold dotted line is a  Maxwellian given by $ P(x) = \sqrt{2/\pi} (x^2/a^3) \exp\left[ -x^2/(2a^2)\right]  $, with $x=B$ and $a^2=B_{rms}^2/3$. }
	\label{fig:pdf_lognormal_B}
\end{figure}

\subsubsection{Intermittency and key scales in the two dynamo phases}\label{sec:kinematic_nonlinear}

In MHD turbulence, magnetic fields generally show a higher degree of intermittency (i.e. deviation from Gaussian statistics) compared to the velocity field. Magnetic intermittency is related to the concept of volume filling by turbulent structures, and manifests in the heavy tails of the probability distribution function (PDF) of $B$. 

To examine this physics in our simulations we plot in Fig.~\ref{fig:pdf_lognormal_B} the PDFs of the magnetic field strength in the kinematic and dynamo nonlinear phases. During the kinematic phase the PDF (red) is well-described by a lognormal distribution (black dotted-dashed), which is often associated with the intermittent character of turbulent dissipation \citep{Mandelbrot1972}, and in this case is derived from a Kazantsev model with coefficient $D=\ln (\langle B^4 \rangle^{1/2} / \langle B^2 \rangle ) \simeq 0.87$ \citep{Schekochihin2004}. The good agreement is perhaps unsurprising as the small scale magnetic field amplification is, to a large extent, independent of the exact mechanism driving the large-scale turbulence.
As the magnetic field strength increases, the heavy tails are gradually suppressed and the degree of intermittency falls away. Nevertheless, in the saturated state (blue) the probability of larger values of $B/B_{rms}$ remains significantly higher than if the field strength was distributed like a Maxwellian (gold dotted line).

\begin{figure}
	\centering
	\includegraphics[width=1.0\columnwidth]{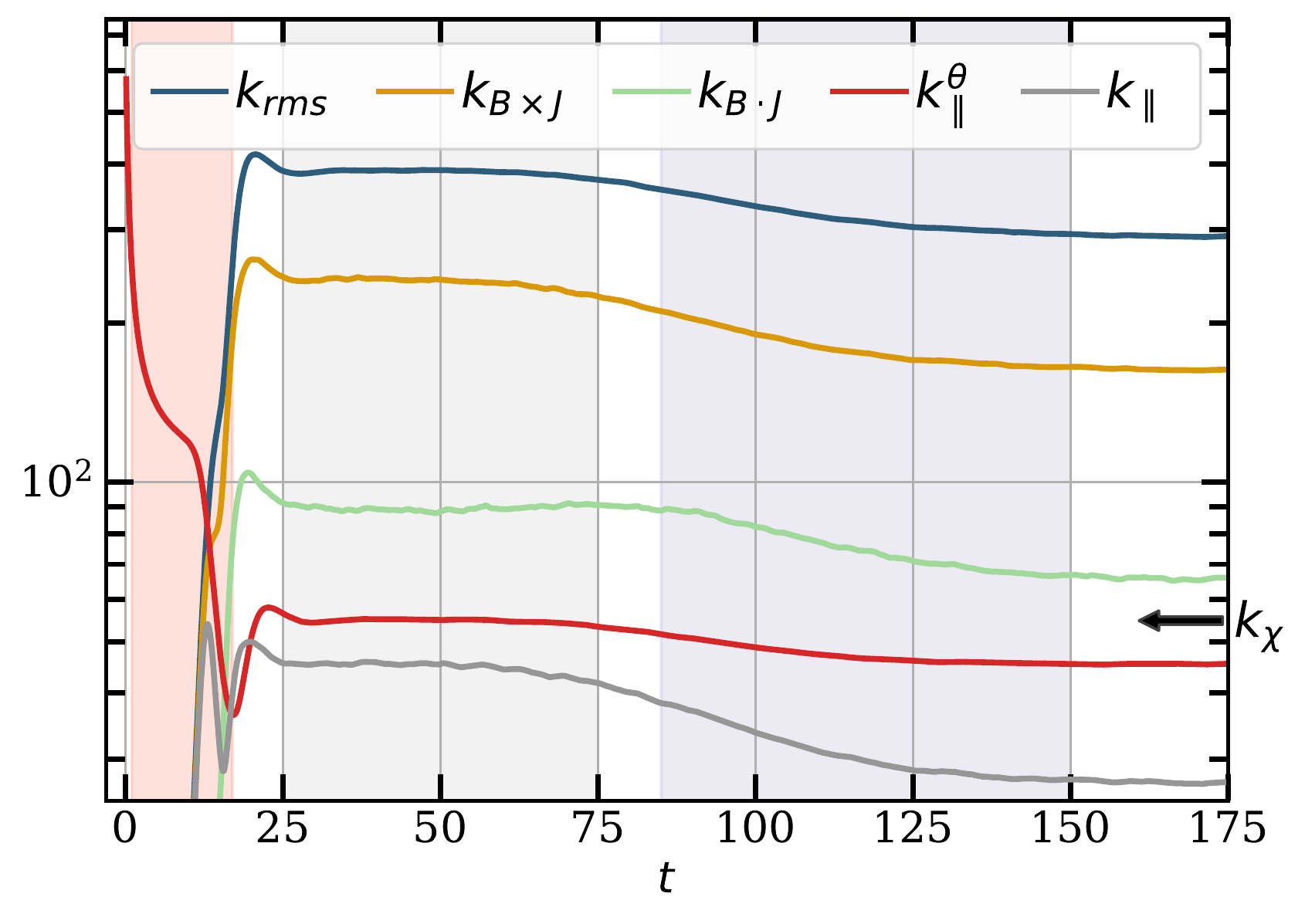}
	\caption{Characteristic wavenumbers of the magnetic field as a function of time for run D0. The normalization differs from \citet{Schekochihin2004} by $2 \pi$. We also plot the evolution of the parallel scale of the buoyancy $k_{\parallel}^{\theta}$. The conduction wavenumber $k_\chi=1/l_\chi$ is represented by the black arrow. }
	\label{fig:3D_wavenumbers}
\end{figure}

The small-scale magnetic fields generated by the dynamo exhibit a folded structure with rapid field reversals in regions of low $B$ strength, and long parallel correlations where the field is strong, thus resembling "ribbons" \citep[][ and references therein]{Schekochihin2001,Schekochihin2004}. This folded structure has been previously observed in simulations of MTI turbulence with and without anisotropic viscosity \citep{Kunz2012}. To describe the structure of the magnetic field, in Fig.~\ref{fig:3D_wavenumbers} we follow the evolution of various magnetic scales  during the kinematic and nonlinear phases of the fluctuation dynamo (cf. Section \ref{sec:scales_numbers}). We find that all the characteristic wavenumbers decrease as the system transitions from the kinematic to the nonlinear phases, which simply reflects the growth of magnetic fields at large scales. But we notice that the scale of variation of $B$ along ($k_{\parallel}$) and across ($k_{\bm B \times \bm J}$) the magnetic field lines are very different, with the former about $4$ times smaller than the latter during the kinematic phase. This ordering of scales is consistent with arrangement of magnetic field in folded structures, for which  $k_{\parallel} < k_{\bm B \cdot \bm J} < k_{\bm B \times \bm J} < k_{rms}$ \citep{Schekochihin2004,StOnge2020}. Similar conclusions were reached by \citet{Seta2020} who determined the morphology of the magnetic field in both the kinematic and the saturated phase through a variety of diagnostics that future work into the MTI dynamo might employ. 

\subsubsection{Impact of the fluctuation dynamo on the MTI}\label{sec:impact_dynamo_MTI}

The fluctuation dynamo feeds back onto the MTI turbulence in several ways.
First, the vertical elongation of velocity and magnetic fluctuations becomes more pronounced when the system enters the nonlinear dynamo phase, as discussed in Section~\ref{sec:vertical_aniso}.
We attribute the increase in this vertical bias to the back-reaction of the magnetic field on the velocity field in a flow that is already dominated by vertical motions: because the Lorentz force becomes progressively larger during the non-linear dynamo phase (and since it is largest in the $xy$ plane), the horizontal motions are increasingly suppressed as the dynamo grows (see the inset in Fig.~\ref{fig:3D_kin_pot_mag_energy}) and the velocity bias increases further. This in turn weakens the stretching of the magnetic field by the horizontal velocity, and thus $\delta_M$ increases concurrently.

Second, this elongation has consequences for the anisotropic conduction of heat. In Fig.~\ref{fig:3D_wavenumbers}, the parallel wavenumber of the density fluctuation $k_\parallel^\theta$ is of the order of $k_{\parallel}$, and also decreases during the nonlinear phase. This follows from the growth of longer correlations parallel to the magnetic field lines, which subsequently permit heat to be exchanged more efficiently over longer distances, with  $k_{\parallel}^\theta$ near the lower bound given by the value of conductivity.  A macroscopic consequence of this is the increase in the potential energy $U$ once we enter the nonlinear dynamo phase (cf. the gold curve in Fig.~\ref{fig:3D_kin_pot_mag_energy}). This is a consequence of the MTI's attempt to enforce isothermality, which fixes (on average) the perturbed thermal gradient: $\langle\nabla\theta \rangle\sim \omega_T^2$ (cf. Eq.~12 in Paper I, and also Section 3.2.2 in Paper II). As $\theta$'s characteristic lengthscales increase, so must the average value of $\theta$ (and hence $U$), in order to keep the gradient fixed.

\subsubsection{Comparison with 2D MTI}\label{sec:comparison_2D}

\begin{figure}	
	\centering
	\includegraphics[width=1.0\columnwidth]{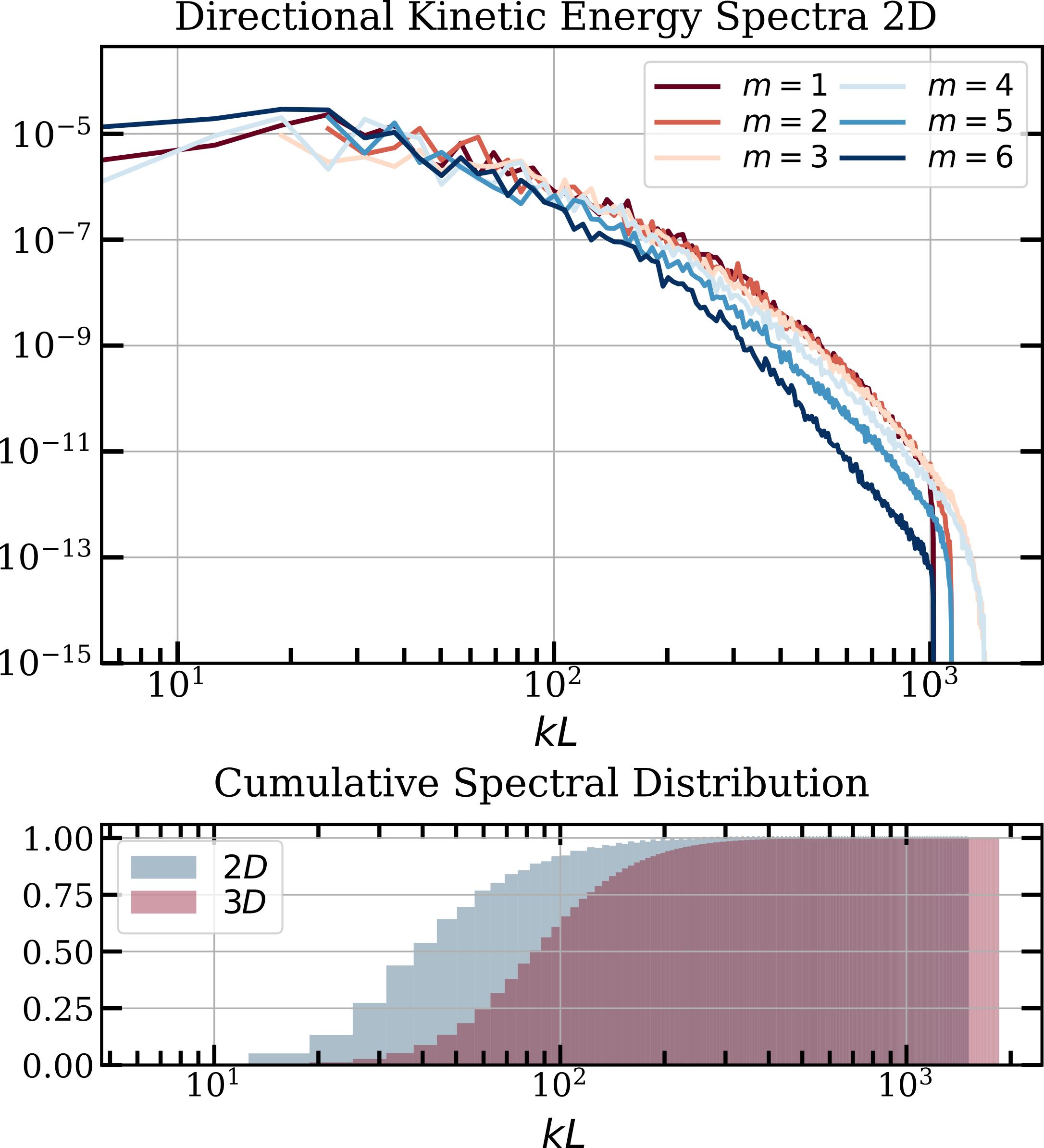}
	\caption{Kinetic directional energy spectra for the 2D run D02D for different latitudinal bands: $m=1$ is near-equatorial ($k_z \approx 0$), $m=6$ is near-polar ($ k \approx k_z$). Lower panel: cumulative spectral distribution of kinetic energy for run D0 and D02D.}
	\label{fig:kinetic_energy_spectra_aniso_2D}
\end{figure}

The two-dimensional cases examined in Paper I fundamentally differ from 3D MTI because of the presence of an inverse cascade, whereby energy is first injected by the MTI at small scales, then transported to larger scales by the inverse cascade, where it ultimately excites g-modes. A consequence of the different saturation mechanism in 2D and 3D is that energy is distributed differently across wavenumbers, as shown in their cumulative spectral distributions (lower panel of  Fig.~\ref{fig:kinetic_energy_spectra_aniso_2D}),  which clearly indicates that the 2D MTI saturates on longer scales than in 3D.

Furthermore, the excitation of g-modes  effectively isotropizes the vertical and horizontal large scales in 2D MTI. This is apparent in the top panel of Fig.~\ref{fig:kinetic_energy_spectra_aniso_2D}, where we plot the directional energy spectra for the 2D reference run: on scales comparable to the integral scale, energy is distributed equally across all latitudinal bands, meaning that there is no favoured orientation of the wavevector. Moving towards small scales, however, we note a progressive vertical elongation, which nonetheless is never as pronounced as in 3D. In 2D simulations that suppress g-modes, either because of strong initial magnetic fields or when the integral scale becomes comparable to (or larger than) the box size (the case when $N^2$ and/or $\text{Pe}$ is small), the large-scale isotropy is broken and such runs are characterized by a clear vertical bias similar in magnitude to what we observe in our 3D (non-dynamo) runs (see below).

\subsection{Non-dynamo runs}\label{sec:non-dynamo-runs}

The 3D MTI does not always stimulate dynamo action, and this is especially so when the stable stratification or the resistivity is strong. In this section, we explore the two regimes of strong stratification and resistivity separately, showing the results of run RsN205, which has $\tilde{N}^2=1.0$ but otherwise the same parameters as D0, and run D0NoDyn, which has a larger resistivity ($\text{Pm}=0.4$) and resolution than D0, but the same thermal conductivity and entropy stratification. Run RsN205, in addition, provides an opportunity to study the high-$N$ regime and thus to touch on lengths beyond the Ozmidov scale, which were inaccessible in the fiducial run D0.

\subsubsection{Low $\text{Pm}$ non-dynamo run}

After the initial MTI growth phase, the magnetic energy in run D0NoDyn saturates at a constant level much less than the kinetic energy.
Its properties are, in fact, very similar to the kinematic phase of the fiducial dynamo run D0, where magnetic tension is dynamically unimportant. In particular, the vertical biases in the kinetic and magnetic fields are similar in size ($\delta_K \simeq 3.8$, $\delta_M \simeq 3.1$ compared to  $\delta_K \simeq 4.3$, $\delta_M \simeq 3.1$, respectively). Though slightly less than those of run D0 once it reaches saturation, the biases are still significantly greatly than in 2D. Similarly, the average vertical magnetic orientation is comparable and approximately $| b_z | \approx 0.66$.

\begin{figure}	
	\centering
	\includegraphics[width=1.0\columnwidth]{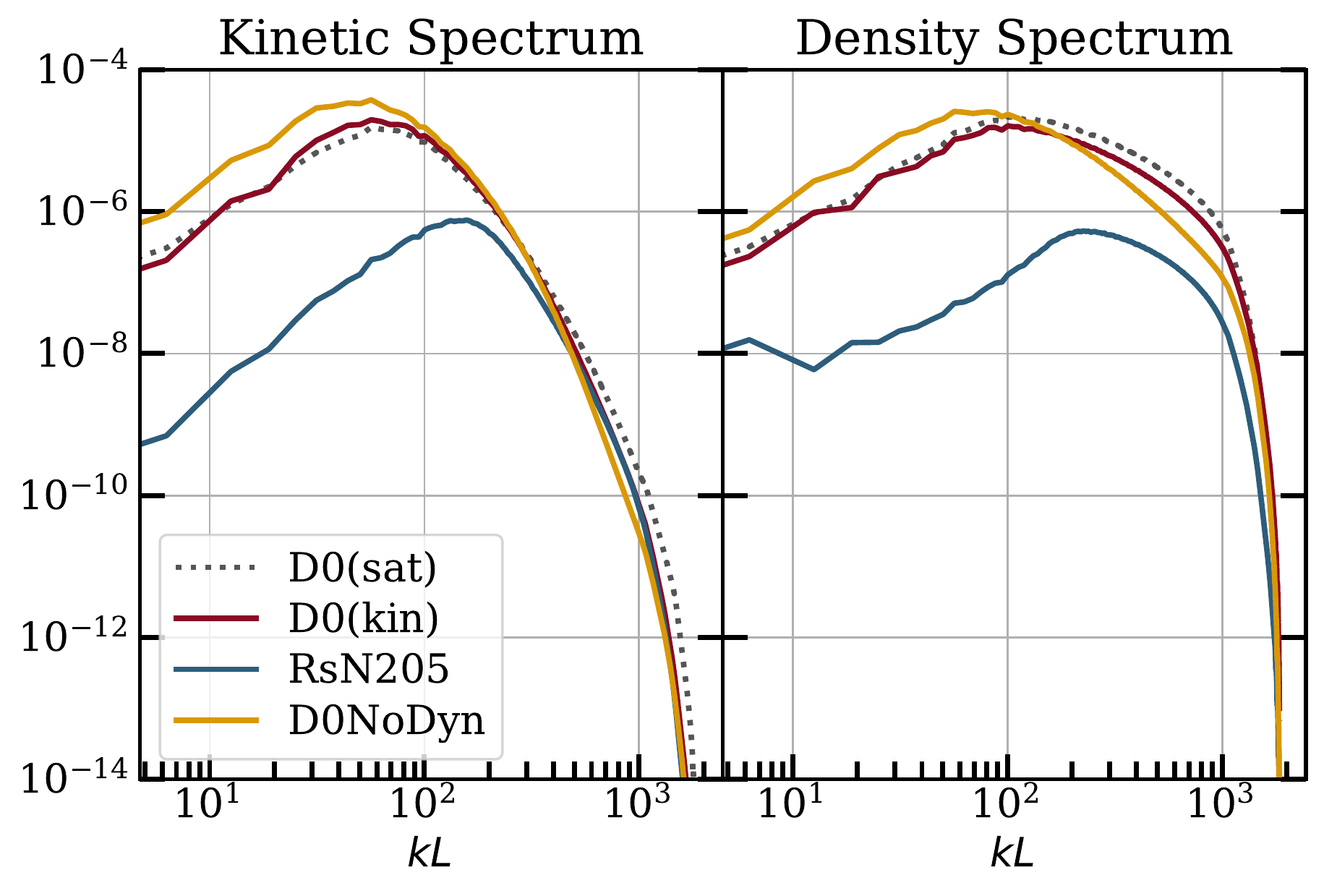}
	\caption{ Kinetic (left) and potential (right) energy power spectra of the fiducial run D0 at saturation (dotted), and during the kinematic dynamo phase at time $t=35-40$ (red), and the saturated phases of run RsN205 and run D0NoDyn. 	}
	\label{fig:energy_spectra_compare_multiplot}
\end{figure}

The distribution of D0NoDyn's and D0's kinetic and potential energies in spectral space, shown in Fig.~\ref{fig:energy_spectra_compare_multiplot}, is also similar, although we note D0NoDyn possesses a slight excess of kinetic and potential energy at large scales compared to the kinematic phase of run D0.
These apparent correspondences can be explained by the fact that, despite the larger resistivity of run D0NoDyn -- which strongly suppresses MTI injection at small scales --  the  "net" injection of energy in the detailed thermal balance (i.e., subtracting anisotropic dissipation) is only slightly modified and pushed towards larger scales. As a result, the turbulence is excited at similar scales through the buoyancy force, and the terms in the detailed kinetic balance present a similar profile in $k$ space, with viscous dissipation and buoyancy now comparable to the nonlinear advection term, while magnetic tension is negligible. 

\subsubsection{Strongly stratified run}\label{sec:strongly-stratified}

Run RsN205 illustrates MTI turbulence in the regime of strong entropy stratification.
Due to the larger $\tilde{ N}^2$, the turbulent energies at saturation  are significantly lower than in the fiducial run and the integral scale is pushed towards smaller scales, as shown in Fig.~\ref{fig:energy_spectra_compare_multiplot}. (This behaviour will be discussed in more detail in Section~\ref{sec:scaling_laws}.) Compared to the fiducial run D0, we do not observe magnetic field amplification following the end of the MTI growth phase.

\begin{figure}	
	\centering
	\includegraphics[width=1.0\columnwidth]{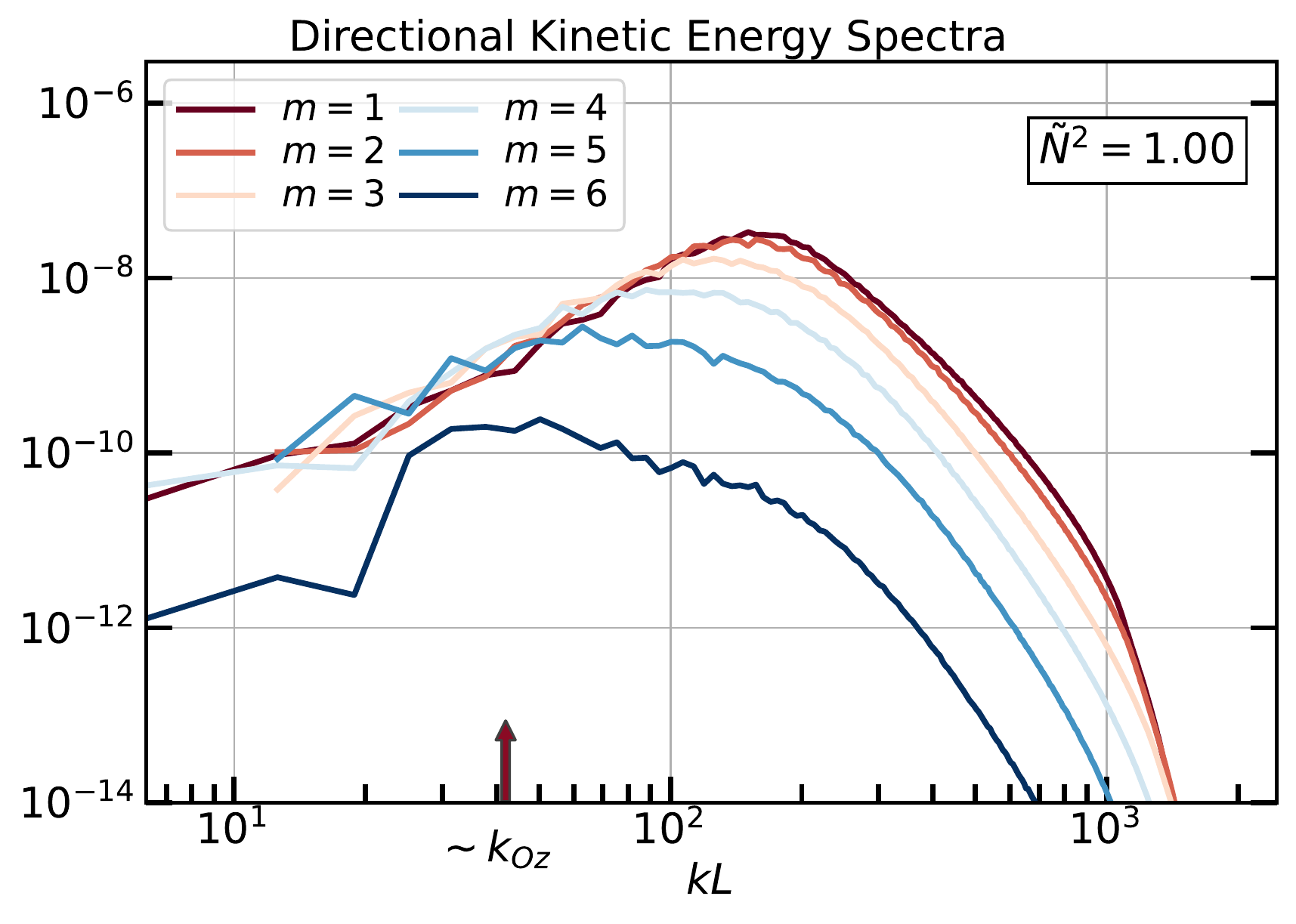}
	\caption{Kinetic directional energy spectra for run RsN205 for different latitudinal bands: $m=1$ is near-equatorial ($k_z \approx 0$), $m=6$ is near-polar ($ k \approx k_z$). The red arrow marks the transition to isotropic turbulence and its location is approximately given by the Ozmidov scale divided by $2 \pi$.}
	\label{fig:3D_kinetic_directional_spectra_N21e0}
\end{figure}

Run RsN205 also allows us to explore the regime beyond the Ozmidov scale. We plot in Fig.~\ref{fig:3D_kinetic_directional_spectra_N21e0} the directional spectra of the kinetic energy. While the small scales are similar to run D0, with the majority of the energy residing in the near-horizontal wavevectors (low $m$), we note a progressive convergence of the latitudinal spectral bands as we move towards larger scales, where for $kL \approx 30-40$ the curves are bunched together, indicating the turbulence has become nearly isotropic. This convergence approximately coincides with the Ozmidov scale (modulo a factor of $2\pi$). It is reasonable to expect that for even stronger stratifications, the turbulence on scales larger than $1/k_{Oz}$ will begin to resemble standard strong stably-stratified turbulence, which forces the flow to exhibit horizontally elongated, pancake-like structures of large $m$ \citep{Lang2019}. Even though RsN205 is a non-dynamo run, this behaviour would presumably carry over to strongly stratified runs with dynamos.

\subsection{Parameter study and scaling laws of 3D MTI}\label{sec:scaling_laws}

We now perform a series of high-resolution runs, varying either the strength of the entropy stratification or the value of the thermal conductivity. Through them we numerically derive scaling laws for the principal properties of MTI turbulence, which we then compare with their 2D counterparts from Paper I. Finally, Section~\ref{sec:onset_dynamo} investigates the onset of dynamo action and establishes a numerical dynamo criterion.

\subsubsection{Energetics}\label{sec:energetics_scalings}

\begin{figure}
	\centering
	\includegraphics[width=1.0\columnwidth]{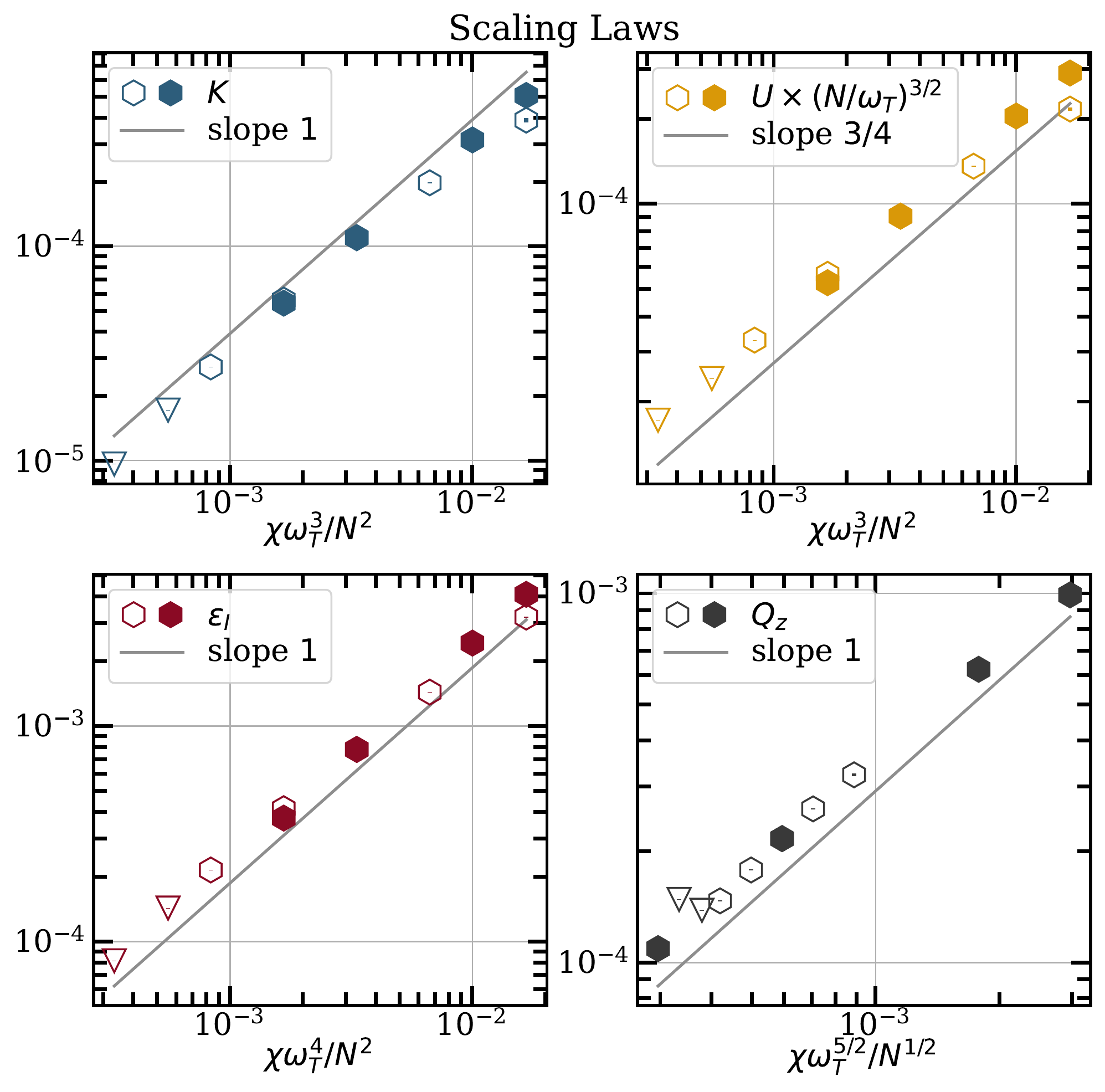}
	\caption{Scaling laws for 3D MTI high-res runs. Empty symbols represent runs RsN, while filled symbols represent runs RsPe. Simulations that do not support a dynamo are marked by triangles, while those with a dynamo are represented by hexagons.}
	\label{fig:3D_energetics_scalings}
\end{figure}

In Fig.~\ref{fig:3D_energetics_scalings} we plot the total kinetic and potential energies in the top left and top right panels, and the energy injection rate and the vertical heat flux in the bottom left and bottom right panels, respectively, for both run series RsPe and RsN, as functions of various parameter groupings. The averages have been computed over a time window of $50$ dynamical times at saturation, which we find sufficient for convergence.

The plots of the kinetic energy and of the energy injection rate confirm that these quantities follow the scaling laws derived in Sec.~\ref{sec:large_scales}, and which also appeared in 2D simulations, namely:
\begin{align}
	K \sim \frac{\chi \omega_T^3}{N^2}, \quad \epsilon_I \sim \frac{\chi \omega_T^4}{N^2}.  
\end{align} 
These scalings seem to be a generic feature of Boussinesq MTI rather than tied to the specific saturation mechanism.

In contrast, the potential energy and the vertical heat flux, obey power-laws slightly different to their two-dimensional counterparts. The potential energy in 3D brings in an additional $(N/\omega_T)^{-1/2}$ factor, while the heat flux depends on $\chi$  more steeply, and also brings in an additional $(N/\omega_T)^{-1/2}$ factor. The scaling law for the heat flux confirms that in 3D the MTI is more effective at transporting heat than in 2D.

\subsubsection{Integral scale and Reynolds number}\label{sec:buoyancy_scale}

\begin{table*}  
	\setlength{\tabcolsep}{0.5em} 
	\centering 
	\caption{Characteristic MTI scales of the 3D, high-resolution simulations in Table~\ref{tab:table_runs}. In addition to the scales defined in Sec.~\ref{sec:scales_numbers}, we also include the wavenumber of maximum growth according to linear theory ($k_{max}$).}  
	\label{tab:table_runs_scales} 
	\begin{tabular}{ccccccccccccc}  
		\hline 
		Run 	 &  $\text{Pe}$ 	 &  $\tilde{N}^2$ 	 &  $\text{Pr}$ 	 &  $k_{max}$ 	 &$k_{\chi}$ &  $k_i$ 	 &  $k_f$ 	 &  $k_{Oz}$ 	 &  $k_{\parallel}^{\theta}$ 	 &  $k_{\nu}$ 	 & $k_{\eta}$ &  $\text{Re}_i$ 	  \\ 
		\hline 
		RsPe00 	    &  $ 6 \times 10^{2} $ 	   &  $ 0.10 $ 	    &  $ 0.012 $ 	    &  $ 69.92 $ 	    & $24.49$ &  $ 44.1(8) $ 	    &  $ 56(2) $ 	    &  $ 2.12(2) $ 	    &  $ 26.4(1) $ 	    &  $ 650(4) $ 	    &  $2060(10)$ &  $ 36(1) $ 	    \\ 
		RsPe01 	    &  $ 1 \times 10^{3} $ 	   &  $ 0.10 $ 	    &  $ 0.02 $ 	    &  $ 78.21 $ 	    & $31.62$ &  $ 51.4(5) $ 	    &  $ 57(1) $ 	    &  $ 2.52(2) $ 	    &  $ 31.1(1) $ 	    &  $ 563(3) $ 	    & $1780(10)$ &  $ 24.3(4) $ 	    \\ 
		RsPe02/D0 	    &  $ 3 \times 10^{3} $ 	   &  $ 0.10 $ 	    &  $ 0.06 $ 	    &  $ 98.13 $ 	    & $54.77$ &  $ 67.6(6) $ 	    &  $ 98(6) $ 	    &  $ 3.76(1) $ 	    &  $ 45.6(1) $ 	    &  $ 407(2) $ 	    & $1287(5)$ &  $ 11.0(1) $ 	    \\ 
		RsPe03 	    &  $ 6 \times 10^{3} $ 	   &  $ 0.10 $ 	    &  $ 0.12 $ 	    &  $ 111.74 $ 	    & $77.46$ &  $ 78.7(5) $ 	    &  $ 110(6) $ 	    &  $ 4.856(8) $ 	    &  $ 58.13(6) $ 	    &  $ 324.7(6) $ 	    & $ 1027(2) $ &  $ 6.62(6) $ 	    \\ 
		\hline 
		RsN200 	    &  $ 3 \times 10^{3} $ 	   &  $ 0.02 $ 	    &  $ 0.06 $ 	    &  $ 95.79 $ 	    & $54.77$ &  $ 35.1(6) $ 	    &  $ 57(1) $ 	    &  $ 0.838(5) $ 	    &  $ 33.55(5) $ 	    &  $ 556(4) $ 	    & $1760(10)$ &  $ 40(1) $ 	    \\ 
		RsN201 	    &  $ 3 \times 10^{3} $ 	   &  $ 0.05 $ 	    &  $ 0.06 $ 	    &  $ 96.69 $ 	    & $54.77$ &  $ 51.7(6) $ 	    &  $ 70(10)  $ 	    &  $ 1.930(4) $ 	    &  $ 39.31(4) $ 	    &  $ 475(2) $ 	    & $1502(5)$ &  $ 19.3(3) $ 	    \\ 
		RsN202 	    &  $ 3 \times 10^{3} $ 	   &  $ 0.20 $ 	    &  $ 0.06 $ 	    &  $ 100.84 $ 	    & $54.77$ &  $ 86.7(4) $ 	    &  $ 117(9) $ 	    &  $ 7.52(1) $ 	    &  $ 53.97(7) $ 	    &  $ 337.0(6) $ 	    & $ 1066(2) $ &  $ 6.11(4) $ 	    \\ 
		RsN203 	    &  $ 3 \times 10^{3} $ 	   &  $ 0.40 $ 	    &  $ 0.06 $ 	    &  $ 105.74 $ 	    & $54.77$ &  $ 108.2(3) $ 	    &  $ 152(9) $ 	    &  $ 15.44(1) $ 	    &  $ 65.27(3) $ 	    &  $ 271.8(4) $ 	    &  $ 860(1) $  &  $ 3.42(1) $ 	    \\ 
		RsN204 	    &  $ 3 \times 10^{3} $ 	   &  $ 0.60 $ 	    &  $ 0.06 $ 	    &  $ 110.07 $ 	    & $54.77$ &  $ 122.4(2) $ 	    &  $ 162(5) $ 	    &  $ 23.92(3) $ 	    &  $ 71.98(6) $ 	    &  $ 235.5(2) $ 	    & $ 744.8(6) $ &  $ 2.392(6) $ 	    \\ 
		RsN205 	    &  $ 3 \times 10^{3} $ 	   &  $ 1.00 $ 	    &  $ 0.06 $ 	    &  $ 117.50 $ 	    & $54.77$ &  $ 138.1(2) $ 	    &  $ 181(7) $ 	    &  $ 41.91(4) $ 	    &  $ 76.44(5) $ 	    &  $ 195.5(1) $ 	    & $ 618.3(3) $ &  $ 1.589(3) $ 	    \\ 
		\hline 
		D0NoDyn 	    &  $ 3 \times 10^{3} $ 	   &  $ 0.10 $ 	    &  $ 0.06 $ 	    &  $ 65.52 $ 	    & $54.77$ &  $ 54.3(4) $ 	    &  $ 58(5) $ 	    &  $ 3.21(1) $ 	    &  $ 42.9(1) $ 	    &  $ 507(2) $ 	    & $ 321(1) $ &  $ 19.6(2) $ 	    \\ 
		D0Bx1e-2$^{\dagger}$ 	    &  $ 3 \times 10^{3} $ 	   &  $ 0.10 $ 	    &  $ 0.06 $ 	    &  $ 48.64 $ 	    & $54.77$ &  $ 60(1) $ 	    &  $ 38(3)  $ 	    &  $ 3.17(6) $ 	    &  $ 11.75(9) $ 	    &  $ 380(20)  $ 	    & $1210(5)$ &  $ 11.9(7) $ 	    \\ 
		D0Bz1e-2 	    &  $ 3 \times 10^{3} $ 	   &  $ 0.10 $ 	    &  $ 0.06 $ 	    &  \dotfill 	    & $54.77$ &  $ 57(2) $ 	    &  $100(10)$ 	    &  $ 4.72(2) $ 	    &  $ 38.1(1) $ 	    &  $ 327(5) $ 	    & $1030(20)$ &  $ 10.4(3) $ 	    \\ 
		\hline 
	\end{tabular} 
\end{table*}  

\begin{figure}
	\centering
	\includegraphics[width=1.0\columnwidth]{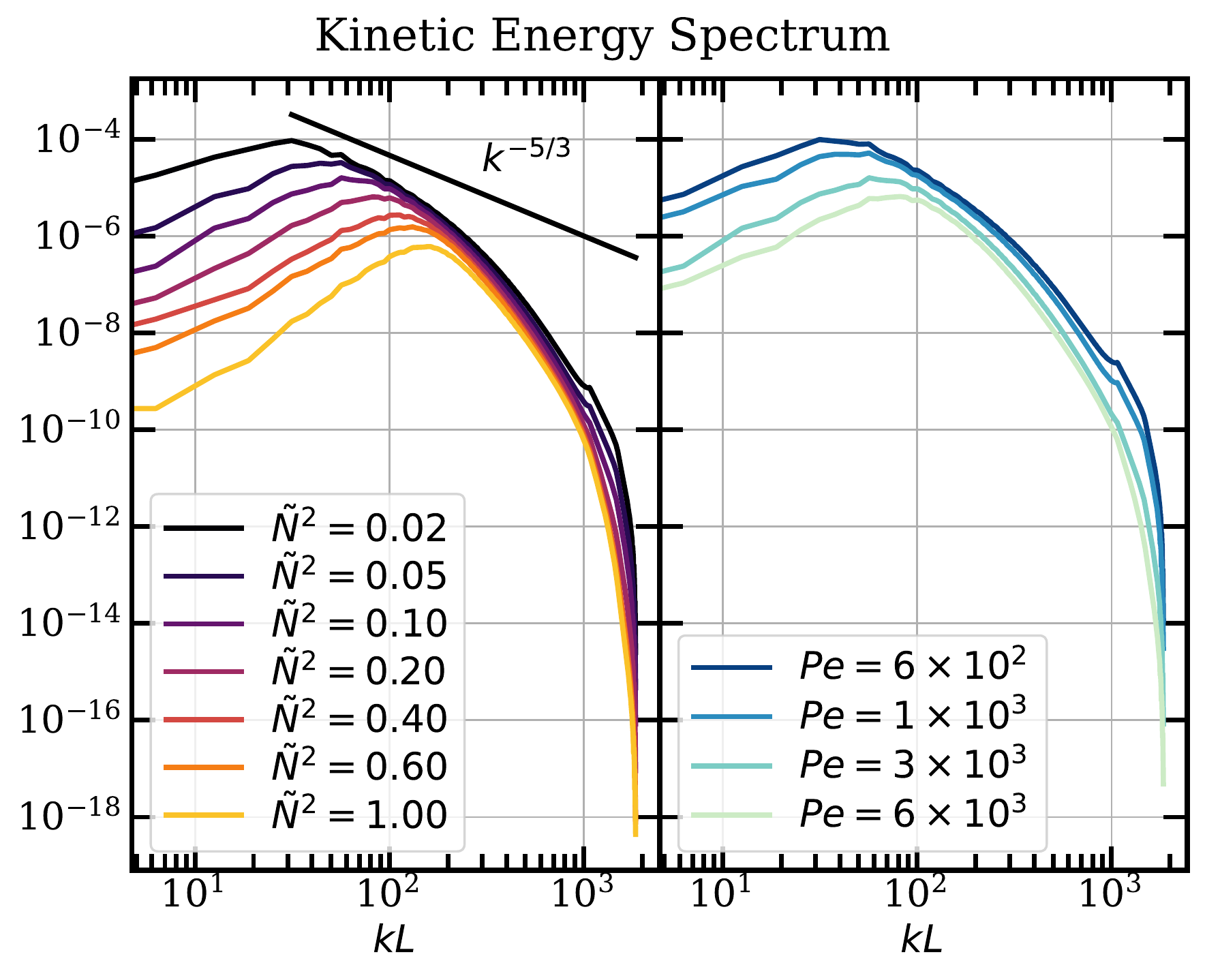}
	\caption{Kinetic power spectra at saturation for run RsN (left panel), and run RsPe, (right panel). The Kolmogorov $-5/3$ slope is shown for reference.}
	\label{fig:3D_all_kin_spectrum}
\end{figure}

We now look at the energy distribution in $k$ space for run series RsPe and RsN, plotting in Fig.~\ref{fig:3D_all_kin_spectrum} their kinetic power spectrum at saturation. As in the 2D cases, the power spectrum is strongly modified when the thermal conductivity and entropy stratification vary: specifically, we observe that the kinetic power spectra peaks at increasingly larger scales as we decrease $N^2$ and/or $\text{Pe}$.

Table~\ref{tab:table_runs_scales} shows how the various turbulent scales vary with  entropy stratification and thermal conduction. 
In particular, we look at the integral scale,  the MTI forcing scale and the Ozmidov scale.

Both the integral and forcing scales move to higher wavenumbers as the Peclet number and the Brunt-V\"ais\"al\"a frequency increase, as indicated already in Fig.~\ref{fig:3D_all_kin_spectrum}. Because the parameter range explored is smaller in our 3D study compared to 2D (due to numerical limitations), we do not compute best-fit power-laws, but their slopes do appear to be shallower than in the 2D runs. The forcing and integral scales are comparable, with no significant transport of energy across scales as was found in 2D.

\begin{figure}
	\centering
	\includegraphics[width=0.85\columnwidth]{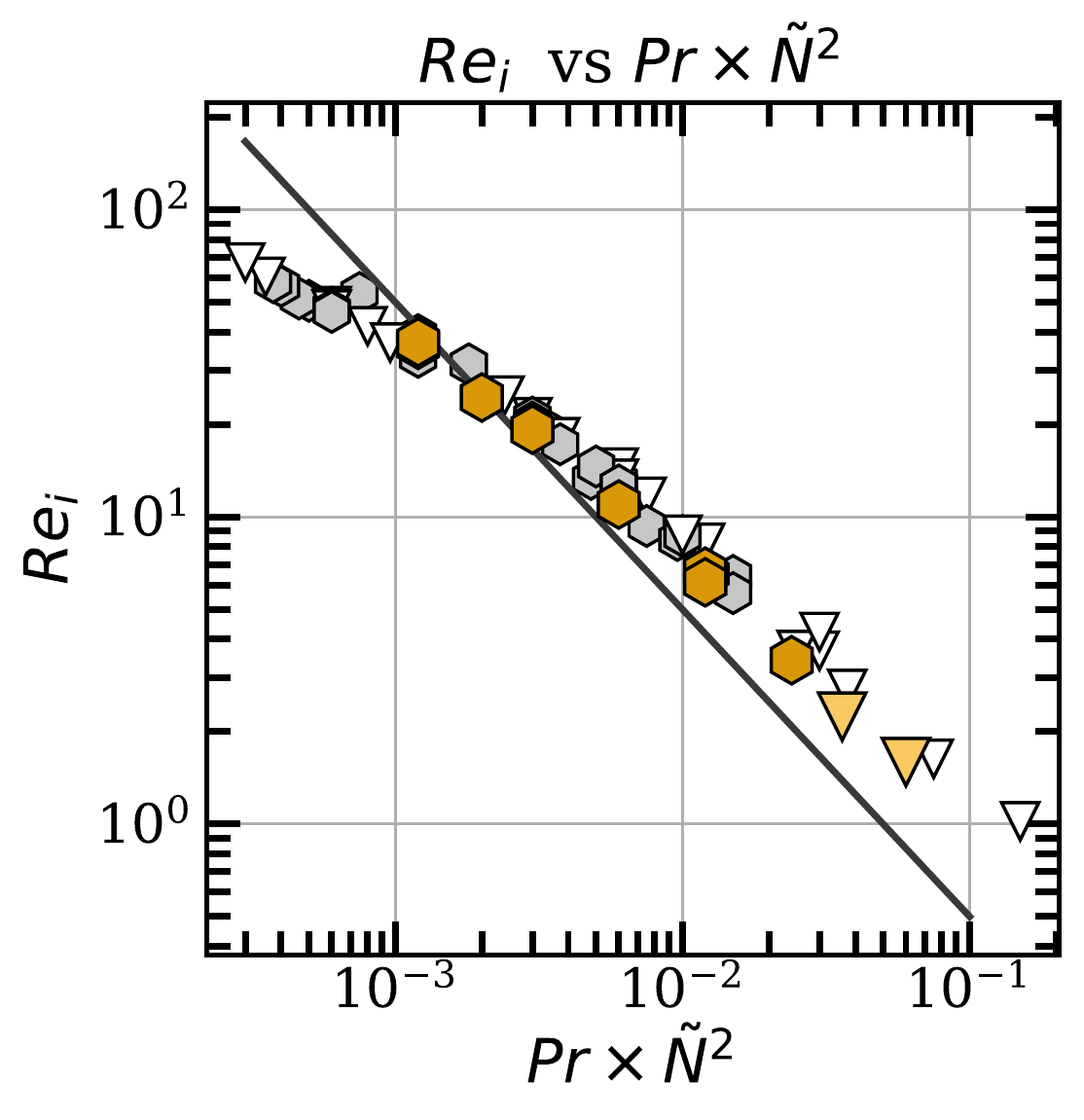}
	\caption{Scaling of $\text{Re}_i$. The data from the high resolution runs is plotted in dark and light gold, in grey and white for the low resolution dynamo runs. Runs that do not support a dynamo are marked by triangles, while those that do are represented by hexagons. }
	\label{fig:scaling_ReB_PrN2}
\end{figure}

We now test the predicted scaling of the Reynolds number at the integral scale ($\text{Re}_i$) shown in Eq.~\eqref{eq:rey_buoy}, which was obtained  by balancing MTI forcing and the buoyancy force, averaging over a time-window of $50 \omega_T^{-1}$. The results of this calculation are shown in Fig.~\ref{fig:scaling_ReB_PrN2}, where we plot the data for the high resolution runs (dark and light gold), together with the data of a number of lower-resolution runs (grey hexagons and white triangles), which we use in Sec.~\ref{sec:onset_dynamo}. As is clear, all the various numerical data does collapse on to a single curve, as expected, but 
this curve is only similar to the theoretical prediction.

\subsubsection{Criterion for dynamo action}\label{sec:onset_dynamo}

We expect the onset of the fluctuation dynamo to be completely determined by our physical parameters, thermal conductivity, entropy and temperature stratification, resistivity, and viscosity. These parameters can be combined to give 4 independent dimensionless numbers $\tilde{ N}$, $\text{Pr}$, $\text{Pm}$, $\text{Re}_i$. We might imagine that there exists a surface in this four-dimensional space that discriminates between growth and decay of the magnetic energy, but sampling this parameter landscape will be laborious. Fortunately, we can greatly reduce the problem by appealing to the physics of MTI-driven turbulence: in Section~\ref{sec:scaling_laws} we found that (a) it is the product $\tilde{ N}^2  \text{Pr}$, rather than the two individual parameters, that determines the large-scale properties of the turbulent flow, and that (b) at saturation the approximate relationship  $\text{Re}_i \sim (\text{Pr} \tilde{N}^2)^{-1}$ holds. These relations project our parameter space onto the plane $(\text{Re}_i, \text{Pm})$. Drawing a parallel with isotropic MHD turbulence, we expect that dynamo action is possible when $\text{Re}_i$ is greater than a certain critical value that depends only on $\text{Pm}$.

We performed an additional large number of runs at lower resolution ($256^3$), sweeping through the physical parameters. All of the dynamo runs, which we summarize in Table~\ref{tab:table_dynamo_runs}, were initialized with a weak horizontal magnetic field sinusoidal in $z$, so as to seed the MTI, and with no-net flux. In each run, we compute the dynamo growth/decay rate $\gamma$ after the initial MTI growth by performing a least-squares fit to $\ln M$ as a function of time.

\begin{figure}
	\centering
	\includegraphics[width=1.0\columnwidth]{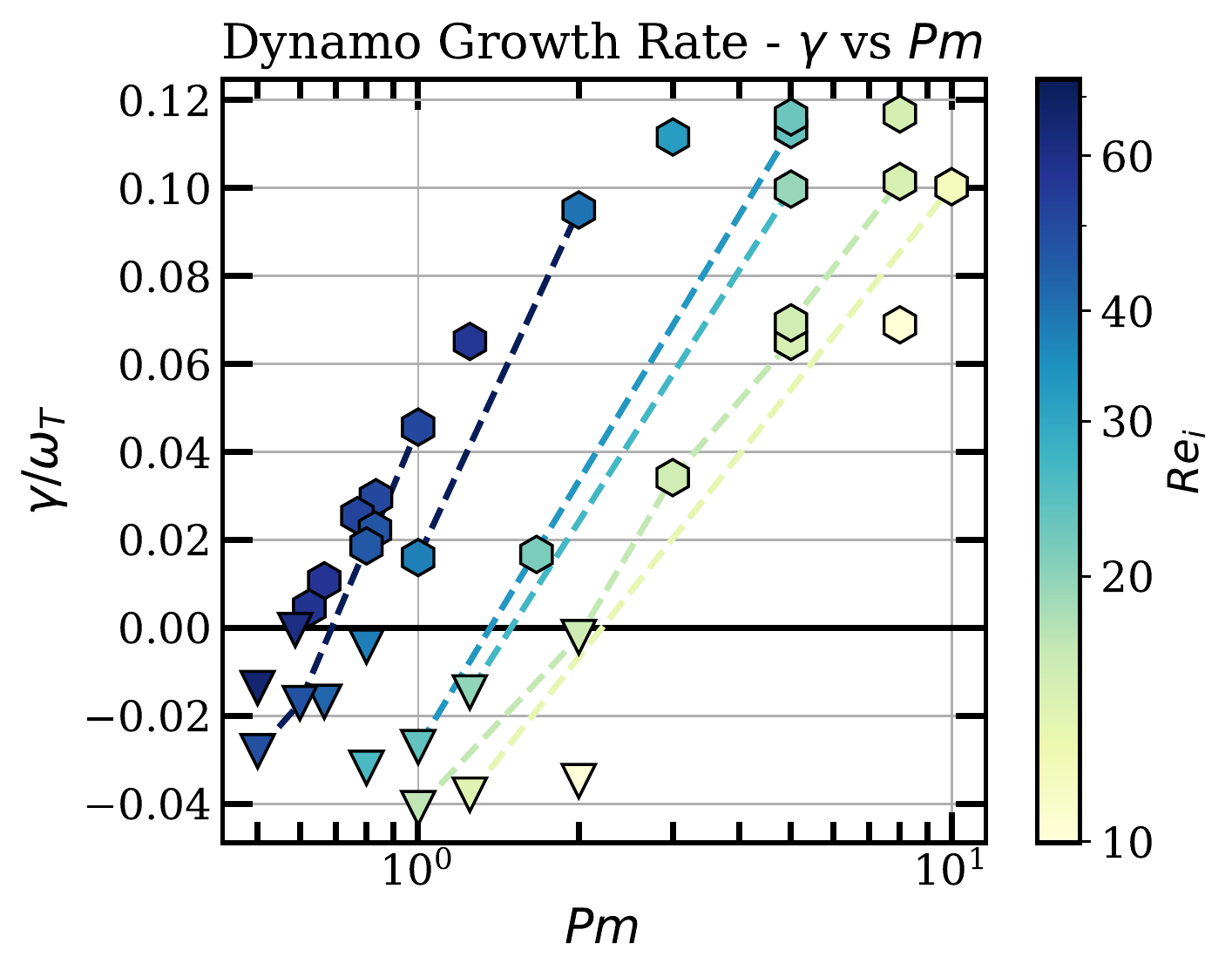}
	\caption{Growth rates of the fluctuation dynamo as a function of $\text{Pm}$ and $\text{Re}_i$. Data points are colored according to the value of $\text{Re}_i$, and runs with the same $\text{Pr}$ are connected by dashed lines. Dynamos are denoted by hexagons, non-dynamos by triangles. All runs have $\text{Pe}=6\times 10^2$ and $\tilde{N}^2=0.1$.}	\label{fig:scaling_dynamo_growth_NNF_gamma_Pm}
\end{figure}

\begin{figure}
	\centering
	\includegraphics[width=1.0\columnwidth]{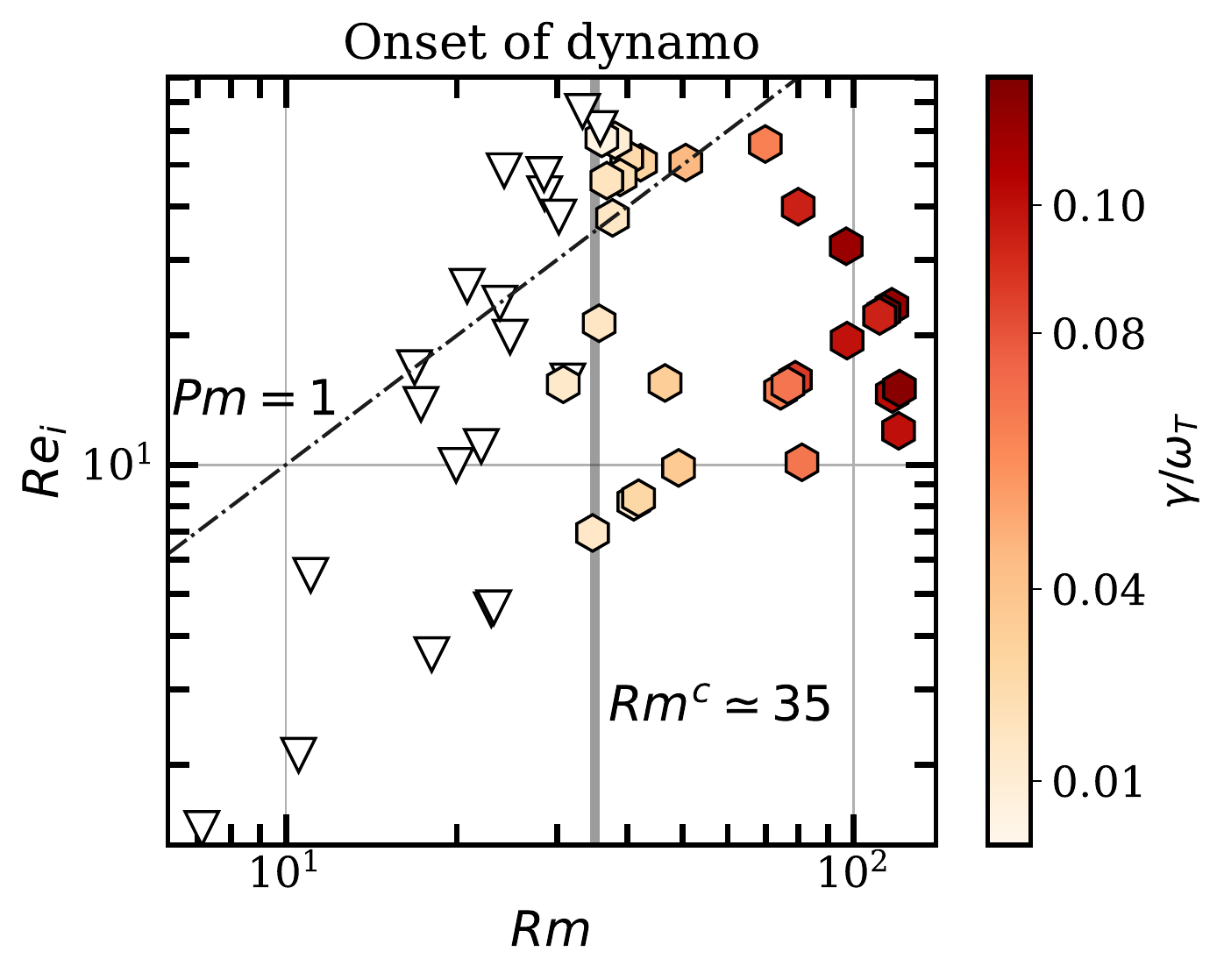}
	\caption{Onset criterion  of the fluctuation dynamo as a function of $\text{Rm}$ and $\text{Re}_i$. The grey  vertical line at $\text{Rm}^c \simeq 35$ represents the boundary between simulations that show dynamo amplification and those that don't. Hexagons indicate dynamo action, with the colour representing the associated growth rate. White triangles indicate non-dynamos. Above the dash-dotted line are runs with $\text{Pm}<1$, while below are those with $\text{Pm}>1$. }
	\label{fig:scaling_dynamo_growth_NNF_ReB_Rm}
\end{figure}

In Fig.~\ref{fig:scaling_dynamo_growth_NNF_gamma_Pm} we plot growth rates versus $\text{Pm}$ for a subset of the runs, all with the same Peclet number $\text{Pe}=6\times 10^2$ and entropy stratification $\tilde{N}^2=0.1$, but different $\text{Re}_i$. Runs with larger $\text{Re}_i$ (lower $\text{Pr}$) show faster dynamo growth rate at any given $\text{Pm}$. We also find evidence of magnetic field amplification for $\text{Pm} \lesssim 1$, but we cannot adequately explore this regime as our resolution struggles to represent the large separation of scales between $l_i$ and $l_{\nu}$. Plotting the growth rates vs $\text{Rm}$ (not shown), we find that all the data points collapse onto a single curve of the form $\gamma \propto \log (\text{Rm}/\text{Rm}^c)$, where $\text{Rm}^c$ is the critical value of the magnetic Reynolds number for dynamo amplification (see Fig.~\ref{fig:scaling_dynamo_growth_NNF_ReB_Rm}). This is consistent with an asymptotic scaling near criticality in a Kazantsev-Kraichnan model at low $\text{Pm}$ \citep{Kleeorin2012}.

The dynamo criterion is more transparently illustrated in the two-dimensional space $(\text{Re}_i, \text{Rm})$ (with $\text{Rm} = \text{Pm} \, \text{Re}_i$ the integral magnetic Reynolds number), which we employ in Fig.~\ref{fig:scaling_dynamo_growth_NNF_ReB_Rm}  
This shows that the runs with positive growth are all distributed in the right portion of the plane and are characterized by a near-monotonic increase in $\gamma$ as $\text{Rm}$ increases (for constant $\text{Re}_i$).
For $\text{Pm} \gtrsim 1$, the vertical line at $\text{Rm}^c \approx 35$ represents the critical value of the magnetic Reynolds number above which fluctuation dynamo is possible, which is in rough agreement with the known value of $\text{Rm}^c \simeq 60$ for the fluctuation dynamo at large $\text{Pm}$ \citep{Meneguzzi1981,Haugen2004,Schekochihin2004}.  
We note a slight increase in the critical magnetic Reynolds number for $\text{Pm} \lesssim 1$ that is consistent with the numerical results on the small $\text{Pm}$ fluctuation dynamo of \citet{Iskakov2007} and \citet{Schekochihin2007} where the value of $\text{Rm}^c$ has been observed to increase with the Reynolds number.

{
\renewcommand{\arraystretch}{0.9}
\begin{table}  
	\setlength{\tabcolsep}{0.5em} 
	\centering 
	\caption{Parameters and growth rates for the 3D dynamo runs. The resolution is $(256)^3$, with an initial horizontal and sinusoidal magnetic field with wavelength $=L$ and strength $B_0=10^{-5}$. }  
	\label{tab:table_dynamo_runs} 
	\begin{tabular}{ccccccc}  
		\hline 
		Run 	 &  $\text{Pe}$ 	 &  $\tilde{N}^2$ 	 &  $\text{Pr}$ 	 &  $\text{Pm}$ 	 &  $\text{Re}_i$ 	 &  $\gamma/\omega_T$ 	  \\ 
		\hline 
		DG00 	    &  $ 6 \times 10^{2} $ 	   &  $ 0.10 $ 	    &  $ 0.006 $ 	    &  $ 0.5 $ 	    &  $ 48.5(5) $ 	    &  $ -0.028 $ 	    \\ 
		DG01 	    &  $ 6 \times 10^{2} $ 	   &  $ 0.10 $ 	    &  $ 0.003 $ 	    &  $ 0.5 $ 	    &  $ 67(1) $ 	    &  $ -0.014 $ 	    \\ 
		DG02 	    &  $ 6 \times 10^{2} $ 	   &  $ 0.10 $ 	    &  $ 0.0035 $ 	    &  $ 0.59 $ 	    &  $ 60.7(7) $ 	    &  $ -0.001 $ 	    \\ 
		DG03 	    &  $ 6 \times 10^{2} $ 	   &  $ 0.10 $ 	    &  $ 0.006 $ 	    &  $ 0.6 $ 	    &  $ 47.5(4) $ 	    &  $ -0.017 $ 	    \\ 
		DG04 	    &  $ 6 \times 10^{2} $ 	   &  $ 0.10 $ 	    &  $ 0.0037 $ 	    &  $ 0.62 $ 	    &  $ 57.5(9) $ 	    &  $ 0.005 $ 	    \\ 
		DG05 	    &  $ 6 \times 10^{2} $ 	   &  $ 0.10 $ 	    &  $ 0.008 $ 	    &  $ 0.67 $ 	    &  $ 42.8(9) $ 	    &  $ -0.017 $ 	    \\ 
		DG06 	    &  $ 6 \times 10^{2} $ 	   &  $ 0.10 $ 	    &  $ 0.004 $ 	    &  $ 0.67 $ 	    &  $ 57(1) $ 	    &  $ 0.011 $ 	    \\ 
		DG07 	    &  $ 6 \times 10^{2} $ 	   &  $ 0.10 $ 	    &  $ 0.0046 $ 	    &  $ 0.77 $ 	    &  $ 51.8(7) $ 	    &  $ 0.026 $ 	    \\ 
		DG08 	    &  $ 6 \times 10^{2} $ 	   &  $ 0.10 $ 	    &  $ 0.024 $ 	    &  $ 0.8 $ 	    &  $ 26(1) $ 	    &  $ -0.032 $ 	    \\ 
		DG09 	    &  $ 6 \times 10^{2} $ 	   &  $ 0.10 $ 	    &  $ 0.0096 $ 	    &  $ 0.8 $ 	    &  $ 37.8(5) $ 	    &  $ -0.004 $ 	    \\ 
		DG10 	    &  $ 6 \times 10^{2} $ 	   &  $ 0.10 $ 	    &  $ 0.006 $ 	    &  $ 0.8 $ 	    &  $ 45.9(6) $ 	    &  $ 0.019 $ 	    \\ 
		DG11 	    &  $ 6 \times 10^{2} $ 	   &  $ 0.10 $ 	    &  $ 0.006 $ 	    &  $ 0.83 $ 	    &  $ 46.7(9) $ 	    &  $ 0.022 $ 	    \\ 
		DG12 	    &  $ 6 \times 10^{2} $ 	   &  $ 0.10 $ 	    &  $ 0.005 $ 	    &  $ 0.83 $ 	    &  $ 50.7(7) $ 	    &  $ 0.030 $ 	    \\ 
		DG13 	    &  $ 6 \times 10^{2} $ 	   &  $ 0.10 $ 	    &  $ 0.06 $ 	    &  $ 1.0 $ 	    &  $ 16.8(3) $ 	    &  $ -0.041 $ 	    \\ 
		DG14 	    &  $ 6 \times 10^{2} $ 	   &  $ 0.10 $ 	    &  $ 0.03 $ 	    &  $ 1.0 $ 	    &  $ 23.4(9) $ 	    &  $ -0.027 $ 	    \\ 
		DG15 	    &  $ 6 \times 10^{2} $ 	   &  $ 0.10 $ 	    &  $ 0.012 $ 	    &  $ 1.0 $ 	    &  $ 37.6(6) $ 	    &  $ 0.016 $ 	    \\ 
		DG16 	    &  $ 6 \times 10^{2} $ 	   &  $ 0.10 $ 	    &  $ 0.006 $ 	    &  $ 1.0 $ 	    &  $ 50(1) $ 	    &  $ 0.046 $ 	    \\ 
		DG17 	    &  $ 6 \times 10^{2} $ 	   &  $ 0.10 $ 	    &  $ 0.075 $ 	    &  $ 1.2 $ 	    &  $ 13.8(3) $ 	    &  $ -0.038 $ 	    \\ 
		DG18 	    &  $ 6 \times 10^{2} $ 	   &  $ 0.10 $ 	    &  $ 0.037 $ 	    &  $ 1.2 $ 	    &  $ 19.8(9) $ 	    &  $ -0.015 $ 	    \\ 
		DG19 	    &  $ 6 \times 10^{2} $ 	   &  $ 0.10 $ 	    &  $ 0.0075 $ 	    &  $ 1.2 $ 	    &  $ 56(2) $ 	    &  $ 0.065 $ 	    \\ 
		DG20 	    &  $ 6 \times 10^{2} $ 	   &  $ 0.10 $ 	    &  $ 0.033 $ 	    &  $ 1.7 $ 	    &  $ 21.4(7) $ 	    &  $ 0.017 $ 	    \\ 
		DG21 	    &  $ 3 \times 10^{3} $ 	   &  $ 0.50 $ 	    &  $ 0.06 $ 	    &  $ 2.0 $ 	    &  $ 5.53(4) $ 	    &  $ -0.025 $ 	    \\ 
		DG22 	    &  $ 6 \times 10^{2} $ 	   &  $ 0.10 $ 	    &  $ 0.12 $ 	    &  $ 2.0 $ 	    &  $ 10.0(8) $ 	    &  $ -0.035 $ 	    \\ 
		DG23 	    &  $ 1 \times 10^{3} $ 	   &  $ 0.50 $ 	    &  $ 0.02 $ 	    &  $ 2.0 $ 	    &  $ 11.04(9) $ 	    &  $ -0.021 $ 	    \\ 
		DG24 	    &  $ 6 \times 10^{2} $ 	   &  $ 0.50 $ 	    &  $ 0.012 $ 	    &  $ 2.0 $ 	    &  $ 15.4(2) $ 	    &  $ 0.014 $ 	    \\ 
		DG25 	    &  $ 6 \times 10^{2} $ 	   &  $ 0.10 $ 	    &  $ 0.06 $ 	    &  $ 2.0 $ 	    &  $ 15.6(8) $ 	    &  $ -0.002 $ 	    \\ 
		DG26 	    &  $ 6 \times 10^{2} $ 	   &  $ 0.10 $ 	    &  $ 0.012 $ 	    &  $ 2.0 $ 	    &  $ 42(3) $ 	    &  $ 0.095 $ 	    \\ 
		DG27 	    &  $ 6 \times 10^{2} $ 	   &  $ 0.10 $ 	    &  $ 0.06 $ 	    &  $ 3.0 $ 	    &  $ 15.5(5) $ 	    &  $ 0.034 $ 	    \\ 
		DG28 	    &  $ 6 \times 10^{2} $ 	   &  $ 0.10 $ 	    &  $ 0.018 $ 	    &  $ 3.0 $ 	    &  $ 32.5(7) $ 	    &  $ 0.112 $ 	    \\ 
		DG29 	    &  $ 6 \times 10^{3} $ 	   &  $ 0.50 $ 	    &  $ 0.3 $ 	    &  $ 5.0 $ 	    &  $ 1.42(1) $ 	    &  $ -0.019 $ 	    \\ 
		DG30 	    &  $ 3 \times 10^{3} $ 	   &  $ 0.50 $ 	    &  $ 0.15 $ 	    &  $ 5.0 $ 	    &  $ 2.11(2) $ 	    &  $ -0.029 $ 	    \\ 
		DG31 	    &  $ 3 \times 10^{3} $ 	   &  $ 0.50 $ 	    &  $ 0.075 $ 	    &  $ 5.0 $ 	    &  $ 3.61(2) $ 	    &  $ -0.022 $ 	    \\ 
		DG32 	    &  $ 6 \times 10^{3} $ 	   &  $ 0.10 $ 	    &  $ 0.3 $ 	    &  $ 5.0 $ 	    &  $ 4.60(6) $ 	    &  $ -0.024 $ 	    \\ 
		DG33 	    &  $ 1 \times 10^{3} $ 	   &  $ 0.50 $ 	    &  $ 0.05 $ 	    &  $ 5.0 $ 	    &  $ 4.65(6) $ 	    &  $ -0.021 $ 	    \\ 
		DG34 	    &  $ 6 \times 10^{2} $ 	   &  $ 0.50 $ 	    &  $ 0.03 $ 	    &  $ 5.0 $ 	    &  $ 6.9(1) $ 	    &  $ 0.015 $ 	    \\ 
		DG35 	    &  $ 3 \times 10^{3} $ 	   &  $ 0.10 $ 	    &  $ 0.15 $ 	    &  $ 5.0 $ 	    &  $ 8.1(3) $ 	    &  $ 0.010 $ 	    \\ 
		DG36 	    &  $ 1 \times 10^{3} $ 	   &  $ 0.50 $ 	    &  $ 0.025 $ 	    &  $ 5.0 $ 	    &  $ 8.36(5) $ 	    &  $ 0.027 $ 	    \\ 
		DG37 	    &  $ 1 \times 10^{3} $ 	   &  $ 0.10 $ 	    &  $ 0.1 $ 	    &  $ 5.0 $ 	    &  $ 9.8(3) $ 	    &  $ 0.036 $ 	    \\ 
		DG38 	    &  $ 6 \times 10^{2} $ 	   &  $ 0.10 $ 	    &  $ 0.06 $ 	    &  $ 5.0 $ 	    &  $ 14.9(7) $ 	    &  $ 0.065 $ 	    \\ 
		DG39 	    &  $ 6 \times 10^{2} $ 	   &  $ 0.10 $ 	    &  $ 0.06 $ 	    &  $ 5.0 $ 	    &  $ 15.3(3) $ 	    &  $ 0.069 $ 	    \\ 
		DG40 	    &  $ 1 \times 10^{3} $ 	   &  $ 0.10 $ 	    &  $ 0.05 $ 	    &  $ 5.0 $ 	    &  $ 15.8(3) $ 	    &  $ 0.087 $ 	    \\ 
		DG41 	    &  $ 6 \times 10^{2} $ 	   &  $ 0.10 $ 	    &  $ 0.037 $ 	    &  $ 5.0 $ 	    &  $ 19.5(2) $ 	    &  $ 0.100 $ 	    \\ 
		DG42 	    &  $ 6 \times 10^{2} $ 	   &  $ 0.05 $ 	    &  $ 0.06 $ 	    &  $ 5.0 $ 	    &  $ 22.2(4) $ 	    &  $ 0.094 $ 	    \\ 
		DG43 	    &  $ 6 \times 10^{2} $ 	   &  $ 0.10 $ 	    &  $ 0.03 $ 	    &  $ 5.0 $ 	    &  $ 22.6(7) $ 	    &  $ 0.116 $ 	    \\ 
		DG44 	    &  $ 6 \times 10^{2} $ 	   &  $ 0.10 $ 	    &  $ 0.03 $ 	    &  $ 5.0 $ 	    &  $ 23.3(7) $ 	    &  $ 0.113 $ 	    \\ 
		DG45 	    &  $ 6 \times 10^{2} $ 	   &  $ 0.10 $ 	    &  $ 0.096 $ 	    &  $ 8.0 $ 	    &  $ 10.1(3) $ 	    &  $ 0.069 $ 	    \\ 
		DG46 	    &  $ 6 \times 10^{2} $ 	   &  $ 0.10 $ 	    &  $ 0.06 $ 	    &  $ 8.0 $ 	    &  $ 14.6(3) $ 	    &  $ 0.101 $ 	    \\ 
		DG47 	    &  $ 6 \times 10^{2} $ 	   &  $ 0.10 $ 	    &  $ 0.048 $ 	    &  $ 8.0 $ 	    &  $ 15.4(6) $ 	    &  $ 0.117 $ 	    \\ 
		DG48 	    &  $ 6 \times 10^{2} $ 	   &  $ 0.10 $ 	    &  $ 0.075 $ 	    &  $ 10.0 $ 	    &  $ 12.0(4) $ 	    &  $ 0.100 $ 	    \\ 
		\hline 
	\end{tabular} 
\end{table}  
}

\subsection{Other field configurations}

To complement the parameter study in Sec.~\ref{sec:scaling_laws}, we perform a number of runs changing the magnetic field initial conditions, focusing on setups with no net-flux and with purely vertical fields. Our main findings confirm the expectation that dynamo runs with no net-flux converge to the same saturated state, irrespective of the orientation or strength of the initial field. Runs with a strong (i.e. comparable to the saturated $u_{rms}$) initial magnetic field, on the other hand,  develop a pronounced vertical anisotropy and structures at the box size. We refer the reader to Appendix~\ref{app:other_field} for more details.

\section{Application to galaxy clusters}\label{sec:applications}

In this section we apply our results to real galaxy clusters, employing the numerically derived scaling laws to estimate the values of typical turbulent quantities in their peripheries.
Apart from our modelling assumptions, the main source of error and uncertainty lies in our limited understanding of the ICM's microphysics, as it bears on its transport processes.
For instance, in dilute magnetized plasmas the presence of kinetic micro-instabilities (particularly the mirror and whistler instabilities) can lead to a suppression of viscosity and thermal conductivity, as charged particles scatter off the stochastic electromagnetic fluctuations.
To take into account the suppression of conductivity, we write $\chi = f \chi_\text{S}$, where $f$ is a suppression factor (summarising micro-instabilities and tangled fields) and $\chi_\text{S}$ is the Spitzer formula for thermal diffusivity:
\begin{align}
	\chi_\text{S} = 4.98 \times 10^{31} \left( \frac{k_B T_e}{5 \si{keV}}\right)^{5/2} \left( \frac{n_e}{10^{-3}\si{cm^{-3}}}\right)^{-1} \si{cm^2.s^{-1}}.
\end{align}
For simplicity, we assume $k_B T_e \simeq 5\si{keV}$ and $n_e \simeq 10^{-4}\si{cm^{-3}}$, and take  $\omega_T^{-1} \sim N^{-1} \sim 600 \si{Myr}$, which is appropriate for the periphery of galaxy clusters\footnote{While the growth rate of the MTI may seem relatively `slow', we stress that the instability will attempt to erase the temperature gradient, curbing its own growth as it develops. Thus this estimate is of a `saturated' growth rate.}, as shown in Paper I. 

Using the best-fit power law $K \simeq 0.03(\chi \omega_T^3/N^2)$, and assuming that the magnetic fields are amplified by a fluctuation dynamo to equipartition with the turbulent kinetic energy ($M \sim 0.5-1 K$) we obtain
\begin{align}\label{eq:galaxy_clusters_estimates}
	u_{rms} \approx 400 f^{1/2} \si{km.s^{-1}}, \, \, \, \, \, \, \beta \approx 10-20 f^{-1}.
\end{align}

We can also estimate the key lengthscales of the MTI turbulence. The integral scale is a fraction of the conduction $l_{\chi} \sim 100 \, f^{1/2} \si{kpc}$, for moderate entropy stratifications.
And as the cluster size is $ R \sim \si{Mpc}$, this offers a partial justification to our adoption of a local model of the ICM (see also discussion in Paper I).
Similarly, using Eqs.~\eqref{eq:rey_buoy}-\eqref{eq:scalings_K_lB} and $\text{Pr} = 0.02$ we can estimate the Reynolds number at the integral scale as $\text{Re}_i \approx 10$. The viscous scale follows and is $l_{\nu} \sim 10 \, f^{1/2} \si{kpc}$. 
Finally, for $N/\omega_T \sim 1$ we estimate the Ozmidov scale to be $l_{Oz} \gtrsim l_{\chi}$ and thus potentially smaller or similar to the cluster size. If this is the case, we would expect turbulent MTI eddies on scales of hundreds of kiloparsecs to not necessarily be vertically elongated, on account of the entropy stratification (see discussion in Sec.~\ref{sec:strongly-stratified}). This is an important point, and contrasts with the usual assumption that a large-scale radial bias in the turbulent energies is evidence of the MTI. It is unlikely, however, that there is sufficient room in the cluster to house the opposite regime of vertically suppressed turbulence on scales $\gg l_{Oz}$.

Lastly, we note that the MTI is capable of driving a heat flux that can be larger than the value of $1/3$ of the Spitzer flux expected from a medium of tangled fields (in other words, $\text{Nu} \gtrsim 1/3$ in all our runs). This is partly due to the contribution of advection, which is a significant fraction of the conductive flux, and which is often neglected in models of galaxy clusters because it is assumed that the ICM is convectively stable \citep{Zakamska2003,Brueggen2003,Kim2003,Pope2006,Voit2011}. With the MTI, however, this is no longer the case, and the role of advective transport must be taken into account.

Typical estimates of the characteristic quantities in galaxy clusters show turbulent velocities of $\sim 300-500 \si{km.s^{-1}}$ \citep{Sanders2011,Churazov2012}, with subthermal magnetic fields ($\beta \sim 100$) at equipartition with the kinetic energy \citep{Carilli2002,Vogt2005}. Moreover, assuming Kolmogorov scaling and the Spitzer value for viscosity, an estimate for the viscous scale in the outskirts of Coma cluster puts it around $\sim 50 \si{kpc}$ \citep{Zhuravleva2019}. However, recent observations of Doppler line broadening with Hitomi in the Perseus cluster favour a lower estimate of $u_{rms} \simeq 100 \si{km.s^{-1}}$ \citep{Hitomi2018}. Future observations will be necessary to obtain more stringent bounds on the turbulent levels in the ICM, but we note that our estimates in Eq.~\eqref{eq:galaxy_clusters_estimates} may be consistent with the lower estimates from Hitomi with a suppression factor of $f \sim 1/10$.

\section{Discussion and Conclusion}\label{sec:conclusions}

\subsection{Summary}

In this work, we extend the 2D study of Paper I by performing a detailed numerical analysis of the MTI in three dimensions, focusing on a parameter regime appropriate for the ICM in the periphery of galaxy clusters.

We find that the MTI produces a state of buoyancy-driven turbulence sustained by energy drawn from the background temperature gradient and composed of density and velocity fluctuations over a wide range of scales. The majority of the energy is dissipated on scales shorter than the conduction length by anisotropic thermal diffusion,  while a fraction is converted into kinetic fluctuations via the buoyancy force at intermediate scales, where it is either viscously dissipated or converted into magnetic fluctuations through the Lorentz force. The latter conversion of energy is connected to a fluctuation dynamo, not possible in 2D, which operates at the viscous scale and on timescales of the order of $ \omega_T^{-1}$. The coupling between density and kinetic fluctuations is most effective at the scale where MTI injection and buoyancy become comparable; this roughly coincides with the peak of the kinetic energy spectrum (the integral scale -- $1/k_i$). On larger scales sits the Ozmidov length, which marks the transition from vertically elongated eddies to horizontally elongated eddies; at the Ozmidov scale itself the turbulent motions are relatively isotropic. The separation between the Ozmidov and integral scales is determined by the strength of the entropy stratification; in realistic ICM models, we generally find that the two scales are roughly of the same order.

At saturation the turbulent properties follow clear power laws; specifically, the kinetic energy obeys $K\sim \chi \omega_T^3/N^2$, and the energy injection rate $\epsilon_I \sim \chi \omega_T^4/N^2$. Despite the fact that the energy flows in spectral space contrast significantly with the 2D MTI (where an inverse cascade transports energy to large scales), the 3D scalings for $K$, $\epsilon_I$, and for the turbulent Reynolds number ($\text{Re}_i \sim  \chi \omega_T^2/N^2 \nu$) are the same. Unlike 2D, however, we fail to find evidence of large-scale g-mode excitation. Through an extensive parameter sweep we numerically determine the criterion for dynamo action, which occurs when the integral scale is larger than the viscous scale, and if $\text{Rm}> \text{Rm}^c \approx 35$, a value similar to previous work on forced fluctuation dynamos.

The scaling laws for the turbulent energy levels derived in this work can be applied to the ICM and compared to actual observations of galaxy clusters. We find $u_{rms} \approx 400 f^{1/2} \si{km.s^{-1}}$ and $\beta \approx 10 f^{-1}$, with $f$ the `suppression factor', quantifying the degree to which microturbulence reduces the thermal diffusivity from Spitzer values. These estimates are in rough agreement with observed values in the periphery of galaxy clusters, if we adopt a realistic suppression factor of $f \sim 1/10$ (see next subsection). Although these results are based on a local Boussinesq model of the ICM, thus preventing us from studying the global dynamics of the MTI on scales comparable to the pressure scale-height, the conduction length (and hence the integral scale) is smaller than the system size $ l_{\chi} \approx 100 \, f^{1/2} \si{kpc} \lesssim  \si{Mpc}$ and thus partially justifies our adoption of a local model. Finally, because the Ozmidov scale appears to sit between the integral and system scales, we do not expect global MTI motions to be radially elongated.  

\subsection{Context and Challenges}

In this paper, we have shown that turbulence may be generated \textit{in situ} in the periphery of galaxy clusters by the MTI. The approximate agreement of our numerical estimates with the actual values of the turbulent energy levels in the ICM is encouraging, but the MTI is only one of several potential source of turbulence. AGN jets can stir the cores of galaxy clusters through the inflation of bubbles (or lobes) \citep{Boehringer1993,McNamara2007,Fabian2012} that can subsequently excite g-modes \citep{Zhang2018,Bambic2018}, shocks \citep{Randall2015,Forman2017}, and sound waves \citep{Fujita2005,Fabian2017}. An alternative scenario may involve the presence of a `sloshing core' that may enhance mixing and facilitate the inflow of heat  \citep{ZuHone2010,ZuHone2013}. However, how the energy produced by these mechanisms is communicated beyond $50-100 \si{kpc}$ from the core is still unclear \citep{Bourne2021}.
On the other hand, merger events and galactic motions could drive large-scale ($\sim \si{Mpc}$) turbulence in the periphery of galaxy clusters directly \citep{Markevitch2001,Churazov2003}, where we expect the MTI to manifest.

Importantly, it has been claimed that these competing sources of turbulence could actively suppress the action of buoyancy instabilities, such as the MTI \citep[e.g.,][]{Bogdanovic2009,Ruszkowski2010,Ruszkowski2011,McCourt2011,Parrish2012,Parrish2012a,Yang2016}. These claims are based, primarily, on the results of numerical simulations rather than observations.  For instance, a number of papers have explored the impact of anisotropic heat conduction in large-scale cosmological style simulations but failed to uncover compelling evidence of MTI action in the outskirts of galaxy clusters \citep{Ruszkowski2010,Ruszkowski2011,ZuHone2013a,Yang2016,Kannan2017,Barnes2019}. This work, and its interpretation require some comment. For one, disentangling the signature of MTI turbulence (or any clear physical mechanism) from the highly disordered flows produced is very challenging \citep{Ruszkowski2011}. But also it is unclear if the key MTI scales are adequately resolved, despite the use of adaptive mesh refinement, and how badly the MTI is (numerically) inhibited if they are not. Moreover, the inevitable and uncontrolled numerical diffusion of heat has been poorly quantified, and if too large will artificially degrade the anisotropic character of the heat conduction (and hence MTI activity). Finally, the diagnostics commonly employed to identify the presence of the MTI, such as the radial bias in the velocity or magnetic field orientation, may not be appropriate on the large-scales resolved by such simulations; as we have shown in this paper, a realistic stable entropy stratification will tend to suppress this radial bias on global ICM scales.

Granted that other sources of (externally driven) turbulence co-exist or compete with the MTI, we might only expect the MTI to be suppressed if the turbulent velocities produced by its competitors are significantly larger than the MTI's at the conduction scale. However, we have demonstrated that the MTI is capable of generating turbulent velocities of the same order as those observed in galaxy clusters, and thus has a good chance of `holding its own' in this hostile environment. Other points concern the volume-filling factor of the turbulence driven by these external sources, which has yet to be quantified, as well as the scale-dependence of the suppression: local studies suggest that the MTI cannot be fully quenched at all scales \citep{McCourt2011}.

So much for the large-scales. But the MTI can also encounter problems if the ICM's anisotropic thermal conductivity is diminished either because of the magnetic field's small-scale tangled geometry or the peculiar micro-instabilities that afflict dilute, weakly-collisional plasmas.
The exponential divergence of tangled magnetic field lines increases the distance travelled by charged particles, restoring diffusion perpendicular to the field \citep{Rechester1978,Chandran1998}.
If the magnetic field is chaotic over a wide range of scales, however, this effect results in a correction of only a factor of $\sim 1/5$ compared to the Spitzer value \citep{Narayan2001,Chandran2004}.
On the other hand, the impact of kinetic micro-scale instabilities on heat transport represents a major source of uncertainty. Recent numerical studies of weakly collisional, magnetized plasmas (not including the MTI) have focused on the mirror and whistler instabilities, suggesting that they can suppress the parallel conductivity by a factor of $\sim 1/5$ \citep{Komarov2016} and  by a factor of $\sim 1/\beta_e$ \citep{Riquelme2016,Roberg2018,Komarov2018,Drake2020} (possibly amounting to $\sim 1/4 - 1$ in the outskirts of galaxy clusters \citep{Komarov2018}).

The interplay between these microinstabilities and the MTI itself has not been examined in detail until relatively recently and represents an area of ongoing research. In the case of the mirror instability, MHD simulations, in which kinetic effects are modelled through a sub-grid prescription for the heat conductivity, have indicated that the overall impact on the MTI may be surprisingly small \citep{Berlok2021}, because regions that are mirror-unstable coincide with those of fastest linear MTI growth. More generally, this demonstrates the intriguing possibility that the MTI can control, to some degree, its own pressure anisotropy so as to minimise the onset and impact of deleterious micro-instabilities. As a consequence, a simple uniform reduction of the thermal conductivity by some fixed factor may be inappropriate. We anticipate a rich vein of research exploring the two way interaction between large and small-scale instabilities, possibly using physically motivated closures.

Finally, \citet{Xu2016} have shown that in a weakly collisional plasmas (such is the ICM) the MTI itself can manifest on the (sub-ion-Larmor) microscales via its kinetic analogue, the electron-MTI, which is driven by electron temperature gradients and grows faster than the fluid (collisional) MTI. The presence of this new kinetic instability, which might act alongside its fluid counterpart, complicates the picture yet again on the very small-scales. In particular, it is not straightforward how the electron-MTI interacts with (possibly impedes?) the other high-$\beta$ microinstabilities. We are thus quite some distance away from clear conclusions on the net suppression of heat conduction on small-scales.

\subsection{Future Work}

To address these open questions, additional physics and more sophisticated physical models are required. On the small scales, the inclusion of anisotropic pressure will alter the character of the fluctuation dynamo, as detailed in \citet{StOnge2020}, with and without pressure limiters, though we anticipate that the magnetic field's folded structure will be largely unaffected. It is also likely the MTI's other saturated properties will alter, though not critically \citep[e.g.,][]{Kunz2012}.
Perhaps the anisotropic pressure's greatest impact will be indirect, via its incitement of microinstabilities (and their impact on heat conduction), which future work should certainly accounted for.  
The clean numerical setup offered by the Boussinesq approximation is uniquely well suited to study the two-way dynamics, and pressure anisotropy regulation in particular, of the large-scale MTI and these microinstabilities.  
Specifically, one might revisit the work of \citet{Berlok2021} on the suppression of heat conduction by the mirror instability -- sidestepping certain numerical issues in their compressible simulations -- and conduct a parameter study in 3D, where there may be more scope for the MTI to marginalise the microinstability. Finally we could also adopt more refined closures so as to model the suppression of heat conductivity by the whistler instability, such as that introduced by \citet{Drake2020}.

An inevitable shortcoming of the Boussinesq approximation, however, is the impossibility of capturing the dynamics of the MTI on scales comparable to the pressure (or density) scale-height of the ICM. This prevents us from determining the interplay between the MTI and other sources of turbulence on large scales. In order to do so, well-constructed global simulations of the ICM in a Boussinesq spherical shell, within the anelastic approximation (where the background pressure/density gradients need not be small), or with fully compressible models (to study the excitation of transonic flows) are required. This may include revisiting the global simulations of \citet{McCourt2011} but with due prominence given to the stable entropy stratification, which we have shown to be a crucial ingredient in the saturation of the MTI.

\section*{Acknowledgements}

The authors thank Fran\c{c}ois Rincon and Thomas Berlok for helpful and generous comments on a draft version of the manuscript.

\section*{Data availability and environmental impact}

The data underlying this article will be shared on reasonable request to the corresponding author. The numerical simulations presented in Paper I and in this work required approximately $2.5 \times 10^6$ CPU-core hours and $10^4 \si{kWh}$ of electrical power to run, emitting about $3\times 10^3 \si{kgCO_2e}$ based on the conversion factor for the UK power grid. This is comparable to the $\si{CO_2}$ emissions per passenger of a return flight London-Perth.




\bibliographystyle{mnras}
\bibliography{MTI_GalaxyClusters} 

\begin{thebibliography}{}
\makeatletter
\relax
\def\mn@urlcharsother{\let\do\@makeother \do\$\do\&\do\#\do\^\do\_\do\%\do\~}
\def\mn@doi{\begingroup\mn@urlcharsother \@ifnextchar [ {\mn@doi@}
  {\mn@doi@[]}}
\def\mn@doi@[#1]#2{\def\@tempa{#1}\ifx\@tempa\@empty \href
  {http://dx.doi.org/#2} {doi:#2}\else \href {http://dx.doi.org/#2} {#1}\fi
  \endgroup}
\def\mn@eprint#1#2{\mn@eprint@#1:#2::\@nil}
\def\mn@eprint@arXiv#1{\href {http://arxiv.org/abs/#1} {{\tt arXiv:#1}}}
\def\mn@eprint@dblp#1{\href {http://dblp.uni-trier.de/rec/bibtex/#1.xml}
  {dblp:#1}}
\def\mn@eprint@#1:#2:#3:#4\@nil{\def\@tempa {#1}\def\@tempb {#2}\def\@tempc
  {#3}\ifx \@tempc \@empty \let \@tempc \@tempb \let \@tempb \@tempa \fi \ifx
  \@tempb \@empty \def\@tempb {arXiv}\fi \@ifundefined
  {mn@eprint@\@tempb}{\@tempb:\@tempc}{\expandafter \expandafter \csname
  mn@eprint@\@tempb\endcsname \expandafter{\@tempc}}}

\bibitem[\protect\citeauthoryear{Alexiades, Amiez  \& Gremaud}{Alexiades
  et~al.}{1996}]{Alexiades1996}
Alexiades V.,  Amiez G.,   Gremaud P.-A.,  1996, \mn@doi [Communications in
  Numerical Methods in Engineering]
  {10.1002/(SICI)1099-0887(199601)12:1<31::AID-CNM950>3.0.CO;2-5}, 12, 31

\bibitem[\protect\citeauthoryear{Balbus}{Balbus}{2000}]{Balbus2000}
Balbus S.~A.,  2000, \mn@doi [\apj] {10.1086/308732}, 534, 420

\bibitem[\protect\citeauthoryear{Balbus}{Balbus}{2001}]{Balbus2001}
Balbus S.~A.,  2001, \mn@doi [\apj] {10.1086/323875}, 562, 909

\bibitem[\protect\citeauthoryear{{Bale}, {Kasper}, {Howes}, {Quataert}, {Salem}
   \& {Sundkvist}}{{Bale} et~al.}{2009}]{Bale2009}
{Bale} S.~D.,  {Kasper} J.~C.,  {Howes} G.~G.,  {Quataert} E.,  {Salem} C.,
  {Sundkvist} D.,  2009, \mn@doi [\prl] {10.1103/PhysRevLett.103.211101}, \href
  {https://ui.adsabs.harvard.edu/abs/2009PhRvL.103u1101B} {103, 211101}

\bibitem[\protect\citeauthoryear{{Bale}, {Pulupa}, {Salem}, {Chen}  \&
  {Quataert}}{{Bale} et~al.}{2013}]{Bale2013}
{Bale} S.~D.,  {Pulupa} M.,  {Salem} C.,  {Chen} C.~H.~K.,   {Quataert} E.,
  2013, \mn@doi [\apjl] {10.1088/2041-8205/769/2/L22}, \href
  {https://ui.adsabs.harvard.edu/abs/2013ApJ...769L..22B} {769, L22}

\bibitem[\protect\citeauthoryear{{Bambic}, {Morsony}  \& {Reynolds}}{{Bambic}
  et~al.}{2018}]{Bambic2018}
{Bambic} C.~J.,  {Morsony} B.~J.,   {Reynolds} C.~S.,  2018, \mn@doi [\apj]
  {10.3847/1538-4357/aab558}, \href
  {https://ui.adsabs.harvard.edu/abs/2018ApJ...857...84B} {857, 84}

\bibitem[\protect\citeauthoryear{{Barnes} et~al.,}{{Barnes}
  et~al.}{2019}]{Barnes2019}
{Barnes} D.~J.,  et~al., 2019, \mn@doi [\mnras] {10.1093/mnras/stz1814}, \href
  {https://ui.adsabs.harvard.edu/abs/2019MNRAS.488.3003B} {488, 3003}

\bibitem[\protect\citeauthoryear{{Batchelor}}{{Batchelor}}{1969}]{Batchelor1969}
{Batchelor} G.~K.,  1969, \mn@doi [Physics of Fluids] {10.1063/1.1692443},
  \href {https://ui.adsabs.harvard.edu/abs/1969PhFl...12C.233B} {12, II}

\bibitem[\protect\citeauthoryear{{Berlok}, {Quataert}, {Pessah}  \&
  {Pfrommer}}{{Berlok} et~al.}{2021}]{Berlok2021}
{Berlok} T.,  {Quataert} E.,  {Pessah} M.~E.,   {Pfrommer} C.,  2021, \mn@doi
  [\mnras] {10.1093/mnras/stab832}, \href
  {https://ui.adsabs.harvard.edu/abs/2021MNRAS.504.3435B} {504, 3435}

\bibitem[\protect\citeauthoryear{{Boehringer}, {Voges}, {Fabian}, {Edge}  \&
  {Neumann}}{{Boehringer} et~al.}{1993}]{Boehringer1993}
{Boehringer} H.,  {Voges} W.,  {Fabian} A.~C.,  {Edge} A.~C.,   {Neumann}
  D.~M.,  1993, \mn@doi [\mnras] {10.1093/mnras/264.1.L25}, \href
  {https://ui.adsabs.harvard.edu/abs/1993MNRAS.264L..25B} {264, L25}

\bibitem[\protect\citeauthoryear{{Bogdanovi{\'c}}, {Reynolds}, {Balbus}  \&
  {Parrish}}{{Bogdanovi{\'c}} et~al.}{2009}]{Bogdanovic2009}
{Bogdanovi{\'c}} T.,  {Reynolds} C.~S.,  {Balbus} S.~A.,   {Parrish} I.~J.,
  2009, \mn@doi [\apj] {10.1088/0004-637X/704/1/211}, \href
  {https://ui.adsabs.harvard.edu/abs/2009ApJ...704..211B} {704, 211}

\bibitem[\protect\citeauthoryear{{Bonafede}, {Feretti}, {Murgia}, {Govoni},
  {Giovannini}, {Dallacasa}, {Dolag}  \& {Taylor}}{{Bonafede}
  et~al.}{2010}]{Bonafede2010}
{Bonafede} A.,  {Feretti} L.,  {Murgia} M.,  {Govoni} F.,  {Giovannini} G.,
  {Dallacasa} D.,  {Dolag} K.,   {Taylor} G.~B.,  2010, \mn@doi [\aap]
  {10.1051/0004-6361/200913696}, \href
  {https://ui.adsabs.harvard.edu/abs/2010A&A...513A..30B} {513, A30}

\bibitem[\protect\citeauthoryear{{Bott} et~al.,}{{Bott}
  et~al.}{2020}]{Bott2020}
{Bott} A.~F.~A.,  et~al., 2020, arXiv e-prints, \href
  {https://ui.adsabs.harvard.edu/abs/2020arXiv200806594B} {p. arXiv:2008.06594}

\bibitem[\protect\citeauthoryear{{Bourne} \& {Sijacki}}{{Bourne} \&
  {Sijacki}}{2021}]{Bourne2021}
{Bourne} M.~A.,  {Sijacki} D.,  2021, \mn@doi [\mnras]
  {10.1093/mnras/stab1662}, \href
  {https://ui.adsabs.harvard.edu/abs/2021MNRAS.506..488B} {506, 488}

\bibitem[\protect\citeauthoryear{Braginskii}{Braginskii}{1965}]{Braginskii1965a}
Braginskii S.~I.,  1965, Reviews of Plasma Physics, 1, 205

\bibitem[\protect\citeauthoryear{{Brandenburg}}{{Brandenburg}}{2011}]{Brandenburg2011}
{Brandenburg} A.,  2011, \mn@doi [\apj] {10.1088/0004-637X/741/2/92}, \href
  {https://ui.adsabs.harvard.edu/abs/2011ApJ...741...92B} {741, 92}

\bibitem[\protect\citeauthoryear{{Brandenburg}, {Haugen}, {Li}  \&
  {Subramanian}}{{Brandenburg} et~al.}{2018}]{Brandenburg2018}
{Brandenburg} A.,  {Haugen} N.~E.~L.,  {Li} X.-Y.,   {Subramanian} K.,  2018,
  \mn@doi [\mnras] {10.1093/mnras/sty1570}, \href
  {https://ui.adsabs.harvard.edu/abs/2018MNRAS.479.2827B} {479, 2827}

\bibitem[\protect\citeauthoryear{{Br{\"u}ggen}}{{Br{\"u}ggen}}{2003}]{Brueggen2003}
{Br{\"u}ggen} M.,  2003, \mn@doi [\apj] {10.1086/376734}, \href
  {https://ui.adsabs.harvard.edu/abs/2003ApJ...593..700B} {593, 700}

\bibitem[\protect\citeauthoryear{{Carilli} \& {Taylor}}{{Carilli} \&
  {Taylor}}{2002}]{Carilli2002}
{Carilli} C.~L.,  {Taylor} G.~B.,  2002, \mn@doi [\araa]
  {10.1146/annurev.astro.40.060401.093852}, \href
  {https://ui.adsabs.harvard.edu/abs/2002ARA&A..40..319C} {40, 319}

\bibitem[\protect\citeauthoryear{{Cattaneo}, {Lenz}  \& {Weiss}}{{Cattaneo}
  et~al.}{2001}]{Cattaneo2001}
{Cattaneo} F.,  {Lenz} D.,   {Weiss} N.,  2001, \mn@doi [\apjl]
  {10.1086/338355}, \href
  {https://ui.adsabs.harvard.edu/abs/2001ApJ...563L..91C} {563, L91}

\bibitem[\protect\citeauthoryear{{Chandran} \& {Cowley}}{{Chandran} \&
  {Cowley}}{1998}]{Chandran1998}
{Chandran} B. D.~G.,  {Cowley} S.~C.,  1998, \mn@doi [\prl]
  {10.1103/PhysRevLett.80.3077}, \href
  {https://ui.adsabs.harvard.edu/abs/1998PhRvL..80.3077C} {80, 3077}

\bibitem[\protect\citeauthoryear{{Chandran} \& {Maron}}{{Chandran} \&
  {Maron}}{2004}]{Chandran2004}
{Chandran} B. D.~G.,  {Maron} J.~L.,  2004, \mn@doi [\apj] {10.1086/380897},
  \href {https://ui.adsabs.harvard.edu/abs/2004ApJ...602..170C} {602, 170}

\bibitem[\protect\citeauthoryear{{Chen}, {Matteini}, {Schekochihin}, {Stevens},
  {Salem}, {Maruca}, {Kunz}  \& {Bale}}{{Chen} et~al.}{2016}]{Chen2016}
{Chen} C.~H.~K.,  {Matteini} L.,  {Schekochihin} A.~A.,  {Stevens} M.~L.,
  {Salem} C.~S.,  {Maruca} B.~A.,  {Kunz} M.~W.,   {Bale} S.~D.,  2016, \mn@doi
  [\apjl] {10.3847/2041-8205/825/2/L26}, \href
  {https://ui.adsabs.harvard.edu/abs/2016ApJ...825L..26C} {825, L26}

\bibitem[\protect\citeauthoryear{{Churazov}, {Sunyaev}, {Forman}  \&
  {B{\"o}hringer}}{{Churazov} et~al.}{2002}]{Churazov2002}
{Churazov} E.,  {Sunyaev} R.,  {Forman} W.,   {B{\"o}hringer} H.,  2002,
  \mn@doi [\mnras] {10.1046/j.1365-8711.2002.05332.x}, \href
  {https://ui.adsabs.harvard.edu/abs/2002MNRAS.332..729C} {332, 729}

\bibitem[\protect\citeauthoryear{{Churazov}, {Forman}, {Jones}  \&
  {B{\"o}hringer}}{{Churazov} et~al.}{2003}]{Churazov2003}
{Churazov} E.,  {Forman} W.,  {Jones} C.,   {B{\"o}hringer} H.,  2003, \mn@doi
  [\apj] {10.1086/374923}, \href
  {https://ui.adsabs.harvard.edu/abs/2003ApJ...590..225C} {590, 225}

\bibitem[\protect\citeauthoryear{{Churazov} et~al.,}{{Churazov}
  et~al.}{2012}]{Churazov2012}
{Churazov} E.,  et~al., 2012, \mn@doi [\mnras]
  {10.1111/j.1365-2966.2011.20372.x}, \href
  {https://ui.adsabs.harvard.edu/abs/2012MNRAS.421.1123C} {421, 1123}

\bibitem[\protect\citeauthoryear{Davidson}{Davidson}{2013}]{Davidson2013}
Davidson P.~A.,  2013, Turbulence in Rotating, Stratified and Electrically
  Conducting Fluids.
Cambridge University Press, \mn@doi{10.1017/CBO9781139208673}

\bibitem[\protect\citeauthoryear{{Delache}, {Cambon}  \& {Godeferd}}{{Delache}
  et~al.}{2014}]{Delache2014}
{Delache} A.,  {Cambon} C.,   {Godeferd} F.,  2014, \mn@doi [Physics of Fluids]
  {10.1063/1.4864099}, \href
  {https://ui.adsabs.harvard.edu/abs/2014PhFl...26b5104D} {26, 025104}

\bibitem[\protect\citeauthoryear{{Drake}, {Pfrommer}, {Reynolds}, {Ruszkowski},
  {Swisdak}, {Einarsson}, {Hassam}  \& {Roberg-Clark}}{{Drake}
  et~al.}{2020}]{Drake2020}
{Drake} J.~F.,  {Pfrommer} C.,  {Reynolds} C.~S.,  {Ruszkowski} M.,  {Swisdak}
  M.,  {Einarsson} A.,  {Hassam} A.~B.,   {Roberg-Clark} G.~T.,  2020, arXiv
  e-prints, \href {https://ui.adsabs.harvard.edu/abs/2020arXiv200707931D} {p.
  arXiv:2007.07931}

\bibitem[\protect\citeauthoryear{{Fabian}}{{Fabian}}{2012}]{Fabian2012}
{Fabian} A.~C.,  2012, \mn@doi [\araa] {10.1146/annurev-astro-081811-125521},
  \href {https://ui.adsabs.harvard.edu/abs/2012ARA&A..50..455F} {50, 455}

\bibitem[\protect\citeauthoryear{{Fabian}, {Walker}, {Russell}, {Pinto},
  {Sanders}  \& {Reynolds}}{{Fabian} et~al.}{2017}]{Fabian2017}
{Fabian} A.~C.,  {Walker} S.~A.,  {Russell} H.~R.,  {Pinto} C.,  {Sanders}
  J.~S.,   {Reynolds} C.~S.,  2017, \mn@doi [\mnras] {10.1093/mnrasl/slw170},
  \href {https://ui.adsabs.harvard.edu/abs/2017MNRAS.464L...1F} {464, L1}

\bibitem[\protect\citeauthoryear{{Forman}, {Churazov}, {Jones}, {Heinz},
  {Kraft}  \& {Vikhlinin}}{{Forman} et~al.}{2017}]{Forman2017}
{Forman} W.,  {Churazov} E.,  {Jones} C.,  {Heinz} S.,  {Kraft} R.,
  {Vikhlinin} A.,  2017, \mn@doi [\apj] {10.3847/1538-4357/aa70e4}, \href
  {https://ui.adsabs.harvard.edu/abs/2017ApJ...844..122F} {844, 122}

\bibitem[\protect\citeauthoryear{{Fujita} \& {Suzuki}}{{Fujita} \&
  {Suzuki}}{2005}]{Fujita2005}
{Fujita} Y.,  {Suzuki} T.~K.,  2005, \mn@doi [\apjl] {10.1086/491649}, \href
  {https://ui.adsabs.harvard.edu/abs/2005ApJ...630L...1F} {630, L1}

\bibitem[\protect\citeauthoryear{{Fusco-Femiano}, {Orlandini}, {Brunetti},
  {Feretti}, {Giovannini}, {Grandi}  \& {Setti}}{{Fusco-Femiano}
  et~al.}{2004}]{Fusco-Femiano2004}
{Fusco-Femiano} R.,  {Orlandini} M.,  {Brunetti} G.,  {Feretti} L.,
  {Giovannini} G.,  {Grandi} P.,   {Setti} G.,  2004, \mn@doi [\apjl]
  {10.1086/382695}, \href
  {https://ui.adsabs.harvard.edu/abs/2004ApJ...602L..73F} {602, L73}

\bibitem[\protect\citeauthoryear{{Ghirardini} et~al.,}{{Ghirardini}
  et~al.}{2019}]{Ghirardini2019}
{Ghirardini} V.,  et~al., 2019, \mn@doi [\aap] {10.1051/0004-6361/201833325},
  \href {https://ui.adsabs.harvard.edu/abs/2019A&A...621A..41G} {621, A41}

\bibitem[\protect\citeauthoryear{{Giovannini}, {Feretti}, {Venturi}, {Kim}  \&
  {Kronberg}}{{Giovannini} et~al.}{1993}]{Giovannini1993}
{Giovannini} G.,  {Feretti} L.,  {Venturi} T.,  {Kim} K.~T.,   {Kronberg}
  P.~P.,  1993, \mn@doi [\apj] {10.1086/172451}, \href
  {https://ui.adsabs.harvard.edu/abs/1993ApJ...406..399G} {406, 399}

\bibitem[\protect\citeauthoryear{{Govoni} \& {Feretti}}{{Govoni} \&
  {Feretti}}{2004}]{Govoni2004}
{Govoni} F.,  {Feretti} L.,  2004, \mn@doi [International Journal of Modern
  Physics D] {10.1142/S0218271804005080}, \href
  {https://ui.adsabs.harvard.edu/abs/2004IJMPD..13.1549G} {13, 1549}

\bibitem[\protect\citeauthoryear{{Haugen}, {Brandenburg}  \& {Dobler}}{{Haugen}
  et~al.}{2004}]{Haugen2004}
{Haugen} N.~E.,  {Brandenburg} A.,   {Dobler} W.,  2004, \mn@doi [\pre]
  {10.1103/PhysRevE.70.016308}, \href
  {https://ui.adsabs.harvard.edu/abs/2004PhRvE..70a6308H} {70, 016308}

\bibitem[\protect\citeauthoryear{{Hitomi Collaboration} et~al.,}{{Hitomi
  Collaboration} et~al.}{2018}]{Hitomi2018}
{Hitomi Collaboration} et~al., 2018, \mn@doi [\pasj] {10.1093/pasj/psx138},
  \href {https://ui.adsabs.harvard.edu/abs/2018PASJ...70....9H} {70, 9}

\bibitem[\protect\citeauthoryear{Iskakov, Schekochihin, Cowley, McWilliams  \&
  Proctor}{Iskakov et~al.}{2007}]{Iskakov2007}
Iskakov A.~B.,  Schekochihin A.~A.,  Cowley S.~C.,  McWilliams J.~C.,   Proctor
  M. R.~E.,  2007, \mn@doi [Physical Review Letters]
  {10.1103/physrevlett.98.208501}, 98

\bibitem[\protect\citeauthoryear{{Kannan}, {Vogelsberger}, {Pfrommer},
  {Weinberger}, {Springel}, {Hernquist}, {Puchwein}  \& {Pakmor}}{{Kannan}
  et~al.}{2017}]{Kannan2017}
{Kannan} R.,  {Vogelsberger} M.,  {Pfrommer} C.,  {Weinberger} R.,  {Springel}
  V.,  {Hernquist} L.,  {Puchwein} E.,   {Pakmor} R.,  2017, \mn@doi [\apjl]
  {10.3847/2041-8213/aa624b}, \href
  {https://ui.adsabs.harvard.edu/abs/2017ApJ...837L..18K} {837, L18}

\bibitem[\protect\citeauthoryear{{Kim} \& {Narayan}}{{Kim} \&
  {Narayan}}{2003}]{Kim2003}
{Kim} W.-T.,  {Narayan} R.,  2003, \mn@doi [\apj] {10.1086/378153}, \href
  {https://ui.adsabs.harvard.edu/abs/2003ApJ...596..889K} {596, 889}

\bibitem[\protect\citeauthoryear{{Kleeorin} \& {Rogachevskii}}{{Kleeorin} \&
  {Rogachevskii}}{2012}]{Kleeorin2012}
{Kleeorin} N.,  {Rogachevskii} I.,  2012, \mn@doi [\physscr]
  {10.1088/0031-8949/86/01/018404}, \href
  {https://ui.adsabs.harvard.edu/abs/2012PhyS...86a8404K} {86, 018404}

\bibitem[\protect\citeauthoryear{{Komarov}, {Churazov}, {Kunz}  \&
  {Schekochihin}}{{Komarov} et~al.}{2016}]{Komarov2016}
{Komarov} S.~V.,  {Churazov} E.~M.,  {Kunz} M.~W.,   {Schekochihin} A.~A.,
  2016, \mn@doi [\mnras] {10.1093/mnras/stw963}, \href
  {https://ui.adsabs.harvard.edu/abs/2016MNRAS.460..467K} {460, 467}

\bibitem[\protect\citeauthoryear{Komarov, Schekochihin, Churazov  \&
  Spitkovsky}{Komarov et~al.}{2018}]{Komarov2018}
Komarov S.,  Schekochihin A.~A.,  Churazov E.,   Spitkovsky A.,  2018, \mn@doi
  [Journal of Plasma Physics] {10.1017/S0022377818000399}, 84, 905840305

\bibitem[\protect\citeauthoryear{Kraichnan}{Kraichnan}{1967}]{Kraichnan1967}
Kraichnan R.~H.,  1967, \mn@doi [The Physics of Fluids] {10.1063/1.1762301},
  10, 1417

\bibitem[\protect\citeauthoryear{{Kulsrud} \& {Anderson}}{{Kulsrud} \&
  {Anderson}}{1992}]{Kulsrud1992}
{Kulsrud} R.~M.,  {Anderson} S.~W.,  1992, \mn@doi [\apj] {10.1086/171743},
  \href {https://ui.adsabs.harvard.edu/abs/1992ApJ...396..606K} {396, 606}

\bibitem[\protect\citeauthoryear{Kumar, Chatterjee  \& Verma}{Kumar
  et~al.}{2014}]{Kumar2014}
Kumar A.,  Chatterjee A.~G.,   Verma M.~K.,  2014, \mn@doi [Phys. Rev. E]
  {10.1103/PhysRevE.90.023016}, 90, 023016

\bibitem[\protect\citeauthoryear{Kunz}{Kunz}{2011}]{Kunz2011}
Kunz M.~W.,  2011, \mn@doi [\mnras] {10.1111/j.1365-2966.2011.19303.x}, 417,
  602

\bibitem[\protect\citeauthoryear{Kunz, Bogdanović, Reynolds  \& Stone}{Kunz
  et~al.}{2012}]{Kunz2012}
Kunz M.~W.,  Bogdanović T.,  Reynolds C.~S.,   Stone J.~M.,  2012, \mn@doi
  [\apj] {10.1088/0004-637X/754/2/122}, 754, 122

\bibitem[\protect\citeauthoryear{{Kunz}, {Schekochihin}  \& {Stone}}{{Kunz}
  et~al.}{2014}]{Kunz2014}
{Kunz} M.~W.,  {Schekochihin} A.~A.,   {Stone} J.~M.,  2014, \mn@doi [\prl]
  {10.1103/PhysRevLett.112.205003}, \href
  {https://ui.adsabs.harvard.edu/abs/2014PhRvL.112t5003K} {112, 205003}

\bibitem[\protect\citeauthoryear{{Lang} \& {Waite}}{{Lang} \&
  {Waite}}{2019}]{Lang2019}
{Lang} C.~J.,  {Waite} M.~L.,  2019, \mn@doi [Physical Review Fluids]
  {10.1103/PhysRevFluids.4.044801}, \href
  {https://ui.adsabs.harvard.edu/abs/2019PhRvF...4d4801L} {4, 044801}

\bibitem[\protect\citeauthoryear{Lappa}{Lappa}{2009}]{Lappa2009}
Lappa M.,  2009, Thermal Convection: Patterns, Evolution and Stability.
John Wiley \& Sons, Ltd, \mn@doi{https://doi.org/10.1002/9780470749982}

\bibitem[\protect\citeauthoryear{{Leccardi} \& {Molendi}}{{Leccardi} \&
  {Molendi}}{2008}]{Leccardi2008}
{Leccardi} A.,  {Molendi} S.,  2008, \mn@doi [\aap]
  {10.1051/0004-6361:200809538}, \href
  {https://ui.adsabs.harvard.edu/abs/2008A&A...486..359L} {486, 359}

\bibitem[\protect\citeauthoryear{{Lesur}}{{Lesur}}{2015}]{Lesur2015}
{Lesur} G.,  2015, {Snoopy: General purpose spectral solver} (\mn@eprint {ascl}
  {1505.022})

\bibitem[\protect\citeauthoryear{{Malyshkin} \& {Boldyrev}}{{Malyshkin} \&
  {Boldyrev}}{2010}]{Malyshkin2010}
{Malyshkin} L.~M.,  {Boldyrev} S.,  2010, \mn@doi [\prl]
  {10.1103/PhysRevLett.105.215002}, \href
  {https://ui.adsabs.harvard.edu/abs/2010PhRvL.105u5002M} {105, 215002}

\bibitem[\protect\citeauthoryear{Mandelbrot}{Mandelbrot}{1972}]{Mandelbrot1972}
Mandelbrot B.~B.,  1972, in Rosenblatt M.,  Van~Atta C.,  eds, Statistical
  Models and Turbulence. Springer Berlin Heidelberg, Berlin, Heidelberg, pp
  333--351

\bibitem[\protect\citeauthoryear{{Markevitch}, {Vikhlinin}  \&
  {Mazzotta}}{{Markevitch} et~al.}{2001}]{Markevitch2001}
{Markevitch} M.,  {Vikhlinin} A.,   {Mazzotta} P.,  2001, \mn@doi [\apjl]
  {10.1086/337973}, \href
  {https://ui.adsabs.harvard.edu/abs/2001ApJ...562L.153M} {562, L153}

\bibitem[\protect\citeauthoryear{{Maron}, {Cowley}  \& {McWilliams}}{{Maron}
  et~al.}{2004}]{Maron2004}
{Maron} J.,  {Cowley} S.,   {McWilliams} J.,  2004, \mn@doi [\apj]
  {10.1086/380504}, \href
  {https://ui.adsabs.harvard.edu/abs/2004ApJ...603..569M} {603, 569}

\bibitem[\protect\citeauthoryear{McCourt, Parrish, Sharma  \& Quataert}{McCourt
  et~al.}{2011}]{McCourt2011}
McCourt M.,  Parrish I.~J.,  Sharma P.,   Quataert E.,  2011, \mn@doi [\mnras]
  {10.1111/j.1365-2966.2011.18216.x}, 413, 1295

\bibitem[\protect\citeauthoryear{{McNamara} \& {Nulsen}}{{McNamara} \&
  {Nulsen}}{2007}]{McNamara2007}
{McNamara} B.~R.,  {Nulsen} P.~E.~J.,  2007, \mn@doi [\araa]
  {10.1146/annurev.astro.45.051806.110625}, \href
  {https://ui.adsabs.harvard.edu/abs/2007ARA&A..45..117M} {45, 117}

\bibitem[\protect\citeauthoryear{{Melville}, {Schekochihin}  \&
  {Kunz}}{{Melville} et~al.}{2016}]{Melville2016}
{Melville} S.,  {Schekochihin} A.~A.,   {Kunz} M.~W.,  2016, \mn@doi [\mnras]
  {10.1093/mnras/stw793}, \href
  {https://ui.adsabs.harvard.edu/abs/2016MNRAS.459.2701M} {459, 2701}

\bibitem[\protect\citeauthoryear{{Meneguzzi}, {Frisch}  \&
  {Pouquet}}{{Meneguzzi} et~al.}{1981}]{Meneguzzi1981}
{Meneguzzi} M.,  {Frisch} U.,   {Pouquet} A.,  1981, \mn@doi [\prl]
  {10.1103/PhysRevLett.47.1060}, \href
  {https://ui.adsabs.harvard.edu/abs/1981PhRvL..47.1060M} {47, 1060}

\bibitem[\protect\citeauthoryear{{Mignone}, {Bodo}, {Massaglia}, {Matsakos},
  {Tesileanu}, {Zanni}  \& {Ferrari}}{{Mignone} et~al.}{2007}]{Mignone2007}
{Mignone} A.,  {Bodo} G.,  {Massaglia} S.,  {Matsakos} T.,  {Tesileanu} O.,
  {Zanni} C.,   {Ferrari} A.,  2007, \mn@doi [\apjs] {10.1086/513316}, \href
  {https://ui.adsabs.harvard.edu/abs/2007ApJS..170..228M} {170, 228}

\bibitem[\protect\citeauthoryear{{Mohapatra}, {Federrath}  \&
  {Sharma}}{{Mohapatra} et~al.}{2020a}]{Mohapatra2020}
{Mohapatra} R.,  {Federrath} C.,   {Sharma} P.,  2020a, \mn@doi [\mnras]
  {10.1093/mnras/staa711}, \href
  {https://ui.adsabs.harvard.edu/abs/2020MNRAS.493.5838M} {493, 5838}

\bibitem[\protect\citeauthoryear{Mohapatra, Federrath  \& Sharma}{Mohapatra
  et~al.}{2020b}]{Mohapatra2020b}
Mohapatra R.,  Federrath C.,   Sharma P.,  2020b, \mn@doi [Monthly Notices of
  the Royal Astronomical Society] {10.1093/mnras/staa3564}, 500, 5072–5087

\bibitem[\protect\citeauthoryear{{Narayan} \& {Medvedev}}{{Narayan} \&
  {Medvedev}}{2001}]{Narayan2001}
{Narayan} R.,  {Medvedev} M.~V.,  2001, \mn@doi [\apjl] {10.1086/338325}, \href
  {https://ui.adsabs.harvard.edu/abs/2001ApJ...562L.129N} {562, L129}

\bibitem[\protect\citeauthoryear{Parrish \& Stone}{Parrish \&
  Stone}{2005}]{Parrish2005}
Parrish I.~J.,  Stone J.~M.,  2005, \mn@doi [\apj] {10.1086/444589}, 633, 334

\bibitem[\protect\citeauthoryear{Parrish \& Stone}{Parrish \&
  Stone}{2007}]{Parrish2007}
Parrish I.~J.,  Stone J.~M.,  2007, \mn@doi [\apj] {10.1086/518881}, 664, 135

\bibitem[\protect\citeauthoryear{Parrish, Stone  \& Lemaster}{Parrish
  et~al.}{2008}]{Parrish2008}
Parrish I.~J.,  Stone J.~M.,   Lemaster N.,  2008, \mn@doi [\apj]
  {10.1086/592380}, 688, 905

\bibitem[\protect\citeauthoryear{Parrish, McCourt, Quataert  \& Sharma}{Parrish
  et~al.}{2012a}]{Parrish2012}
Parrish I.~J.,  McCourt M.,  Quataert E.,   Sharma P.,  2012a, \mn@doi [\mnras]
  {10.1111/j.1745-3933.2011.01171.x}, 419, L29

\bibitem[\protect\citeauthoryear{Parrish, McCourt, Quataert  \& Sharma}{Parrish
  et~al.}{2012b}]{Parrish2012a}
Parrish I.~J.,  McCourt M.,  Quataert E.,   Sharma P.,  2012b, \mn@doi [\mnras]
  {10.1111/j.1365-2966.2012.20650.x}, 422, 704

\bibitem[\protect\citeauthoryear{{Pope}, {Pavlovski}, {Kaiser}  \&
  {Fangohr}}{{Pope} et~al.}{2006}]{Pope2006}
{Pope} E. C.~D.,  {Pavlovski} G.,  {Kaiser} C.~R.,   {Fangohr} H.,  2006,
  \mn@doi [\mnras] {10.1111/j.1365-2966.2006.10032.x}, \href
  {https://ui.adsabs.harvard.edu/abs/2006MNRAS.367.1121P} {367, 1121}

\bibitem[\protect\citeauthoryear{Prakash, Martínez~Mercado, van Wijngaarden,
  Mancilla, Tagawa, Lohse  \& Sun}{Prakash et~al.}{2016}]{Prakash2016}
Prakash V.~N.,  Martínez~Mercado J.,  van Wijngaarden L.,  Mancilla E.,
  Tagawa Y.,  Lohse D.,   Sun C.,  2016, \mn@doi [Journal of Fluid Mechanics]
  {10.1017/jfm.2016.49}, 791, 174–190

\bibitem[\protect\citeauthoryear{Quataert}{Quataert}{2008}]{Quataert2008}
Quataert E.,  2008, \mn@doi [\apj] {10.1086/525248}, \href
  {https://ui.adsabs.harvard.edu/abs/2008ApJ...673..758Q} {673, 758}

\bibitem[\protect\citeauthoryear{{Randall} et~al.,}{{Randall}
  et~al.}{2015}]{Randall2015}
{Randall} S.~W.,  et~al., 2015, \mn@doi [\apj] {10.1088/0004-637X/805/2/112},
  \href {https://ui.adsabs.harvard.edu/abs/2015ApJ...805..112R} {805, 112}

\bibitem[\protect\citeauthoryear{{Rechester} \& {Rosenbluth}}{{Rechester} \&
  {Rosenbluth}}{1978}]{Rechester1978}
{Rechester} A.~B.,  {Rosenbluth} M.~N.,  1978, \mn@doi [\prl]
  {10.1103/PhysRevLett.40.38}, \href
  {https://ui.adsabs.harvard.edu/abs/1978PhRvL..40...38R} {40, 38}

\bibitem[\protect\citeauthoryear{{Rincon}, {Ligni{\`e}res}  \&
  {Rieutord}}{{Rincon} et~al.}{2005}]{Rincon2005}
{Rincon} F.,  {Ligni{\`e}res} F.,   {Rieutord} M.,  2005, \mn@doi [\aap]
  {10.1051/0004-6361:200400130}, \href
  {https://ui.adsabs.harvard.edu/abs/2005A&A...430L..57R} {430, L57}

\bibitem[\protect\citeauthoryear{{Riquelme}, {Quataert}  \&
  {Verscharen}}{{Riquelme} et~al.}{2016}]{Riquelme2016}
{Riquelme} M.~A.,  {Quataert} E.,   {Verscharen} D.,  2016, \mn@doi [\apj]
  {10.3847/0004-637X/824/2/123}, \href
  {https://ui.adsabs.harvard.edu/abs/2016ApJ...824..123R} {824, 123}

\bibitem[\protect\citeauthoryear{{Roberg-Clark}, {Drake}, {Reynolds}  \&
  {Swisdak}}{{Roberg-Clark} et~al.}{2016}]{Roberg2016}
{Roberg-Clark} G.~T.,  {Drake} J.~F.,  {Reynolds} C.~S.,   {Swisdak} M.,  2016,
  \mn@doi [\apjl] {10.3847/2041-8205/830/1/L9}, \href
  {https://ui.adsabs.harvard.edu/abs/2016ApJ...830L...9R} {830, L9}

\bibitem[\protect\citeauthoryear{{Roberg-Clark}, {Drake}, {Reynolds}  \&
  {Swisdak}}{{Roberg-Clark} et~al.}{2018}]{Roberg2018}
{Roberg-Clark} G.~T.,  {Drake} J.~F.,  {Reynolds} C.~S.,   {Swisdak} M.,  2018,
  \mn@doi [\prl] {10.1103/PhysRevLett.120.035101}, \href
  {https://ui.adsabs.harvard.edu/abs/2018PhRvL.120c5101R} {120, 035101}

\bibitem[\protect\citeauthoryear{{Roediger}, {Lovisari}, {Dupke}, {Ghizzardi},
  {Br{\"u}ggen}, {Kraft}  \& {Machacek}}{{Roediger}
  et~al.}{2012}]{Roediger2012}
{Roediger} E.,  {Lovisari} L.,  {Dupke} R.,  {Ghizzardi} S.,  {Br{\"u}ggen} M.,
   {Kraft} R.~P.,   {Machacek} M.~E.,  2012, \mn@doi [\mnras]
  {10.1111/j.1365-2966.2011.20287.x}, \href
  {https://ui.adsabs.harvard.edu/abs/2012MNRAS.420.3632R} {420, 3632}

\bibitem[\protect\citeauthoryear{{Ruszkowski} \& {Oh}}{{Ruszkowski} \&
  {Oh}}{2010}]{Ruszkowski2010}
{Ruszkowski} M.,  {Oh} S.~P.,  2010, \mn@doi [\apj]
  {10.1088/0004-637X/713/2/1332}, \href
  {https://ui.adsabs.harvard.edu/abs/2010ApJ...713.1332R} {713, 1332}

\bibitem[\protect\citeauthoryear{{Ruszkowski}, {Lee}, {Br{\"u}ggen}, {Parrish}
  \& {Oh}}{{Ruszkowski} et~al.}{2011}]{Ruszkowski2011}
{Ruszkowski} M.,  {Lee} D.,  {Br{\"u}ggen} M.,  {Parrish} I.,   {Oh} S.~P.,
  2011, \mn@doi [\apj] {10.1088/0004-637X/740/2/81}, \href
  {https://ui.adsabs.harvard.edu/abs/2011ApJ...740...81R} {740, 81}

\bibitem[\protect\citeauthoryear{{Sanders}, {Fabian}  \& {Smith}}{{Sanders}
  et~al.}{2011}]{Sanders2011}
{Sanders} J.~S.,  {Fabian} A.~C.,   {Smith} R.~K.,  2011, \mn@doi [\mnras]
  {10.1111/j.1365-2966.2010.17561.x}, \href
  {https://ui.adsabs.harvard.edu/abs/2011MNRAS.410.1797S} {410, 1797}

\bibitem[\protect\citeauthoryear{{Sarazin}}{{Sarazin}}{1988}]{Sarazin1988}
{Sarazin} C.~L.,  1988, {X-ray emission from clusters of galaxies}

\bibitem[\protect\citeauthoryear{{Schekochihin} \& {Cowley}}{{Schekochihin} \&
  {Cowley}}{2006}]{Schekochihin2006}
{Schekochihin} A.~A.,  {Cowley} S.~C.,  2006, \mn@doi [Physics of Plasmas]
  {10.1063/1.2179053}, \href
  {https://ui.adsabs.harvard.edu/abs/2006PhPl...13e6501S} {13, 056501}

\bibitem[\protect\citeauthoryear{{Schekochihin}, {Cowley}, {Maron}  \&
  {Malyshkin}}{{Schekochihin} et~al.}{2001}]{Schekochihin2001}
{Schekochihin} A.,  {Cowley} S.,  {Maron} J.,   {Malyshkin} L.,  2001, \mn@doi
  [\pre] {10.1103/PhysRevE.65.016305}, \href
  {https://ui.adsabs.harvard.edu/abs/2001PhRvE..65a6305S} {65, 016305}

\bibitem[\protect\citeauthoryear{{Schekochihin}, {Cowley}, {Hammett}, {Maron}
  \& {McWilliams}}{{Schekochihin} et~al.}{2002}]{Schekochihin2002}
{Schekochihin} A.~A.,  {Cowley} S.~C.,  {Hammett} G.~W.,  {Maron} J.~L.,
  {McWilliams} J.~C.,  2002, \mn@doi [New Journal of Physics]
  {10.1088/1367-2630/4/1/384}, \href
  {https://ui.adsabs.harvard.edu/abs/2002NJPh....4...84S} {4, 84}

\bibitem[\protect\citeauthoryear{{Schekochihin}, {Cowley}, {Taylor}, {Maron}
  \& {McWilliams}}{{Schekochihin} et~al.}{2004}]{Schekochihin2004}
{Schekochihin} A.~A.,  {Cowley} S.~C.,  {Taylor} S.~F.,  {Maron} J.~L.,
  {McWilliams} J.~C.,  2004, \mn@doi [\apj] {10.1086/422547}, \href
  {https://ui.adsabs.harvard.edu/abs/2004ApJ...612..276S} {612, 276}

\bibitem[\protect\citeauthoryear{{Schekochihin}, {Cowley}, {Kulsrud}, {Hammett}
   \& {Sharma}}{{Schekochihin} et~al.}{2005}]{Schekochihin2005}
{Schekochihin} A.~A.,  {Cowley} S.~C.,  {Kulsrud} R.~M.,  {Hammett} G.~W.,
  {Sharma} P.,  2005, \mn@doi [\apj] {10.1086/431202}, \href
  {https://ui.adsabs.harvard.edu/abs/2005ApJ...629..139S} {629, 139}

\bibitem[\protect\citeauthoryear{{Schekochihin}, {Iskakov}, {Cowley},
  {McWilliams}, {Proctor}  \& {Yousef}}{{Schekochihin}
  et~al.}{2007}]{Schekochihin2007}
{Schekochihin} A.~A.,  {Iskakov} A.~B.,  {Cowley} S.~C.,  {McWilliams} J.~C.,
  {Proctor} M.~R.~E.,   {Yousef} T.~A.,  2007, \mn@doi [New Journal of Physics]
  {10.1088/1367-2630/9/8/300}, \href
  {https://ui.adsabs.harvard.edu/abs/2007NJPh....9..300S} {9, 300}

\bibitem[\protect\citeauthoryear{{Schekochihin}, {Cowley}, {Kulsrud}, {Rosin}
  \& {Heinemann}}{{Schekochihin} et~al.}{2008}]{Schekochihin2008}
{Schekochihin} A.~A.,  {Cowley} S.~C.,  {Kulsrud} R.~M.,  {Rosin} M.~S.,
  {Heinemann} T.,  2008, \mn@doi [\prl] {10.1103/PhysRevLett.100.081301}, \href
  {https://ui.adsabs.harvard.edu/abs/2008PhRvL.100h1301S} {100, 081301}

\bibitem[\protect\citeauthoryear{{Schober}, {Schleicher}, {Bovino}  \&
  {Klessen}}{{Schober} et~al.}{2012}]{Schober2012}
{Schober} J.,  {Schleicher} D.,  {Bovino} S.,   {Klessen} R.~S.,  2012, \mn@doi
  [\pre] {10.1103/PhysRevE.86.066412}, \href
  {https://ui.adsabs.harvard.edu/abs/2012PhRvE..86f6412S} {86, 066412}

\bibitem[\protect\citeauthoryear{{Schuecker}, {Finoguenov}, {Miniati},
  {B{\"o}hringer}  \& {Briel}}{{Schuecker} et~al.}{2004}]{Schuecker2004}
{Schuecker} P.,  {Finoguenov} A.,  {Miniati} F.,  {B{\"o}hringer} H.,   {Briel}
  U.~G.,  2004, \mn@doi [\aap] {10.1051/0004-6361:20041039}, \href
  {https://ui.adsabs.harvard.edu/abs/2004A&A...426..387S} {426, 387}

\bibitem[\protect\citeauthoryear{{Seta}, {Bushby}, {Shukurov}  \&
  {Wood}}{{Seta} et~al.}{2020}]{Seta2020}
{Seta} A.,  {Bushby} P.~J.,  {Shukurov} A.,   {Wood} T.~S.,  2020, arXiv
  e-prints, \href {https://ui.adsabs.harvard.edu/abs/2020arXiv200307997S} {p.
  arXiv:2003.07997}

\bibitem[\protect\citeauthoryear{{Skoutnev}, {Squire}  \&
  {Bhattacharjee}}{{Skoutnev} et~al.}{2021}]{Skoutnev2021}
{Skoutnev} V.,  {Squire} J.,   {Bhattacharjee} A.,  2021, \mn@doi [\apj]
  {10.3847/1538-4357/abc8ee}, \href
  {https://ui.adsabs.harvard.edu/abs/2021ApJ...906...61S} {906, 61}

\bibitem[\protect\citeauthoryear{Spitzer}{Spitzer}{1962}]{Spitzer1962}
Spitzer L.,  1962, Physics of Fully Ionized Gases

\bibitem[\protect\citeauthoryear{{St-Onge}, {Kunz}, {Squire}  \&
  {Schekochihin}}{{St-Onge} et~al.}{2020}]{StOnge2020}
{St-Onge} D.~A.,  {Kunz} M.~W.,  {Squire} J.,   {Schekochihin} A.~A.,  2020,
  \mn@doi [Journal of Plasma Physics] {10.1017/S0022377820000860}, \href
  {https://ui.adsabs.harvard.edu/abs/2020JPlPh..86e9003S} {86, 905860503}

\bibitem[\protect\citeauthoryear{{Verma}, {Kumar}  \& {Pandey}}{{Verma}
  et~al.}{2017}]{Verma2017}
{Verma} M.~K.,  {Kumar} A.,   {Pandey} A.,  2017, \mn@doi [New Journal of
  Physics] {10.1088/1367-2630/aa5d63}, \href
  {https://ui.adsabs.harvard.edu/abs/2017NJPh...19b5012V} {19, 025012}

\bibitem[\protect\citeauthoryear{{Vikhlinin}, {Markevitch}, {Murray}, {Jones},
  {Forman}  \& {Van Speybroeck}}{{Vikhlinin} et~al.}{2005}]{Vikhlinin2005}
{Vikhlinin} A.,  {Markevitch} M.,  {Murray} S.~S.,  {Jones} C.,  {Forman} W.,
  {Van Speybroeck} L.,  2005, \mn@doi [\apj] {10.1086/431142}, \href
  {https://ui.adsabs.harvard.edu/abs/2005ApJ...628..655V} {628, 655}

\bibitem[\protect\citeauthoryear{{Vogt} \& {En{\ss}lin}}{{Vogt} \&
  {En{\ss}lin}}{2005}]{Vogt2005}
{Vogt} C.,  {En{\ss}lin} T.~A.,  2005, \mn@doi [\aap]
  {10.1051/0004-6361:20041839}, \href
  {https://ui.adsabs.harvard.edu/abs/2005A&A...434...67V} {434, 67}

\bibitem[\protect\citeauthoryear{{Voigt} \& {Fabian}}{{Voigt} \&
  {Fabian}}{2004}]{Voigt2004}
{Voigt} L.~M.,  {Fabian} A.~C.,  2004, \mn@doi [\mnras]
  {10.1111/j.1365-2966.2004.07285.x}, \href
  {https://ui.adsabs.harvard.edu/abs/2004MNRAS.347.1130V} {347, 1130}

\bibitem[\protect\citeauthoryear{{Voit}}{{Voit}}{2011}]{Voit2011}
{Voit} G.~M.,  2011, \mn@doi [\apj] {10.1088/0004-637X/740/1/28}, \href
  {https://ui.adsabs.harvard.edu/abs/2011ApJ...740...28V} {740, 28}

\bibitem[\protect\citeauthoryear{Xu \& Kunz}{Xu \& Kunz}{2016}]{Xu2016}
Xu R.,  Kunz M.~W.,  2016, \mn@doi [Journal of Plasma Physics]
  {10.1017/S0022377816000908}, 82, 905820507

\bibitem[\protect\citeauthoryear{{Yang} \& {Reynolds}}{{Yang} \&
  {Reynolds}}{2016}]{Yang2016}
{Yang} H. Y.~K.,  {Reynolds} C.~S.,  2016, \mn@doi [\apj]
  {10.3847/0004-637X/818/2/181}, \href
  {https://ui.adsabs.harvard.edu/abs/2016ApJ...818..181Y} {818, 181}

\bibitem[\protect\citeauthoryear{{Zakamska} \& {Narayan}}{{Zakamska} \&
  {Narayan}}{2003}]{Zakamska2003}
{Zakamska} N.~L.,  {Narayan} R.,  2003, \mn@doi [\apj] {10.1086/344641}, \href
  {https://ui.adsabs.harvard.edu/abs/2003ApJ...582..162Z} {582, 162}

\bibitem[\protect\citeauthoryear{Zeldovich}{Zeldovich}{1957}]{Zeldovich1957}
Zeldovich Y.~B.,  1957, Sov. Phys. JETP, 4, 460

\bibitem[\protect\citeauthoryear{{Zhang}, {Churazov}  \&
  {Schekochihin}}{{Zhang} et~al.}{2018}]{Zhang2018}
{Zhang} C.,  {Churazov} E.,   {Schekochihin} A.~A.,  2018, \mn@doi [\mnras]
  {10.1093/mnras/sty1269}, \href
  {https://ui.adsabs.harvard.edu/abs/2018MNRAS.478.4785Z} {478, 4785}

\bibitem[\protect\citeauthoryear{{Zhuravleva} et~al.,}{{Zhuravleva}
  et~al.}{2014}]{Zhuravleva2014b}
{Zhuravleva} I.,  et~al., 2014, \mn@doi [\nat] {10.1038/nature13830}, \href
  {https://ui.adsabs.harvard.edu/abs/2014Natur.515...85Z} {515, 85}

\bibitem[\protect\citeauthoryear{{Zhuravleva}, {Churazov}, {Schekochihin},
  {Allen}, {Vikhlinin}  \& {Werner}}{{Zhuravleva}
  et~al.}{2019}]{Zhuravleva2019}
{Zhuravleva} I.,  {Churazov} E.,  {Schekochihin} A.~A.,  {Allen} S.~W.,
  {Vikhlinin} A.,   {Werner} N.,  2019, \mn@doi [Nature Astronomy]
  {10.1038/s41550-019-0794-z}, \href
  {https://ui.adsabs.harvard.edu/abs/2019NatAs...3..832Z} {3, 832}

\bibitem[\protect\citeauthoryear{{ZuHone}}{{ZuHone}}{2011}]{ZuHone2011}
{ZuHone} J.~A.,  2011, \mn@doi [\apj] {10.1088/0004-637X/728/1/54}, \href
  {https://ui.adsabs.harvard.edu/abs/2011ApJ...728...54Z} {728, 54}

\bibitem[\protect\citeauthoryear{{ZuHone}, {Markevitch}  \& {Johnson}}{{ZuHone}
  et~al.}{2010}]{ZuHone2010}
{ZuHone} J.~A.,  {Markevitch} M.,   {Johnson} R.~E.,  2010, \mn@doi [\apj]
  {10.1088/0004-637X/717/2/908}, \href
  {https://ui.adsabs.harvard.edu/abs/2010ApJ...717..908Z} {717, 908}

\bibitem[\protect\citeauthoryear{{ZuHone}, {Markevitch}, {Ruszkowski}  \&
  {Lee}}{{ZuHone} et~al.}{2013a}]{ZuHone2013a}
{ZuHone} J.~A.,  {Markevitch} M.,  {Ruszkowski} M.,   {Lee} D.,  2013a, \mn@doi
  [\apj] {10.1088/0004-637X/762/2/69}, \href
  {https://ui.adsabs.harvard.edu/abs/2013ApJ...762...69Z} {762, 69}

\bibitem[\protect\citeauthoryear{{ZuHone}, {Markevitch}, {Brunetti}  \&
  {Giacintucci}}{{ZuHone} et~al.}{2013b}]{ZuHone2013}
{ZuHone} J.~A.,  {Markevitch} M.,  {Brunetti} G.,   {Giacintucci} S.,  2013b,
  \mn@doi [\apj] {10.1088/0004-637X/762/2/78}, \href
  {https://ui.adsabs.harvard.edu/abs/2013ApJ...762...78Z} {762, 78}

\makeatother
\end{thebibliography}




\appendix

\section{\textsc{PLUTO} simulations}\label{app:PLUTO_runs}

We briefly present the results of a set of simulations performed with the fully compressible MHD code PLUTO. Our aim is to bolster our main findings obtained with SNOOPY, using a different numerical method -- finite volume rather than pseudo-spectral -- and a different physical model. Our comparison mainly focuses on 2D simulations.
The ideal MHD equations solved by PLUTO in the presence of anisotropic heat conduction, are given in Appendix A in Paper I, namely (A1)-(A4), in a plane parallel compressible atmosphere with  $\bm g = - g(z) \bm e_z$.

We look for an equilibrium profile where both $N^2$ and $\omega_T^2$ are constant (and positive) across the domain. From the definitions of the Brunt-V\"ais\"al\"a and MTI frequencies, the problem reduces to finding two functions $T(z)$ and $g(z)$ that satisfy the following set of relations:
\begin{align}
	\omega_T^2 = g(z) \left| \frac{d \ln T}{d z} \right|, \quad \omega_T^2 \left(1 + \tilde{N}^2 \right) = \frac{2 g(z)^2}{5 T(z)},
\end{align}
where $\tilde{N}^2 = N^2 / \omega_T^2$. One such choice for the temperature profile and the gravitational acceleration is 
\begin{align}\label{eq:app_grav_T}
	g(z) &=  n  g_0\left( 1 -  \frac{z}{2 H_p}\right), \quad T(z) = c_{s0}^2 \left(1 - \frac{z}{2 H_p}\right)^2,
\end{align}
where we normalized the vertical coordinate to the pressure scale-height $H_p = c_{s0}^2/g_0$, and $n = 5 (\tilde{N}^2 + 1)/2$ is a dimensionless constant. Assuming hydrostatic equilibrium, we solve for the density and pressure, which yields:
\begin{align}\label{eq:app_prs_rho}
	\rho(z) = \rho_0 \left(  1 -  \frac{z}{2 H_p} \right)^{2n -2}, \quad p(z) = \rho_0 c_{s0}^2 \left(  1 -  \frac{z}{2 H_p} \right)^{2n}.
\end{align}
With these profiles for the background quantities, we can choose to set $\omega_T = H_p = \rho_0 = 1$, from which it follows that $c_{s0}^2 = g_0 = 1 / n$, with $n$ a free parameter. One advantage of this choice over previous numerical setups for local MTI simulations \citep[e.g.,][]{Parrish2005,Parrish2007,McCourt2011} is that we are allowed to vary the relative strength of the entropy to the temperature gradient, given by $\tilde{N}$, which has proven to be an essential parameter in understanding the dynamics of the MTI. Moreover, setting the vertical size of the box to be $L_z = 0.1 \ll 1$, we ensure that the background fluid quantities do not change significantly over the domain, thus allowing us to compare local simulations with PLUTO with the results obtained with SNOOPY.

We perform a set of 2D simulations varying both the thermal conductivity and the entropy stratification. The runs were performed at a resolution of $(512)^2$, using a RK3 timestepping algorithm, and WENO3 method for spatial reconstruction. Calculations were considerably sped-up with the use of a Runge-Kutta Legendre time-stepping scheme for anisotropic thermal conduction. We did not compute viscous and resistive dissipation explicitly. We chose periodic boundary conditions in the horizontal direction, and reflecting boundary conditions in the vertical direction, fixing the temperature at the top and bottom surfaces and extrapolating the pressure and density in the ghost cells to maintain hydrostatic equilibrium. Similarly to previous simulations of the MTI, we find that, in order to keep a steady heat flux through the domain, it is necessary to "sandwich" the MTI-unstable layer, characterized by the profiles in Eqs.~\eqref{eq:app_grav_T}-\eqref{eq:app_prs_rho} and endowed with purely anisotropic heat conduction, between two stable layers where the temperature is constant and equal to the value at the boundary and where thermal conduction is purely isotropic. Choosing the isotropic conductivity to be $5 \kappa_{\parallel}$ ensures that the stable layers remain isothermal throughout the simulation. The vertical extent of these "boundary layers" is $1/16$ of the domain each, which we have found to be sufficient.

\begin{figure}
	\centering
	\includegraphics[width=0.9\columnwidth]{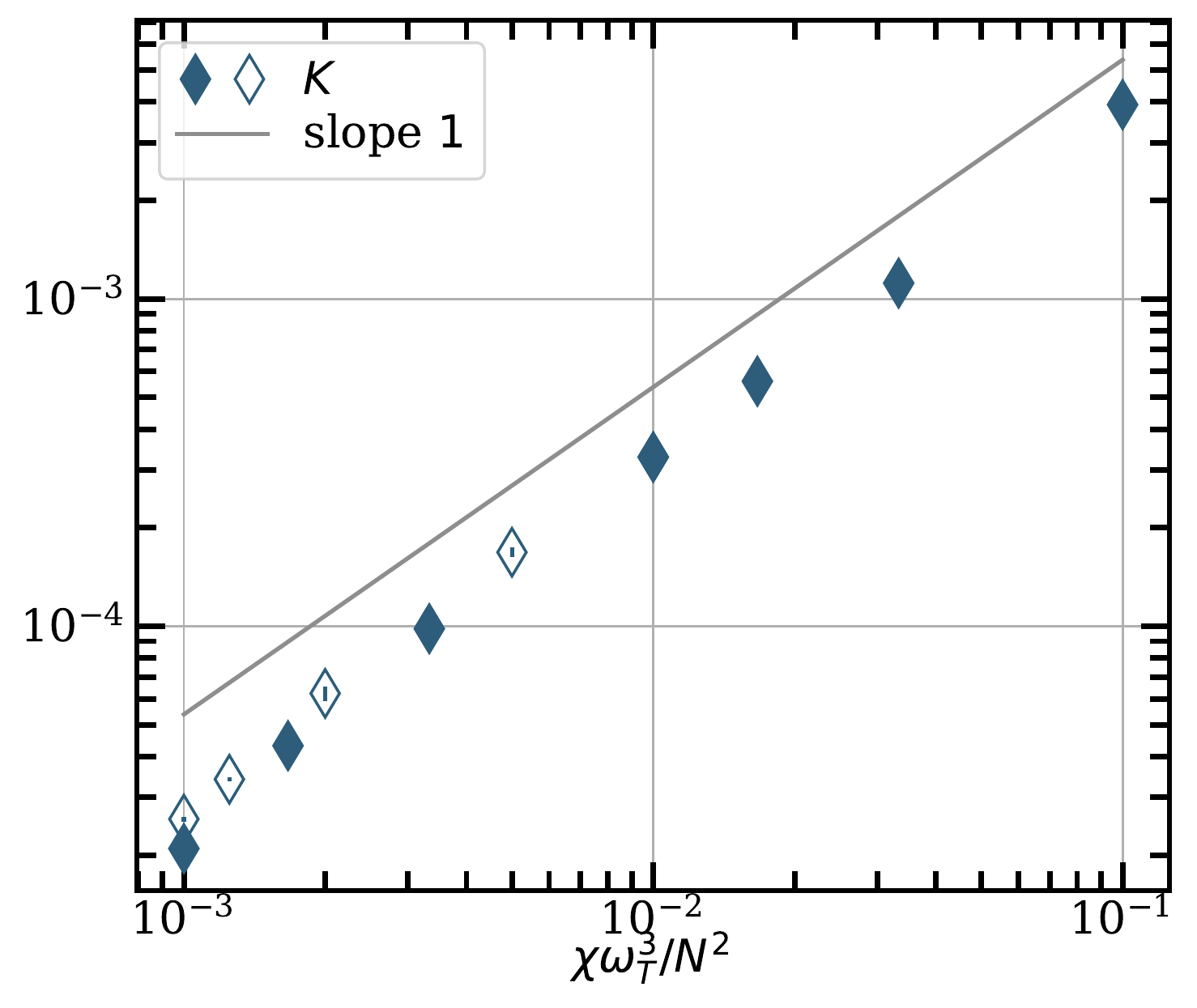}
	\caption{ Plot of the turbulent kinetic energy of the MTI at saturation for PLUTO runs varying either the thermal conductivity (full diamonds) or the entropy stratification (empty diamonds). The kinetic energy has been normalized to the size of the box, while the grey solid line is plotted for reference and represents the theoretical scalings derived in Paper I.  }
	\label{fig:app_pluto_scaling}
\end{figure}

We initialize the simulations with a homogeneous horizontal magnetic field of $\beta = 10^{10}$, and with random velocity fluctuations with amplitude $\delta u = 10^{-4} c_s$. The results of our parameter sweep are shown in Fig.~\ref{fig:app_pluto_scaling}, where we show the saturated kinetic energy as a function of $\chi \omega_T^3 / N^2$, the theoretical scaling derived in Paper I in the Boussinesq approximation. As we can see, there is very good agreement with the Boussinesq scaling law, with a slight departure visible at lower values of $\chi$, where grid effects become more important.

\section{Other field configurations }\label{app:other_field}

We show the results of dynamo runs with a variety of initial magnetic field  conditions. This extends the exploration in Sec. 4.5 in Paper I to three dimensions.

\subsection{No net-flux and weak vertical magnetic field}

\begin{figure}
	\centering
	\includegraphics[width=1.0\columnwidth]{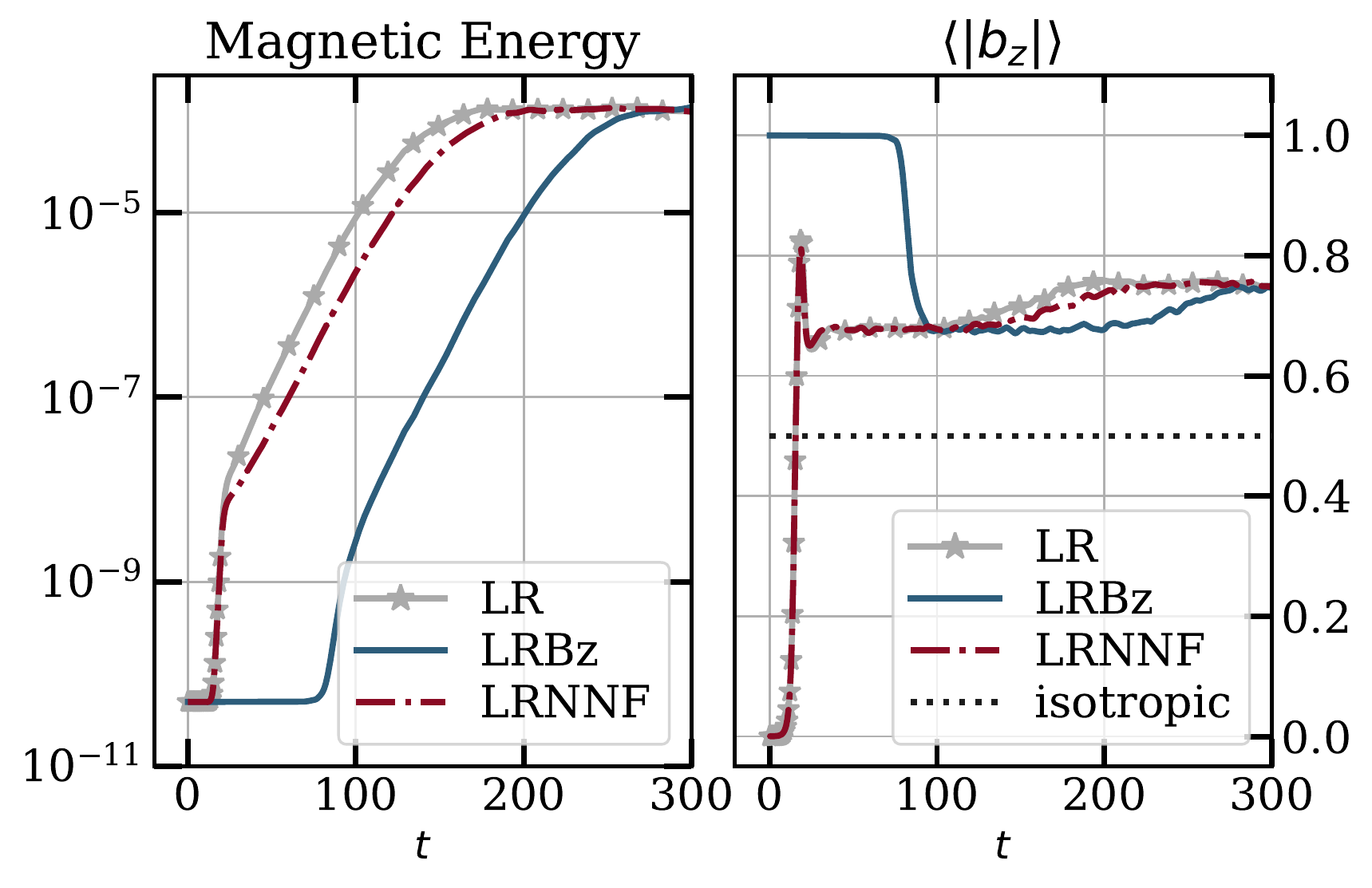}
	\caption{Magnetic energy (left panel) and average orientation of the magnetic field (right panel) in runs LR (weak uniform horizontal field; gray stars), LRBz (weak uniform vertical field; solid blue line) and LRNNF (no-net-flux horizontal field; dash-dotted red line).}
	\label{fig:other_compare_Bz_mag_energy}
\end{figure}

We first compare the results of three lower-resolution runs ($(256)^3$): the first run (LR) is initialized similarly to the fiducial run, with a weak and uniform horizontal magnetic field; the second (LRNNF) with no-net flux horizontal field; and the third (LRBz) with a weak and uniform vertical field. The initial strength of the magnetic field is the same in all runs and equal to $B_0 = 10^{-5}$. We recall that a vertical magnetic field configuration, while being linearly stable to the MTI, can be destabilized with a sufficiently large initial perturbation \citep{McCourt2011}. In Paper I, we showed that in 2D all these initial conditions eventually arrived at the same saturated turbulent state, apart from the zero-net flux run, in which the magnetic field decayed. Fig.~\ref{fig:other_compare_Bz_mag_energy} indicates that this is also the case in 3D: the magnetic energy levels and the orientation of the field converge to the same values, even if their time evolution follows different trajectories. The saturated kinetic and potential energies are also equivalent.

Next, we present a set of runs with no net-flux and with varying initial strengths of the vertical or horizontal magnetic field: $B_0 = \left\lbrace 10^{-4}, 10^{-3}, 10^{-2} \right\rbrace$. We find that at saturation the energy levels are practically equivalent across all the runs.

These results confirm (a) that a small initial net flux does not substantially affect the saturation properties of the MTI in three dimensions; and (b) the generality of the dynamo-dominated state as the end state of the MTI under a variety of initial conditions. The only exception, as we shall see in the next section, is the case of a strong uniform magnetic field.

\subsection{Strong initial magnetic field}

\begin{figure}
	\centering
	\includegraphics[width=1.0\columnwidth]{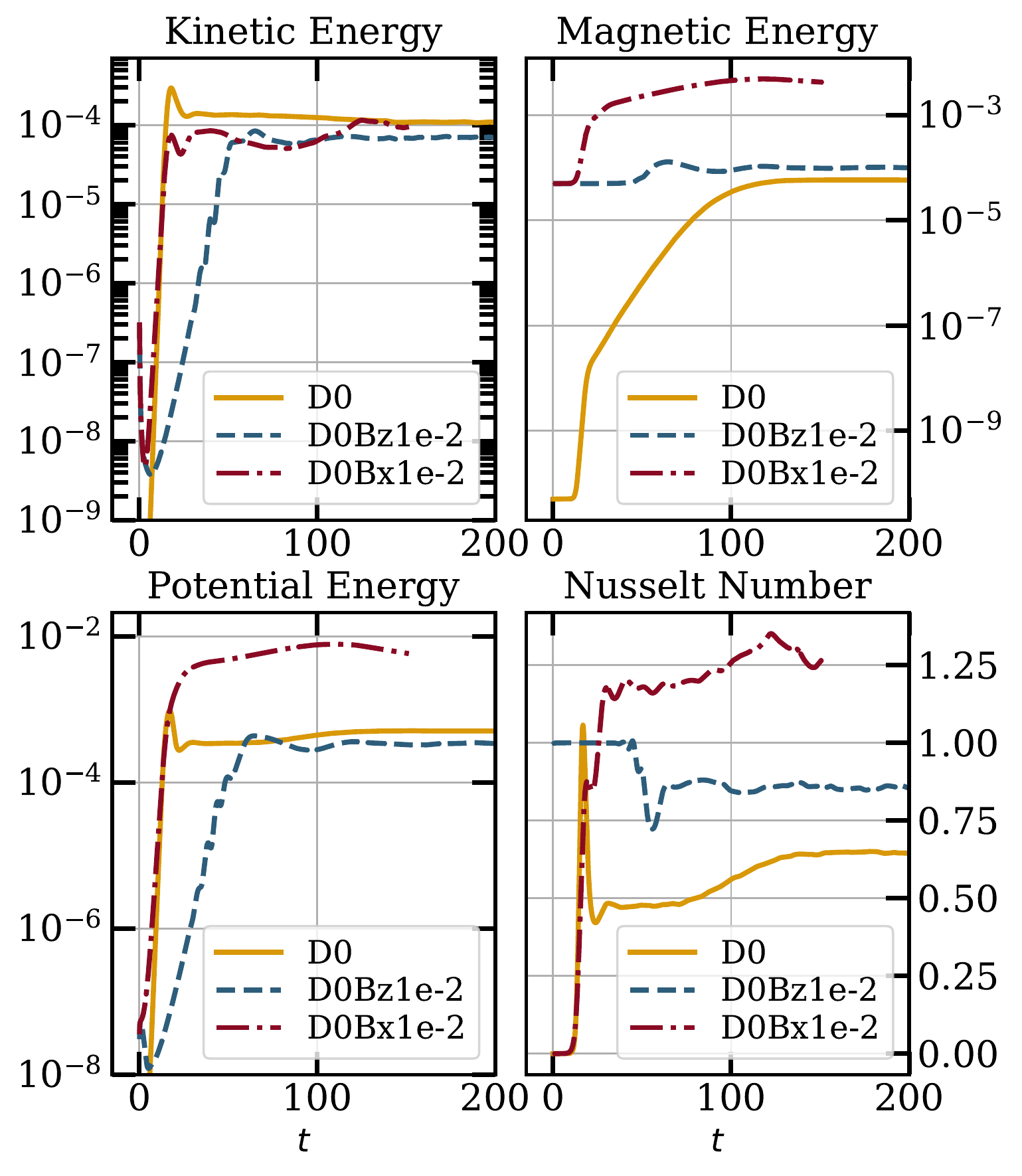}
	\caption{Time evolution of the kinetic energy (top-left), magnetic energy (top-right), potential energy (bottom-left), and Nusselt number (bottom-right) of run D0Bx1e-2 and D0Bz1e-2, in the dash-dotted red line and the dashed line, respectively. For comparison, we also show the fiducial run D0 in the solid gold line.  }
	\label{fig:strongB_compare_kin_mag_energy}
\end{figure}

We perform two high-resolution simulations, run D0Bx1e-2 and run D0Bz1e-2, with the same parameters as run D0 with the exception of the magnetic field, which has now strength $B_0 = 10^{-2}$ and is aligned in the $x$ and $z$ directions, respectively. As shown in Fig.~\ref{fig:strongB_compare_kin_mag_energy}, the two runs are characterized by a very different evolution: after the initial MTI growth phase, run D0Bx1e-2 further amplifies its magnetic fields to roughly $\times 100$ times the kinetic energy, with a significant increase in the potential energy; run D0Bz1e-2, on the other hand, exhibits magnetic and potential energy at levels comparable to run D0. The kinetic energies of the three runs, in contrast, are roughly comparable throughout. The behaviour of run D0Bx1e-2 is reminiscent of the strong $B$ case in 2D, which we examined in Paper I, where the stronger initial horizontal field produces large-scale vertical structures and enhances the vertical anisotropy of the flow ($\langle |b_z| \rangle \gtrsim 0.9$ in both runs). The vertical bias is accompanied by an increased efficiency of vertical heat transport compared to the fiducial run, as can be seen from the plot of $Nu$ in Fig.~\ref{fig:strongB_compare_kin_mag_energy}.

\begin{figure*}
	\begin{subfigure}{.5\linewidth}
		\centering
		\includegraphics[width=1.0\columnwidth,trim=0 0 400 0,clip]{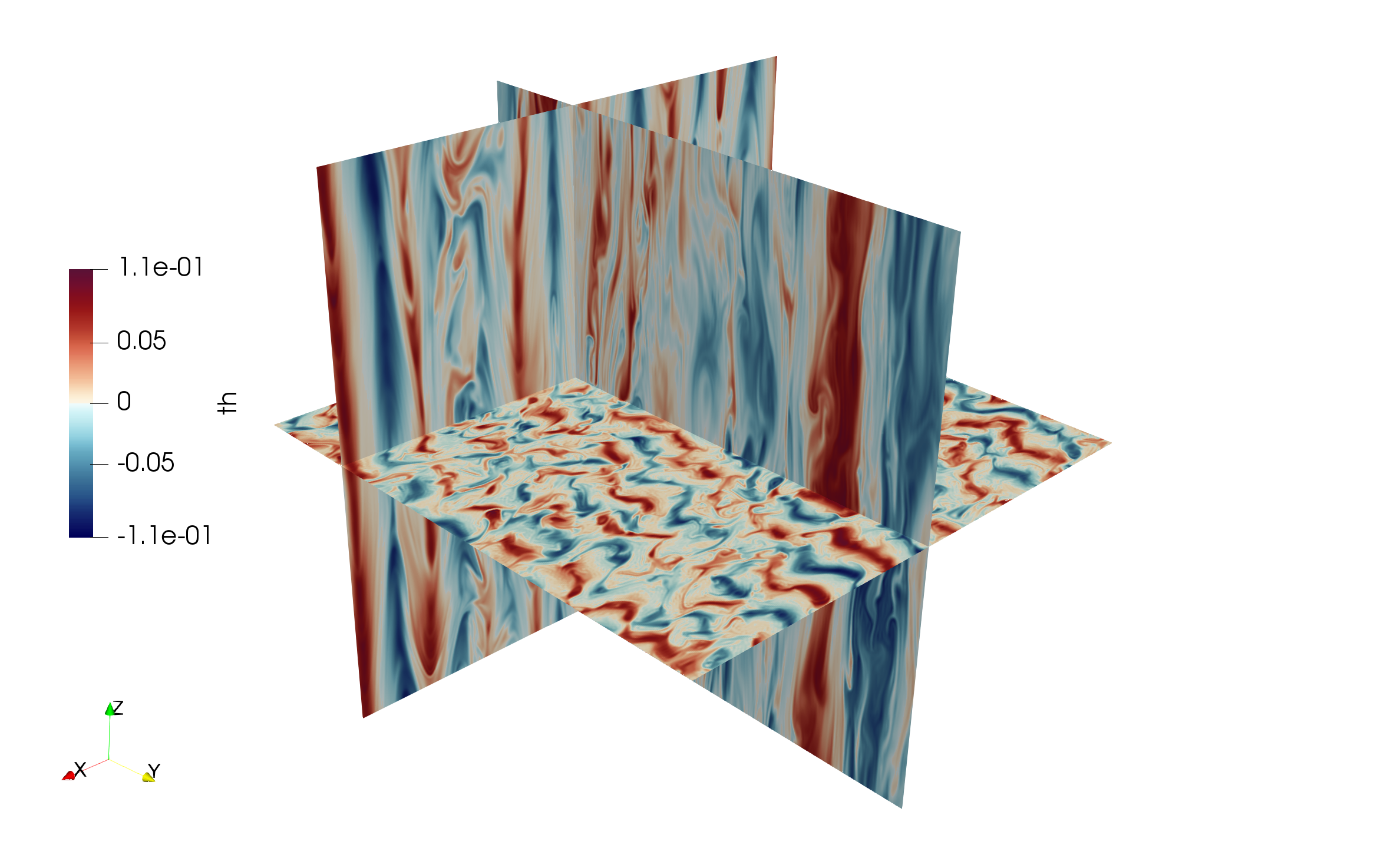}
		\caption{}
		\label{fig:other_strongB_density_horizontal}
	\end{subfigure}%
	\begin{subfigure}{.5\linewidth}
		\centering
		\includegraphics[width=1.0\columnwidth,trim=0 0 450 0,clip]{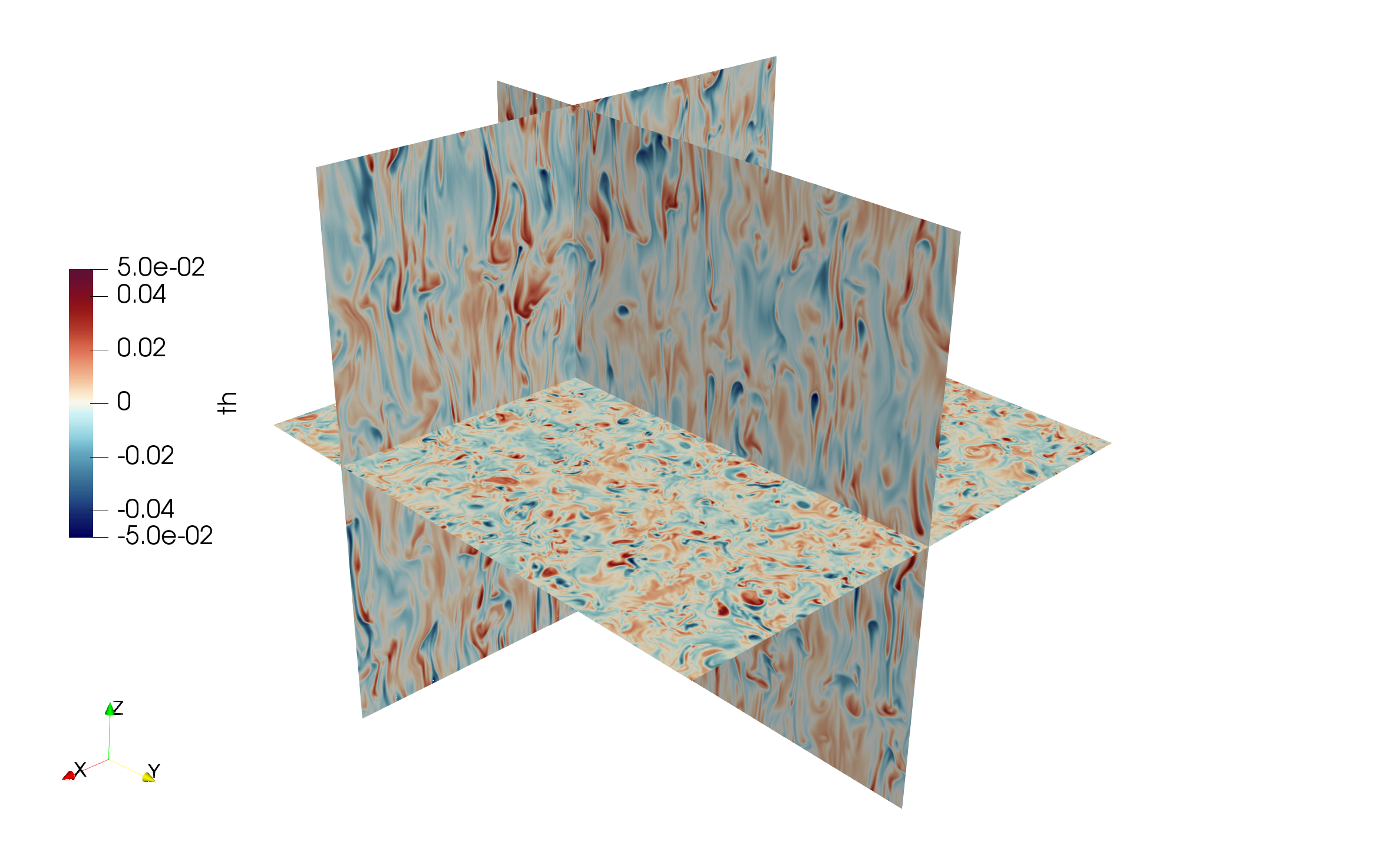}
		\caption{}
		\label{fig:other_strongB_density_vertical}
	\end{subfigure}%
	\caption{Three dimensional visualization of the buoyancy field $\theta$ for run D0Bx1e-2 at time $t=60$ (left panel) and D0Bz1e-2 at time $t=140$ (right panel).}
	\label{fig:other_strongB_density}
\end{figure*}

After the MTI growth phase, run D0Bx1e-2 features structures at the box size that show very little variation in the $z$ direction, and that resemble alternating upwards and downwards streams of fluid (see Fig.~\ref{fig:other_strongB_density}). This peculiar arrangement is the result of the initial MTI phase, during which the horizontal magnetic field is stretched into a vertical undulated pattern: after the MTI has saturated, the magnetic tension is large enough that the subsequent turbulence cannot significantly perturb the field. Note, however, that run D0Bx1e-2 has not reached saturation yet and it is likely that this regime cannot adequately be captured by our numerical setup due to the presence of structures at the box size, requiring global simulations.

The presence of long-lived magnetic structures at the box size allows the formation of  large-scale temperature correlations along the magnetic field lines ($k_{\parallel}^{\theta} \approx 12$, see Table~\ref{tab:table_runs_scales}), that explain the larger potential energy of run D0Bx1e-2 compared to the fiducial run. Run D0Bz1e-2, on the other hand, does not show evidence of such large-scale structures in the magnetic field.

\bsp	
\label{lastpage}
\end{document}